\documentclass[superscriptaddress,a4paper,twocolumn, floatfix, longbibliography, aps, prx]{revtex4-2}
\usepackage[english]{babel}
\usepackage[utf8]{inputenc}
\usepackage{amsmath}
\usepackage{amssymb}
\usepackage[colorinlistoftodos, color=green!40, prependcaption]{todonotes}
\usepackage{amsthm}
\usepackage{mathtools}
\usepackage{physics}
\usepackage{braket}
\usepackage{xcolor}
\usepackage{graphicx}
\usepackage{subcaption}
\usepackage[left=23mm,right=13mm,top=35mm,columnsep=15pt]{geometry} 
\usepackage{adjustbox}
\usepackage{placeins}
\usepackage[T1]{fontenc}
\usepackage{newtxtext, newtxmath, newtxtt}
\usepackage{lipsum}
\usepackage{csquotes}

\usepackage{xcolor}
\definecolor{codegreen}{rgb}{0,0.6,0}
\definecolor{codegray}{rgb}{0.5,0.5,0.5}
\definecolor{codepurple}{rgb}{0.58,0,0.82}
\definecolor{tqblue}{HTML}{08293d}
\definecolor{backcolour}{HTML}{fefdf5}
\usepackage{listings}
\lstdefinestyle{mystyle}{
    backgroundcolor=\color{backcolour},   
    commentstyle=\color{codegreen},
    keywordstyle=\color{magenta},
    numberstyle=\tiny\color{codegray},
    stringstyle=\color{codepurple},
    basicstyle=\ttfamily\footnotesize\color{tqblue},
    breakatwhitespace=false,         
    breaklines=true,
    postbreak=\mbox{\textcolor{magenta}{$\hookrightarrow$}\space},                 
    captionpos=b,                    
    keepspaces=true,                 
    numbers=left,                    
    numbersep=5pt,                  
    showspaces=false,                
    showstringspaces=false,
    showtabs=false,                  
    tabsize=2
}

\usepackage[pdftex, pdftitle={Article}, pdfauthor={Author}]{hyperref} 

\hypersetup{
    colorlinks,
    linkcolor={red!50!black},
    citecolor={blue!50!black},
    urlcolor={blue!80!black}
}

\usepackage{booktabs}

\usepackage{import}
\usepackage{xifthen}
\usepackage{transparent}

\newcommand{\trot}{\ensuremath{[2]_{\textrm{R12}}~}}
\newcommand{\e}{\ensuremath{\mathrm{e}}}
\newcommand{\bas}[1]{\texttt{#1}}

\newcommand{\rv}[2]{{#2}}

\begin{document}
\title{Improving the Accuracy of the Variational Quantum Eigensolver for Molecular Systems by the Explicitly-Correlated Perturbative \trot-Correction}

\author{Philipp Schleich}\textbf{}
\email[E-Mail:]{philipps@cs.toronto.edu}
\affiliation{Department of Computer Science, University of Toronto, Canada.}
\affiliation{Applied and Computational Mathematics, Department of Mathematics, RWTH Aachen University, Aachen, Germany}
\affiliation{Vector Institute for Artificial Intelligence, Toronto, Canada.}

\author{Jakob~S.~Kottmann}\textbf{}
\email[E-mail:]{jakob.kottmann@utoronto.ca}
\affiliation{Department of Computer Science, University of Toronto, Canada.}
\affiliation{Chemical Physics Theory Group, Department of Chemistry, University of Toronto, Canada.}

\author{Al\'{a}n Aspuru-Guzik}
\email[E-mail:]{aspuru@utoronto.ca}
\affiliation{Department of Computer Science, University of Toronto, Canada.}
\affiliation{Chemical Physics Theory Group, Department of Chemistry, University of Toronto, Canada.}
\affiliation{Vector Institute for Artificial Intelligence, Toronto, Canada.}
\affiliation{Canadian  Institute  for  Advanced  Research  (CIFAR)  Lebovic  Fellow,  Toronto,  Canada}

\date{\today} 

\begin{abstract}
We provide an integration of the universal, perturbative explicitly correlated \trot-correction in the context of the Variational Quantum Eigensolver (VQE). This approach is able to increase the accuracy of the underlying reference method significantly while requiring no additional quantum resources. 
The proposed approach only requires knowledge of the one- and two-particle reduced density matrices (RDMs) of the reference wavefunction; these can be measured after having reached convergence in the VQE. 
This computation comes at a cost that scales as the sixth power of the number of electrons.
We explore the performance of the VQE+\trot approach using both conventional Gaussian basis sets and our recently proposed directly determined pair-natural orbitals obtained by multiresolution analysis (MRA-PNOs). Both Gaussian orbital and PNOs are investigated as a potential set of complementary basis functions in the computation of \trot. In particular the combination of MRA-PNOs with \trot has turned out to be very promising -- persistently throughout our data, this allowed very accurate simulations at a quantum cost of a minimal basis set. Additionally, we found that the deployment of PNOs as complementary basis can greatly reduce the number of complementary basis functions that enter the computation of the correction at a complexity.  
\end{abstract}

\maketitle
\section{Introduction}

Quantum computing is an emergent computational paradigm with the potential to disrupt some areas of scientific and financial computing. While recent advances in the development of the hardware bring the time of error-corrected quantum computers closer~\cite{Arute2019,AIQuantum2020,Zhong2021}, current quantum computers, often called NISQ (noisy intermediate scale quantum computers), still do not offer practically useful computational advantage given their lack of noise resilience, short coherence times and a rather low, yet steadily increasing, available number of qubits~\cite{Preskill2018}.

Popular approaches to NISQ~computing are constituted by hybrid quantum-classical procedures, which rely on the interplay of a classical and a quantum computer. One important class of such hybrid quantum-classical algorithms is given by variational quantum algorithms (VQAs) that rely on the definition of a parametric cost function, which is evaluated on a quantum computer, and then optimized numerically on a classical computer.\cite{cerezo2020variational, bharti2021noisy} In the context of quantum chemistry, particularly ground-state calculations, this cost function is set to be the energy expectation value of a molecular system.

To make algorithms more suitable for NISQ~hardware, its weaknesses on the hardware end are usually addressed by error mitigation techniques, more efficient mappings of the underlying problem to the quantum computer's language or compilation strategies; see Refs.~\cite{bharti2021noisy,endo2021hybrid} for an extensive overview.
As an alternative, or in particular as an addition to these efforts, one can -- just as in classical computing -- exploit some physical knowledge to make the simulation more efficient. In this context, one aims to reduce quantum resources (in terms of qubits needed and gate count / circuit depth), usually coming at the cost of some additional computational efforts on a classical machine.\cite{McClean2017,Takeshita2020,Motta2020,Kottmann2020a,zhang2021mutual} 
Efforts to do so include \emph{a priori} and \emph{a posteriori} computations, which either pre-modify the input for an existing quantum procedure or make use of the output to obtain an external correction, both with the ultimate goal to increase accuracy while maintaining the amount of quantum resources necessary. Conversely, this can be translated into a reduction of quantum resources for a certain level of accuracy.

In this work, we demonstrate a procedure that allows to reduce the \rv{}{number of} qubits required for a certain quantum chemical  quantum calculation involving electronic systems, that allow for an efficient output of the one- and two-particle reduced density matrices (1- and 2-RDM). This perturbative correction is based on methods usually summarized as ``R12'', ``F12'' or explicitly correlated methods, cf. section~\ref{sec:expcorrel}.
Recently, \emph{a priori} modifications inspired by Boys' and Handy's transcorrelated Hamiltonian~\cite{Boys1969a,Boys1969b} -- \rv{a similarity-transformed $ H' = \e^{-G} H \e^G $ with $G$ such that $H'$ is cusp-free}{where a similarity-transformation of $H$ renders $ H' = \e^{-G} H \e^G $ cusp-free given an appropriate $G$} -- have been adapted for the use in near-term quantum simulations. McArdle et. al.~\cite{McArdle2020} apply Boys' and Handy's original formulation with a non-Hermitian $H'$ within the imaginary time evolution ansatz~\cite{McArdle2019}, whereas Motta et. al.~\cite{Motta2020} implement a regularized Hamiltonian straightforwardly in a VQE simulation. The latter makes use of the canonical transcorrelation introduced in Ref.~\cite{Yanai2012}, which modifies the Hamiltonian as $H'=\e^{A^\dagger} H \e^{A}$ with some anti-Hermitian $A$ such that $H'$ is still Hermitian; the Baker-Campbell-Hausdorff-expansion of $H'$ is truncated so that only one- and two-body terms are kept.

On the contrary, we opt to use an \emph{a posteriori} correction based on a perturbative \emph{explicitly correlated} method, namely the spin-free variant of \trot~\cite{Torheyden2009,Kong2011,Roskop2014}. This approach is to be contrasted with other post-corrections in form of the family of quantum subspace expansion \rv{}{(QSE)} techniques~\cite{McClean2017,McClean2020,Takeshita2020}; an application of VQE together with the transcorrelated approach, \trot or a so called ``CABS~singles'' correction~\cite{Kong2010a} has already been suggested in Ref.~\cite{Cao2018}. 
Brief comments on the distinction of our approach with this one will follow in Section~\ref{sec:trot}.

Aside of providing a framework for the application of the explicitly correlated \trot-method within the variational quantum eigensolver, the goal of this work is to raise additional awareness of explicitly correlated methods within the quantum computing community. 
This work builds on a recent Master's thesis~\cite{schleich2020regularization} and contains updated data.

\paragraph*{A comment on notation}
Within this work, we will make use of the Einstein summation convention, where the indices are clarified in Tab.~\ref{tab:indices}. Frequently we label orbitals only with their indices $ \{p\}\equiv\{\phi_p\} $.  For second-quantized creation and annihilation operators as well as matrix elements, we follow the notation of the works by Kutzelnigg and Mukherjee, which is prevalent in the literature of explicitly correlated methods (defined e.g. in Refs.~\cite{Kutzelnigg1997,Kutzelnigg2003}). Excitation operators and matrix elements up to order two can be found in Tab.~\ref{tab:indices} as well.
Higher-order excitations and matrix elements can be defined in a similar manner but are not needed in this context.
If not stated otherwise, we make use of atomic units $\hslash=e=m_\mathrm{e}=1$ except for {\AA}ngstrom for the unit of length. 

\section{Background}
\rv{}{For the sake of self-containment, we will first give a high-level description of explicitly correlated techniques and a historical overview before presenting and motivating the \trot-correction in more detail. After that, we will propose a workflow to embed \trot into the hybrid quantum-classical variational quantum eigensolver routine. 
}
\subsection{Explicit Correlation}\label{sec:expcorrel}
\subsubsection{Basics}
It is well-known that the wavefunction of an electronic system exhibits a \emph{correlation cusp} at the points of inter-electronic coalescence~\cite{Kato1957,Lakin1965,Pack1966,Kutzelnigg1992}. A generic numerical representation of such kinks in the functions with smooth functions imposes problems and leads to a rather high number of basis functions to represent the true wavefunction up to a certain accuracy. Yet it is rather obvious that an explicit incorporation of the inter-electronic distance $r_{12}=\lvert r_1 - r_2 \rvert$ helps to overcome this slow convergence, as demonstrated for the simple case of the ground-state of Helium by Hylleraas~\cite{Hylleraas1929},   later again proposed~\cite{Kato1957,Hirschfelder1963} and supported by rigorous work~\cite{Schwartz1963,Hill1985,Kutzelnigg1992}.
 The essential take-away message can be formulated as follows: Upon occurrence of electron-electron correlation, the deployment of explicitly correlated basis function leads to more accurate energies, given that both the partial-wave expansion~\cite{Schwartz1963,Kutzelnigg1992} as well as a natural-orbital expansion~\cite{Kutzelnigg2008}  decay faster in such a basis (\rv{$\leftrightarrow$}{thus} fewer basis functions needed for the same level of accuracy). 
For a comprehensive overview of explicitly correlated methods, we refer to the reviews~\cite{Kong2012,Klopper2006,Ten-no2012,Gruneis2017}.

However, making use of an explicitly correlated basis in the existing methods (such as CI, leading to so called Hylleraas CI~\cite{sims1971hyllci}) does not allow an efficient implementation due to the appearance of very high-dimensional integrals~\cite{Cances2003,Kong2012}. One of the two most prominent ways to exploit explicit correlation is given by perturbative descendants of Kutzelniggs R12-theory~\cite{Kutzelnigg1985,Klopper1987,Klopper1991,Kutzelnigg1991,Kutzelnigg1992}, where MP2-R12 can be seen as the basis for all upcoming, more sophisticated, perturbation methods such as \trot or the approach in Ref.~\cite{Ten-no2007correction}. The other one consists of regularizing the system throughout a similarity-transformed Hamiltonian via Boys and Handy's transcorrelated approach~\cite{Boys1969,Boys1969a}, made more applicable throughout approximations such that the effective Hamiltonian is still Hermitian in Refs.~\cite{Luo2010,Yanai2012}. In a recent benchmark~\cite{Kersten2016}, both the perturbative \trot-correction, that we will explain in detail in section~\ref{sec:trot} and the canonical transcorrelated approach from Ref.~\cite{Yanai2012} performed comparably well.
Here, we focus on the universal perturbative \trot method and will provide a discussion to distinguish the approach from the transcorrelated Hamiltonian when applied to quantum algorithms.

Initially, the explicitly correlated basis has been built by correlating one-electron basis functions, achieved through multiplication with the inter-electronic distance $r_{12}$. Motivated by that fact, that a simple factor of $r_{12}$ does not inherit the correct asymptotics, Gaussian correlation factors $\exp{-r_{12}^2}$ (\emph{Gaussian geminals}~\cite{boys1960integral,singer1960gaussian}) and Slater-type correlation factors $\exp\{ -r_{12}\}$~\cite{Kutzelnigg1985,Ten-no2004slater} have been introduced. Within a study that investigated the performance of different correlation factors, it was found that Slater-type factors often have the most favorable behaviour~\cite{Johnson2017}. 
Hence following the \trot-literature, we use correlation factors and its matrix elements given as follows:
\begin{equation} \label{eq:f12-integ-correlfactor}
    f_{12} = -\frac{1}{\gamma} \e^{-\gamma r_{12}}, \quad r^{\kappa \lambda}_{\mu \nu} = \braket{\mu \nu | f_{12} | \kappa \lambda}. 
\end{equation}
Note that when using \texttt{Psi4}~\cite{Smith2020} in our simulations, the correlation factor $f_{12}$ is represented as a linear combination of Gaussians.

\subsubsection{\trot-Correction}\label{sec:trot}
Further on, we provide a brief outline to \trot as in Refs.~\cite{Torheyden2009,Kong2011,Roskop2014}, recalling that this is a perturbative explicitly correlated method.  

Let $\mathcal{H}$ be the Hilbert space of the true wavefunction of a certain electronic system with $N_\mathrm{el}$ \rv{}{electrons} in Born-Oppenheimer approximation, described by its Hamiltonian operator 
\begin{equation}
H=h^q_p a^p_q + \frac{1}{2} g^{rs}_{pq} a^{pq}_{rs},
\end{equation}
where $h^p_q$ denotes one-electron integrals and $g^{pq}_{rs}$ two-electron integrals over the Coulomb interaction:
\begin{equation}\label{eq:coulomb-int}
    g^{pq}_{rs} = \braket{rs | {(r_{12})}^{-1} | pq}.
\end{equation}

\newcommand{\kvref}{\ket{\Phi_\mathrm{ref}}}
\newcommand{\vref}{\Phi_\mathrm{ref}}
Further we have a \textit{reference} solution $\kvref$ using an arbitrary method to find an approximate solution to the ground-state problem for $H$  using a certain basis consisting of a \emph{finite} number of one-electron orbitals $\{p\}_{p=1}^{ N_\mathrm{OBS}}$ that span $\mathcal{P}\subset \mathcal{H}$.
The perturbative correction we use here aims to correct for the incompleteness of this one-electron basis by excitations into the complementary space $\mathcal{Q} = \mathcal{H}\setminus\mathcal{P}$.
Note that formally, up to this point, the procedure \rv{would be}{is} equivalent to the Quantum Subspace Expansion~\cite{McClean2017,McClean2020,Takeshita2020}.
Yet here, the basis for the complementary space is enforced to be explicitly dependent on the inter-electronic coordinate and thus is deemed to exhibit faster convergence; it is tailored to account specifically for completeness due to the correlation cusp. Conversely, we expect that for systems where the dominant error originates from insufficient representation of the correlation cusp, the present method needs less basis functions for a qualitative representation of the complementary space than  QSE-based procedures with a primitive basis and is thus likely to be more efficient. On the other hand, the application of QSE is not limited to quantum chemistry~\cite{McClean2020}.
It is worth mentioning that all \rv{these}{} perturbative methods relying on the RDMs require the evaluation of complex equations (e.g. via the generalized Wick's theorem~\cite{Kutzelnigg1997}), as well as a set of non-trivial cumulant approximations to avoid measurement of higher-order RDMs~\cite{Kutzelnigg1999,Kutzelnigg2010,Torheyden2009,McClean2017}. While the QSE depends on the solution of a generalized eigenvalue problem, \trot amounts in the \textsc{sp}-ansatz~\cite{Ten-No2004}, which fixes amplitudes so that they fulfill the cusp condition as shown in Eq.~\eqref{eq:sp-ansatz}, to a set of tensor contractions.

\begin{table}[ht]
    \centering
    \caption{Glossary (operator and index conventions, acronyms sorted by subject matter)}
    \begin{footnotesize} 
    \begin{tabular}{p{4.5cm}p{3.5cm}}
    \hline
        Indices ($\{p\} \equiv \{\phi_p\}, \ket{p}\equiv \ket{\phi_p}$) &  Explanation \\ 
        \hline
         $\{p,q,r,s\}$ & Orbital basis set (OBS) for orbital space $\mathcal{P}$ / reference\\
         $\{i,j\}$ & Occupied within OBS  \\
         $\{a,b\}$ & Unoccupied within OBS  \\
         $\{\kappa, \lambda\} \rightsquigarrow \{\kappa', \lambda'\}$ & Formally complete basis and finite RI basis to approximate full Hilbert space $\mathcal{H}$ with RI space \\
         $\{\alpha, \beta\}\rightsquigarrow \{\alpha',\beta'  \}$ & Complementary basis and finite complementary auxiliary basis (CABS) to resolve $\mathcal{Q} = \mathcal{H}\setminus\mathcal{P}$  \\
         \hline
         Operators and integrals & Following e.g.~\cite{Kutzelnigg2003} \\
         \hline
         $a^p = a_p^\dagger , \quad a_p$ & Fermionic creation and annihilation operators \\
         $a^p_q = a^p a_q,  {  } a^{pq}_{rs} = a^pa^q a_sa_r, \ldots $ & Particle-number conserving excitation operators \\
         $X^p_q = \braket{q(1) | X(1) | p(1)}$ &  One-electron integrals \\
         $X^{pq}_{rs} = \braket{r(1)s(2) | X(1,2) | p(1)q(2)}$ & Two-electron integrals  \\
         $g^{pq}_{rs}$ & Coulomb integrals, cf. Eq.~\eqref{eq:coulomb-int} \\
         $r^{pq}_{rs} $ & F12 integrals, cf.~Eq.~\eqref{eq:f12-integ-correlfactor} \\
         $\gamma^p_q, \gamma^{pq}_{rs}$ & One- and two-particle reduced density matrices (1- and 2-RDM), cf.~Eq.~\eqref{eq:rdms} \\
         $f^\kappa_\lambda$ & Generalized Fock operator, cf.~Eq.~\eqref{eq:genfock}\\
         \hline
         Acronyms & Explanation and Reference \\
         \hline
         NISQ & Noisy intermediate-scale quantum (computing) \cite{Preskill2018,bharti2021noisy} \\
         VQA/VQE & Variational Quantum Algorithms/Eigensolver \cite{cerezo2020variational,Peruzzo2014} \\
         PNO & Pair-natural orbitals \cite{Kottmann2020,Kottmann2020a} \\
         MRA & Multiresolution analysis \cite{Harrison2016} \\
         RDM & Reduced density matrix \\
         QSE & Quantum subspace expansion \cite{McClean2017,McClean2020,Takeshita2020} \\
         OBS & Orbital basis set \cite{Valeev2004a} \\
         CBS & Complete basis set \cite{Valeev2004a}\\
         CABS & Complementary auxiliary basis set \cite{Valeev2004a}\\
         RI & Resolution of the identity \cite{Valeev2004a}\\
         \hline
    \end{tabular}
    \end{footnotesize} 
    \label{tab:indices}
\end{table}

The \trot correction then can formally be derived\footnote{This is one way to derive it. One could also blindly start with some perturbation and the second-order Hylleraas functional, interpreting the RI-approximation in the usual sense to break down higher-order integrals. But since this approach well showcases where the correction for \emph{basis-set incompleteness} comes from, we choose to sketch it here. A more thorough outline can be found in Ref.~\cite{schleich2020regularization}.} using L\"owdin partitioning~\cite{Lowdin1962,Valeev2008,Torheyden2008,Torheyden2009}. Then, one can approximate the full complementary space $\mathcal{Q}=\mathrm{span}\{\alpha\}$ using a complementary auxiliary basis set (CABS)  $\{\alpha'\}_{\alpha'=1}^{N_\mathrm{CABS}}$ with finite $N_\mathrm{CABS}$ -- this CABS basis can be an arbitrary basis of functions and does not need to come from a Gaussian basis set, even if the OBS is built from Gaussians.
Note that here, this approximation is formally equivalent to the resolution of the identity approximation in the evaluation of molecular integrals of usual R12/F12 methods (usually abbreviated by RI, insertion of $I =  \ket{\kappa}\bra{\kappa} \approx  \ket{\kappa'}\bra{\kappa'} $ to avoid higher-order integrals). 
We construct the RI basis as $\{\kappa'\}=\{p\}\cup\{\alpha'\}$, i.e., as the union of the orbitals $\{p\}$ resulting from the regular orbital basis and $\{\alpha'\}$ coming from the complementary auxiliary basis. The partitioning into a reference $\kvref\in\mathcal{P}$ and a perturbation (to account for incompleteness due to $\mathcal{Q}$) then leads to a correction given by the minimization of a second-order Hylleraas functional 
\begin{equation}
  J_H^{(2)}=\braket{\psi^{(1)} | H^{(0)} | \psi^{(1)}   } + 2 \braket{\psi^{(0)} | H^{(1)}| \psi^{(1)}  },  
\end{equation}
named after Hylleraas' general approach to perturbation theory~\cite{Hylleraas1929}.
The reference Hamiltonian corresponds to the model Hamiltonian $H^{(0)}=H$, and as zeroth-order state, the solution obtained by the reference method $\ket{\psi^{(0)}}=\ket{\Phi_\text{ref}}$ is picked.
Further, the first-order Hamiltonian is chosen to be a normalized, generalized Fock operator 
\begin{equation}\label{eq:genfock}
    H^{(1)}=F = f^\kappa_\lambda a^\lambda_\kappa - E_0 = (h^\kappa_\lambda + \bar{g}^{\lambda r}_{\kappa s} \gamma^s_r) a^\lambda_\kappa - E_0,
\end{equation}
while the appearing 1-RDM corresponds to the reference method (not HF as in the standard Fock operator)~\cite{Kutzelnigg2003}. The bar over the Coulomb-tensor denotes antisymmetrization. This induces maximal resemblance to the MP2-R12 method~\cite{Torheyden2009} .
Explicit correlation is brought into play with the first-order perturbation
\begin{equation}
  \ket{\psi^{(1)}}= QR \ket{\Phi_\text{ref}},
\end{equation}
where $Q$ denotes a projector ensuring that contributions of OBS are projected out \rv{as well}{} as well as only semi-internal first-order excitations are kept (which was found to be important in contrast to Ref.~\cite{Ten-no2007correction}).\cite{Torheyden2009,Kong2011,Kong2012} Further excitations from occupied to CABS orbitals are excluded, being the origin of one-particle incompleteness, which are to be treated by a ``CABS~Singles'' correction as in Ref.~\cite{Kong2010a}. The correlation operator $R$ is defined here as
\begin{equation}
  R=\frac{1}{2} d^{rs}_{pq} r^{pq}_{\kappa\lambda} a^{\kappa\lambda}_{rs} 
\end{equation}
and generates two-particle correlation, performing excitations into the complementary space $\mathcal{Q}$, which is represented by a set of explicitly correlated functions.
Using a set of approximations\footnote{Cumulant approximations~\cite{Kutzelnigg1999} as well as the standard, extended and generalized Brillouin conditions~\cite{Kutzelnigg1991} and a set of so called screening approximations~\cite{Valeev2008,Torheyden2009,Kong2011}.}, one could then seek to minimize the Hylleraas functional; this would typically result in an ill-conditioned linear system~\cite{Kong2012}, which usually is avoided by making use of the \textsc{sp}-Ansatz by Ten-No~\cite{Ten-No2004}.
In this formulation, the amplitudes $d^{rs}_{ps}$ are fixed such that the cusp conditions both for singlet~($\sim$\textsc{s}) and triplet~($\sim$\textsc{p}) states are fulfilled.
Using spin-free orbitals, the \textsc{sp}-ansatz reads 
\begin{equation}\label{eq:sp-ansatz}
    d^{rs}_{pq}=\frac{3}{8}\delta^r_{p}\delta^s_{q}+\frac{1}{8}\delta^r_{q}\delta^s_{p}.
\end{equation}
This way, \trot amounts to a set of tensor contractions, whose evaluation scales as $\mathcal{O}{(N_{\mathrm{el}}^6)}$ assuming $N_\mathrm{OBS}, N_\mathrm{CABS} \in \order{N_{\mathrm{el}}}$ within the framework of the so called  approximation C, introduced in Ref.~\cite{Kedzuch2005} and applied to \trot in Ref.~\cite{Roskop2014}.
Instead of restating the rather complex equations, we refer to the original works~\cite{Torheyden2009,Kong2011,Roskop2014}. We further point out that a reduction of the cost can be achieved by correlating only a subset of the reference orbitals that may be chosen according to their occupation number~\cite{Kong2011,Roskop2014}\rv{}{, as well as by following ideas from Ref.~\cite{pavovsevic2016sparsemaps} to lower the order of dependency on the RI dimension}. Diagonalization of the 1-RDM to obtain these comes at almost no cost, since they need to be available anyway. 
The framework of approximation C of \trot requires evaluation of additional molecular integrals, as well as only  the 1- and 2-RDM with respect to the given reference 
\begin{align}\label{eq:rdms}
\gamma^p_q = \braket{\Phi_\text{ref}|a^{p}_{q}|\Phi_\text{ref}}, \quad \gamma^{pq}_{rs}=\braket{\Phi_\text{ref} |   a^{pq}_{rs}| \Phi_\text{ref}}.
\end{align}
Insofar, this correction can be applied  to universal references, as long as the 1- and 2-RDM can be made available.
Here, we exclusively use the spin-free formulation, called SF-\trot. Consequently, excitations, RDMs etc. are to be understood as spin-free obtained by summation over spin -- for ease of notation, we treat this implicitly.
A more detailed interpretation of \trot can be found in Ref.~\cite{schleich2020regularization}.

We note that the application of the correction is not necessarily restricted to ground state calculations but can be applied to excited states in a similar manner.

\subsection{VQE and VQE + \trot}
In this work, we rely on the Variational Quantum Eigensolver~\cite{Peruzzo2014,McClean2016} that can be assigned to the class of variational quantum algorithms, which is one of the most promising classes of algorithms in the NISQ~era of quantum computing~\cite{Cao2018,McArdle2018,cerezo2020variational,bharti2021noisy}. Here, one defines an objective function in form of the expected value of the Hamiltonian operator with respect to a parametrized quantum state. The quantum state is composed by a sequence of parametrized unitary operations in form of quantum gates (often called {ansatz}). Minimization of the objective function yields an approximation to the ground-state energy of the system; the accuracy of this approach greatly depends on the capability of the ansatz to express the true ground state~\cite{Romero2019, anand2021quantum, Lee2019, Grimsley2019, Grimsley2020, Izmaylov2020,ryabinkin2018qubit}.

This way, we obtain a reference for the perturbative approach in form of $\ket{\Phi_\text{ref}} = U(\theta^\star) \ket{0}$ with 
\begin{equation}
    \theta^\star = \arg\min_\theta \braket{0|U^\dagger(\theta) {H} U(\theta)|0}   
\end{equation}
 and $U(\theta)$ a sequence of parametrized quantum gates. Throughout this work, we will restrict ourselves on Unitary Coupled Cluster-type ans\"atze (UCC)~\cite{anand2021quantum,Peruzzo2014,Romero2019,Kottmann2020a,Kottmann2020b,kottmann2021feasible}.
Given a basis set $\{p\}$, a generic UCC-type wavefunction can be obtained by
\begin{equation}
  \ket{\psi_\mathrm{UCC}(\theta)} = \e^{\hat{T} - \hat{T}^\dagger} \ket{\Phi_\mathrm{HF}}  ,
\end{equation}
where $\hat{T}= t^i_a a^a_i + \frac{1}{4}  t^{ij}_{ab} a^{ab}_{ij}$ is the usual cluster operator. 
The amplitudes $t^i_a, t^{ij}_{ab}$ serve as variational parameters in this approach.
In practice, a wide range of approximations is raised in order to implement or approximate such a wavefunction~\cite{anand2021quantum}.

\begin{figure*}
    \centering
    \begin{adjustbox}{width=.6\linewidth}
        \begin{footnotesize}
        \def\svgwidth{\columnwidth}
        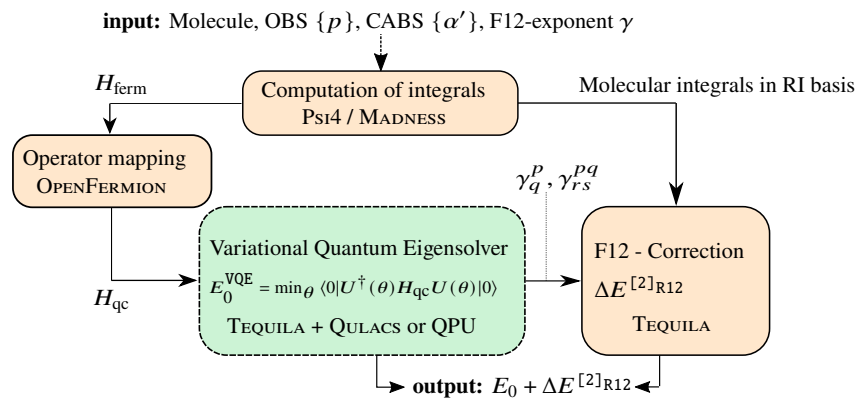

        \end{footnotesize}
    \end{adjustbox}
\caption{Workflow of VQE+\trot. Modules marked by green colored can be performed on a quantum processing unit or a classical simulator, while orange modules denote classical pre- and post-operations. Used software packages are marked in the above illustration, and reiterated in section~\ref{sec:software}.}
\label{fig:workflow}
\end{figure*}

In Fig.~\ref{fig:workflow}, we summarized the proposed workflow of a VQE+\trot computation. Upon having chosen the molecule of desire along with a choice of orbital basis and CABS basis, one can build the fermionic Hamiltonian $H_\mathrm{ferm}$ in the orbital basis as well as molecular integrals needed for the F12-correction (see Ref.~\cite{Kong2011,Roskop2014}) in the RI-basis. 
Additionally, one needs to provide a F12-exponent $\gamma$, cf. Eq.~\eqref{eq:f12-integ-correlfactor}.
Within this work, we fixed $\gamma=1.4$, which overall led to good results. When aiming for the best energy possible, one might perform computations with a set of values and then pick the lowest one if the CBS-limit is not available. In general we found that \trot is not very likely to yield energies below the CBS-limit despite it being a non-variational method. Next, one needs to map the fermionic Hamiltonian to a set of qubit operators using an operator mapping. Here, we used Jordan-Wigner, but other mappings can be employed in a similar manner; see e.g. Refs.~\cite{Cao2018,McArdle2019, bharti2021noisy}. The qubit Hamiltonian $H_\mathrm{qc}$ then serves as input for the VQE computation. After VQE has reached convergence, the 1- and 2-RDM need to be measured, which then are processed together with the molecular integrals in the RI basis to a scalar correction in the energy. 


To obtain  the necessary 1-~and~2-RDM, the VQE measurement procedure needs to be adapted to resolving the full reduced density matrices, coming at a cost $\order{N_\mathrm{OBS}^4}$ for the 2-RDM in the worst case (symmetry conditions and spin-free formulation only allow for a reduction by a constant factor). 
A wide set of measurement reduction techniques has been developed for the evaluation of Hamiltonian expectation values. These dominated by two classes of such techniques: One relies on the identification of sets of commuting operators that can be measured simultaneously, e.g. Refs.~\cite{izmaylov2019unitary, Verteletskyi2020, crawford2021measurement}, which allow  for an approximately cubic cost. The other one determines basis rotations that result in a formulation with less non-mutually commutative terms, leading to less than quadratic number of terms that need to be measured separately~\cite{Huggins2019a,Yen2020}.
A comprehensive overview of advances in this area can be found in Ref.~\cite{bonet2020measurement}.
However, as pointed out in Ref.~\cite{bonet2020measurement}, most of these methods are not directly applicable for the measurement of general fermionic $k$-RDMs.
Using a basis of Majorana operators, Ref.~\cite{bonet2020measurement} shows that smart operator partitioning can yield a reduction of measurement cost up to $\Omega(N_\mathrm{OBS}^2)$.
An advisable strategy for a VQE+\trot routine then would be to implement one of the measurement reduction techniques that reliably can suppress the cost to lower than quadratic order, and make use of the strategy outlined in \rv{}{Ref.}~\cite{bonet2020measurement} for the final RDM-evaluation. 
An additional set of interesting approaches improving the measurement of RDMs  based on the $n$-representability conditions is given in Ref.~\cite{Rubin2018}.
The efficacy of these approaches has been demonstrated in Ref.~\cite{gonthier2020identifying}.  
Such techniques are in particular important in light of stochastic errors when sampling the RDMs. As it was investigated in Ref.~\cite{booth2012explicitly} who applied \trot in FCIQMC calculations, sampling noise in the RDMs spreads as well to the correction; such behaviour can be expected to appear as well when extracting RDMs from VQE and thus high fidelity in doing so at moderate cost is desirable.

\section{Results}

\subsection{Computational Setup}

\subsubsection{Construction of Complementary Basis}
Within this work, we follow the CABS+ procedure introduced in Ref.~\cite{Valeev2004a} to generate a complementary basis. This means, given an orbital (reference) basis and complementary basis, the complementary space is specfied by orthogonalization of the joint RI basis $\{\kappa'\}$, obtained by the union of orbital and complementary basis, and projecting out the orbital space. Formally, one constructs a coefficient matrix $C_{\alpha'\kappa'}$, which maps from the RI to the complementary set, such that $\sum_{\kappa'} C_{\alpha'\kappa'}\braket{\kappa'|p} = 0~\forall p$.

In contrast to that, we comment on the possibility of a ``quick \& dirty''-approach via the choice of an active space.
Here, one starts with a given basis set, that is employed in an active space fashion.
That is, there is a set of frozen virtual orbitals which do not contribute to electron correlation effects.
Consequently, we find ourselves in the same setting as the work by Takeshita et. al.~\cite{Takeshita2020} who make use of QSE to account for the ``active space incompleteness''.
However, we do not present results for this situation since we found, in accordance to Valeev's arguments~\cite{Valeev2004a}, that corrections obtained by CABS+ are more accurate and are not sensitive to the choice of active space. Still this does depend on the specific intention behind the computation -- if a valid active space is chosen, this procedure allows in the case of the virtual QSE to retrieve the expressibility of the full orbital basis~\cite{urbanek2020chemistryqse}, and a similar behaviour can be expected for \trot.   

Accordingly it is advised to follow the CABS+ procedure, as done in Ref.~\cite{Motta2020}.
But, if one were to perform an active space calculation (and an appropriate set of orbitals is chosen), the application of \trot is expected to be beneficial.
While we have no numerical evidence to support this, we expect \trot to perform better than similar methods that excite into conventional basis functions, such as the subspace expansion-based approach in Ref.~\cite{Takeshita2020}, whenever electron-electron cusp effects are dominant, re-emphasizing that the F12 method constructs an explicitly correlated basis that is expected to ``converge faster'' and hence should require fewer basis functions for a given energy correction.

As basis functions, we will mostly resort to the recently introduced MRA-PNOs (globally orthonormalized, directly determined pair natural orbitals obtained by multiresolution analysis)~\cite{Kottmann2020,Kottmann2020a}. This enables us to, on one hand, build up on the existing data thereof, and on the other hand, investigate the potential of a combination of PNOs together with a F12-method.
Even the most simple MRA-PNO generation at this time generates at least $4N_{\mathrm{diag}}+8N_{\mathrm{offdiag}}$ orbitals given a number of diagonal and off-diagonal pairs, which means that in most cases there are a few unused PNOs. This raises the question whether PNOs are suited for a deployment as CABS-basis. 
Building on the theory, it is to be expected that explicitly correlated methods also deliver faster convergence in terms of natural orbitals~\cite{Kutzelnigg2008}, which means we would expect that PNOs built from a MP2-R12 surrogate, as already done in Ref.~\cite{Kottmann2020}, inherit this behaviour. \rv{}{Such behaviour was also found in Refs.~\cite{tew2011local,schmitz2014explicitly}.} Further, a basis of natural orbitals can be seen as a least-squares approximation~\cite{Lowdin1955,Kong2012} and we again expect the PNOs to be somewhat near-optimal despite the pair-approximation. This means that PNOs in general and in particular MP2-F12 PNOs might be interesting candidates. To that end, we provide some data from Ref.~\cite{schleich2020regularization} and additional discussion in appendix~\ref{sec:pnocabs}. Based on these results and the discussion, hereinafter, we will pair VQE-results based on MRA-PNOs for the reference with a PNO-CABS (all with plain MP2 for the surrogate computation), given that they have turned out to outperform PNOs obtained by MP2-F12 if obtained in a self-contained procedure and are also far more compact as an RI-approximation than Gaussian basis functions (\rv{}{The reason herefore is that the reference computation when using PNOs generated by R12 performs worse. While this is expected, we found that in our case, this is so significant that the correction does not fully account for that, although the correction is higher; also } see appendix~\ref{sec:pnocabs}). We emphasize here however that our RI by PNOs is not necessarily perfectly accurate but rather more compact and accurate than a GBS of a comparable size. Additionally the point we want to make is that any PNOs generated that are not used in OBS can increase accuracy when used for the complementary basis.

Of course any set of basis functions -- be it the most prominent choice of orbitals from Gaussian basis sets (GBS), PNOs from GBS or plane-waves -- can be used to apply \trot. 
As already mentioned, we employ MRA-PNOs here to be able to compare with previous results in Ref.~\cite{Kottmann2020a}. Further, as specified therein and in Ref.~\cite{kottmann2021optimized}, the MRA-PNOs yield surprisingly good results in bond-breaking regimes despite the MP2 dependency. Additionally, in contrast to Gaussian basis sets, they exhibit faster convergence so that even for small basis set sizes, the dominant error can already be expected due to the cusp. 
In section~\ref{sec:software} we will point to open-source, freely available software that we employed and wherein we implemented the computation of the correction; both MRA-PNOs and GBS are ready to be employed. 

\subsubsection{Choice of {Ansatz} in VQE-Calculations}\label{sec:compu-ansatz}
In the subsequent results, we use a classical FCI calculation for systems with Gaussian basis sets (GBS); for the PNO-basis, we focus on UCC-type ans\"atze. However, the use of a specialized F12-basis set in the case of GBS is recommended, as this enables the reference computation to focus on the representation of the cusp-free part of the wavefunction and the correction takes care of the cusp.
This work can be understood as an extension to Ref.~\cite{Kottmann2020a} and thus we aim to show the potential of MRA-PNOs together with \trot. Consequently the results we show mostly deploy MRA-PNOs as a basis but we showcase a few exemplary simulations using FCI/GBS as a verification that general bases can be made use of.

For all MRA-PNO-based computations, we exploit the pair-structure in form of the SPA-ansatz from Ref.~\cite{kottmann2021optimized}, which is \rv{a generalization of the PNO-UpCCG(S)D ansatz in Ref.~\cite{Kottmann2020a} (inspired by Ref.~\cite{Lee2019}) and is}{} repeated here in brevity; also see Ref.~\cite{anand2021quantum} for a description thereof. Given a set of HF reference orbitals $\{i\}$, for each $(ij)$-pair we generate orthonormalized PNOs  $\tilde{\mathcal{S}}_{ij} = \bigcup_{\tilde{a}_{ij}=1}^{r_{ij}} \{ \ket{\tilde{a}_{ij}} \}$. As an orthonormalization procedure, we use a Cholesky decomposition, which has turned out deliver superior results as opposed to e.g. symmetric orthonormalization because it changes the ``most important'' PNOs, ordered by occupation number, the least. Then, we restrict to pair-excitations from the references to the associated PNOs ($U_{\tilde{\mathrm{ D}}}$) and generalized excitations within each PNO ($U_{\mathrm{G}\tilde{\mathrm{D}}}$). One might also think of single excitations, which we do not consider in this work. We can express the \rv{PNO-UpCCGD or}{} SPA-UpCCGD wavefunction as
\begin{equation}
    \ket{\text{SPA-UpCCGD}} = U_{\mathrm G \tilde{\mathrm D}} U_{\tilde{\mathrm D}} U_{\mathrm{HF}} \ket{0}
\end{equation}
where $\ket{\Phi_0} = \ket{\Phi_{\mathrm{HF}}} = U_{\mathrm{HF}} \ket{0}$  and 
\begin{align}
    U_{\tilde{\mathrm D}}  &= \prod_{i=1}^{N_{\mathrm{el}/2}} \prod_{a\in \tilde{\mathcal{S}}_{ii}} \exp \left\{ \frac{\theta}{2} \tilde{G}^{aa}_{ii} \right\} \\
    U_{\mathrm G \tilde{\mathrm D}}  &= \prod_{i=1}^{N_{\mathrm{el}/2}} \prod_{a,b\in \tilde{\mathcal{S}}_{ii}} \exp\left\{ \frac{\theta}{2} \tilde{G}^{aa}_{bb} \right\}.
\end{align}
The pair-excitation generator $\tilde G$ is defined as
\begin{equation}
    \tilde{G}^{aa}_{ii} = a^{a_\uparrow a_\downarrow}_{i_\uparrow i_\downarrow} - \mathrm{h.c.}
\end{equation}
In what follows, whenever we use PNOs as a basis, we will employ the SPA-UpCCD ansatz and use ``SPA'' to denote this.
Ref.~\cite{Kottmann2020a} demonstrated that whenever the PNO approximation is a good representation of the molecular system behind, this PNO-restricted parametrization provides potential for a drastic reduction in number of parameters and CNOT-counts as opposed to a UpCCGSD-wavefunction, which itself is compact in comparison to a full UCCSD-parametrization~\cite{Lee2019}. \rv{}{More detailed demonstrations of the performance of SPA can be found in Ref.~\cite{kottmann2021optimized}, aiding to classify it in comparison to popular classical methods.}

Further we provide results for a \emph{cheap} and \emph{good} set of PNOs obtained by MRA-PNO computations.
The \emph{cheap} set signifies a minimal version of PNOs, with the least rich class of excitations from Ref.~\cite{Kottmann2020} (``dipole+'') that uses only the spare orbitals from the generation which one does not opt to use for the orbital space.
In this sense, this procedure would be equivalent to using an active space of some GBS and exploiting the spare orbitals in a correction scheme.
Apart from that, the \emph{good} set of PNOs  allows ten macro-iterations in their generation -- convergence in the PNO-generation usually is achieved beforehand when a certain maximum number of orbitals is specified -- and uses a richer excitation ansatz (``multipole'') with a prescribed goal of the number of PNOs.
We set this number accordingly such that \trot converges, which for our purposes can be seen as a measure that the identity is resolved well enough. This method then would be somewhat analogous to adding a RI-optimized CABS to some orbital basis set. 

\subsubsection{Software}\label{sec:software}
We have outlined the software packages that have been used in Fig.~\ref{fig:workflow}. 
The leading framework organizing the workflow for the work presented here is \textsc{Tequila}~\cite{Kottmann2020b}, while we rely on the methodology in Ref.~\cite{kottmann2021feasible} to obtain analytical gradients at the cost of only two energy expectation values, independent of excitation rank. 

As a quantum chemistry backend for classical CI calculations and the construction of a CABS made up by Gaussian orbitals, we use \textsc{Psi4}~\cite{Smith2020}. 
To compute MRA-PNOs, we employed the software package \textsc{Madness}~\cite{Harrison2016}. 
For the CABS-functionality, our implementation uses modified forks of \textsc{Psi4} and \textsc{Madness}~\footnote{
\textsc{Psi4}: Currently, CABS-functionality is not available in the main repository; a hacky but ready-to-use implementation is accessible at \url{https://github.com/philipp-q/psi4/tree/ri_space}.
For \textsc{Madness}, we use the forked branch \url{https://github.com/kottmanj/madness/tree/pno_integrals_cabs}, which is described in more detail in Ref.~\cite{Kottmann2020a}. Installation instructions for \textsc{Madness} together with \textsc{Tequila} can be found at \url{https://github.com/kottmanj/madness/tree/tequila}. Integration into the main repository is planned here.  
}. F12-integrals in Eq.~\eqref{eq:f12-integ-correlfactor} from \textsc{Madness} are computed directly as Slater-type ($\e^{-\gamma r_{12}}$), when obtained from \textsc{Psi4} via GBS they are approximated by a linear combination of Gaussians.
 
Further dependencies include \textsc{OpenFermion}~\cite{McClean2017a} for the handling of fermionic and qubit operations as well as \textsc{Qulacs}~\cite{suzuki2020qulacs} as quantum computing backend. Further, automatic differentiation is taken care of by \textsc{Jax}~\cite{jax2018github}. \textsc{SciPy}~\cite{virtanen2020scipy} is used for the classical optimization procedures (BFGS).
 
In section~\ref{sec:codesample}, there is a code sample for the computation of the correction, both in the case of Gaussian orbitals and MRA-PNOs. Further, we made available a tutorial notebook on the matter on the github repository for \textsc{Tequila}~\cite{Kottmann2020b}.

\subsection{Atomic systems}
We start by looking at the atomic systems Helium ($N_{\mathrm{el}}  = 2$) and Beryllium ($N_{\mathrm{el}}  = 4$) and consider a choice of different bases to obtain an intuition as how well MRA-PNOs+\trot perform against a naive procedure with a GBS. This should give us an indication regarding possible savings of quantum resources.

\begin{figure*}[t]
    \centering
    \begin{subfigure}{.45\linewidth}
        \begin{adjustbox}{width=\linewidth}
            \input{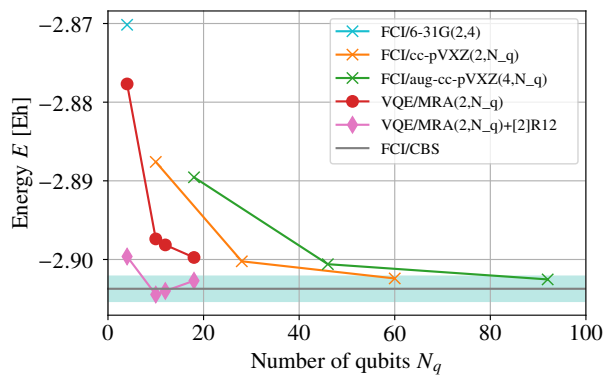}
        \end{adjustbox}
        \caption{He, $E_{\mathrm{CBS}} = -2.9037$ \cite{dehesa1992study,nakatsuji2012discovery,Bischoff2014a}. }
    \end{subfigure}
    \hfill
    \begin{subfigure}{.45\linewidth}
        \begin{adjustbox}{width=\linewidth}
            \input{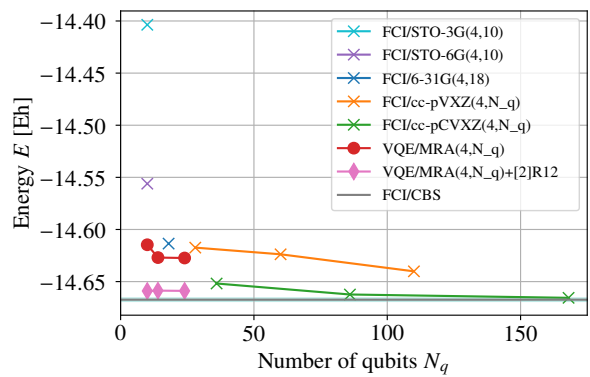}
        \end{adjustbox}
        \caption{Be, $E_{\mathrm{CBS}} = -14.667$ \cite{nakatsuji2012discovery}}
    \end{subfigure}
    \caption{Atomic systems. 
    Comparison of classical FCI with corrected basis-set-free VQE+\trot energies using \emph{good} PNOs. We use a SPA-UpCCD ansatz for VQE~\cite{kottmann2021optimized}. Data  for classical computations is from \cite{Kottmann2020a}.  \rv{}{Note that the corrected results of He are in parts lower than FCI and thus show the non-variational nature of the perturbative correction.} }  
        \label{fig:atomicsys}
    \end{figure*}

Except for the corrected VQE/MRA-energies, Fig.~\ref{fig:atomicsys} is identical to Fig.~2 in \cite{Kottmann2020a}. Compared to the uncorrected energy, \trot provides significantly lower energies, which slightly ``overshoot'' but still within a chemical accuracy of $1.6$~mEh, for the simulations with ten and twelve qubits (we denote number of qubits as $N_q$). This showcases the non-variational nature of the perturbation method.
In the case of Beryllium -- this is a behaviour that carries through the results in particular for $N_\mathrm{el}>2$, although in general no assumptions can be made that this needs to be true -- the calculations with 10 qubits are equivalently accurate as the one with 24 qubits.
Note that as anticipated in Ref.~\cite{Kottmann2020a}, the result for Beryllium ``saturates'' at a certain level. 
This way, the combination of MRA-PNOs as a system-adapted basis and the \trot-correction enables a significant reduction in the necessary number of qubits for both Helium and Beryllium.

\subsection{Molecules}\label{subsec:res-mol}
Again the shown results here widely follow Ref.~\cite{Kottmann2020a} to demonstrate the potential of a combination of \trot with MRA-PNOs.

In the following, we present potential energy surfaces for the dissociation of H$_2$, LiH, BH and BeH$_2$ (symmetric) in Fig.~\ref{fig:pes-mra}. Recall from section~\ref{sec:compu-ansatz}  that we mostly present outcomes using a MRA/PNO-basis with a distinction in \emph{cheap} and \emph{good} PNOs, i.e., using the left-over, unused orbitals vs. creating a CABS on purpose. 
In addition, Fig.~\ref{fig:h2lihgbs} shows potential energy surfaces for H$_2$ and LiH in Gaussian basis sets with a specific CABS basis.  
Moving on, we will discuss the results shown in the figures. 

\textbf{H$_\mathbf{2}$} First, look at the hydrogen molecule with $N_\mathrm{el}=2$ in Figs.~\ref{fig:h21},\ref{fig:h22},\ref{fig:h23}. For the \emph{cheap} PNOs, only $5$ PNOs are available, which leaves only one orbital for the CABS in the 4-qubit computation.
Hence, the 8~qubit computation (with 4 spatial orbitals) is left with only one PNO to represent the complementary space. As expected, the correction is very small in this case, however it still makes sense to apply it given it only improves the result. 
Given that there is only one electron pair in this case, the PNO generation can be seen as quite efficient, and the correction for the \emph{cheap} set is rather low. On the other hand, the 8~qubit computation using \emph{good} PNOs almost reaches the accuracy of the 120~qubit FCI-calculation in the regime of small bond distances. We note that for longer distances, the correction tends towards zero because in case of two basically separated hydrogen atoms, there is no more electron-electron correlation present. Further we point out an additional anomaly in Fig.~\ref{fig:h21}: At a distance of 3~\AA{}ngstrom, the 8~qubit result should lie lower (roughly -1 Eh) -- this is due to the MP2 surrogate model for the PNO generation, which has not been able to produce orbitals of the correct symmetry in this case (for more details, see Ref.~\cite{schleich2020regularization}). 

\textbf{LiH}
The previous findings mostly can be confirmed by looking at LiH in Figs.~\ref{fig:lih1},\ref{fig:lih2},\ref{fig:lih3}. Note that since there are two electron pairs now, there would be 8~PNOs left to represent the CABS in the larger OBS for the \emph{cheap} PNOs. However, the associated correction almost vanished since all these orbitals stem from an off-diagonal pair -- since the LiH wavefunction very well separates in a product structure of pair one and two, these orbitals barely contribute, also visible by their low occupation number.  The \emph{good} PNOs allow for a considerable correction, with the corrected energy of both the 12~and 20~qubit computation at a similar level.   
Within the dissociation curve for \emph{cheap} PNOs, the energies for both 12 and 20 qubits at $-0.6$~\AA{} are far too high -- this anomaly has the same origin as the $3.0$~\AA{}-point for H$_2$, and can again be resolved by allowing for another macro-iteration in the orbital determination.

In contrast to diatomic hydrogen, the corrections for systems containing atoms with more than two electrons do not vanish for long bond distances because there is still electron-electron correlation embodied in the individual atoms.

Furthermore, we observe that for both sets of PNOs, the lower-qubit computation reaches the same level as the computation with more qubits. This behaviour shows again in the case of BH.

\textbf{BH}
Considering Figs.~\ref{fig:bh1},\ref{fig:bh2},\ref{fig:bh3}, we additionally observe that while the uncorrected energies with both 12 and 22 qubits are very close (they differ by roughly $0.01~Eh$, the corrected energies are almost equal, and except for around the equilibrium differ only in the order of milli-Hartrees, where the fewer qubits computation actually performs better because of the higher-dimensional CABS. 
Additionally, the results for \emph{cheap} and \emph{good} PNOs are much closer than before.
This rationale behind this is that with $N_{\mathrm{el}}=6$, with three diagonal and three off-diagonal pairs, there are $4N_{\mathrm{diag}} + 8N_{\mathrm{offdiag}} = 36$ PNOs readily generated within one macro-iteration, and given they do not differ too much in ``quality'', this already yields a decent RI-representation. For higher bond distance, this almost levels out, while for lower distances with stronger electron-electron correlation, the larger CABS still pays off.

\textbf{BeH$_\mathbf{2}$}
Results for BeH$_2$ are shown in Figs.~\ref{fig:beh1},\ref{fig:beh2},\ref{fig:beh3}; these have been obtained in a frozen-core approximation with four active reference orbitals. We point out that for larger bond distances, the PNO-approximation in the construction of the quantum circuit does not hold anymore as good. 
This was already noticed in \cite{Kottmann2020a} and overcome using an adaptively enriched parametrization in the style of ADAPT-VQE~\cite{Grimsley2020}. 
Here, we abstain from doing so and investigate the impact of the correction in such a regime. We see that -- as expected -- the explicitly correlated correction is not able to account for weak properties of the underlying ansatz.

\newcommand{\hhhrule}{\vspace{4pt}\hrule\vspace{4pt}}
\begin{figure*}[h]
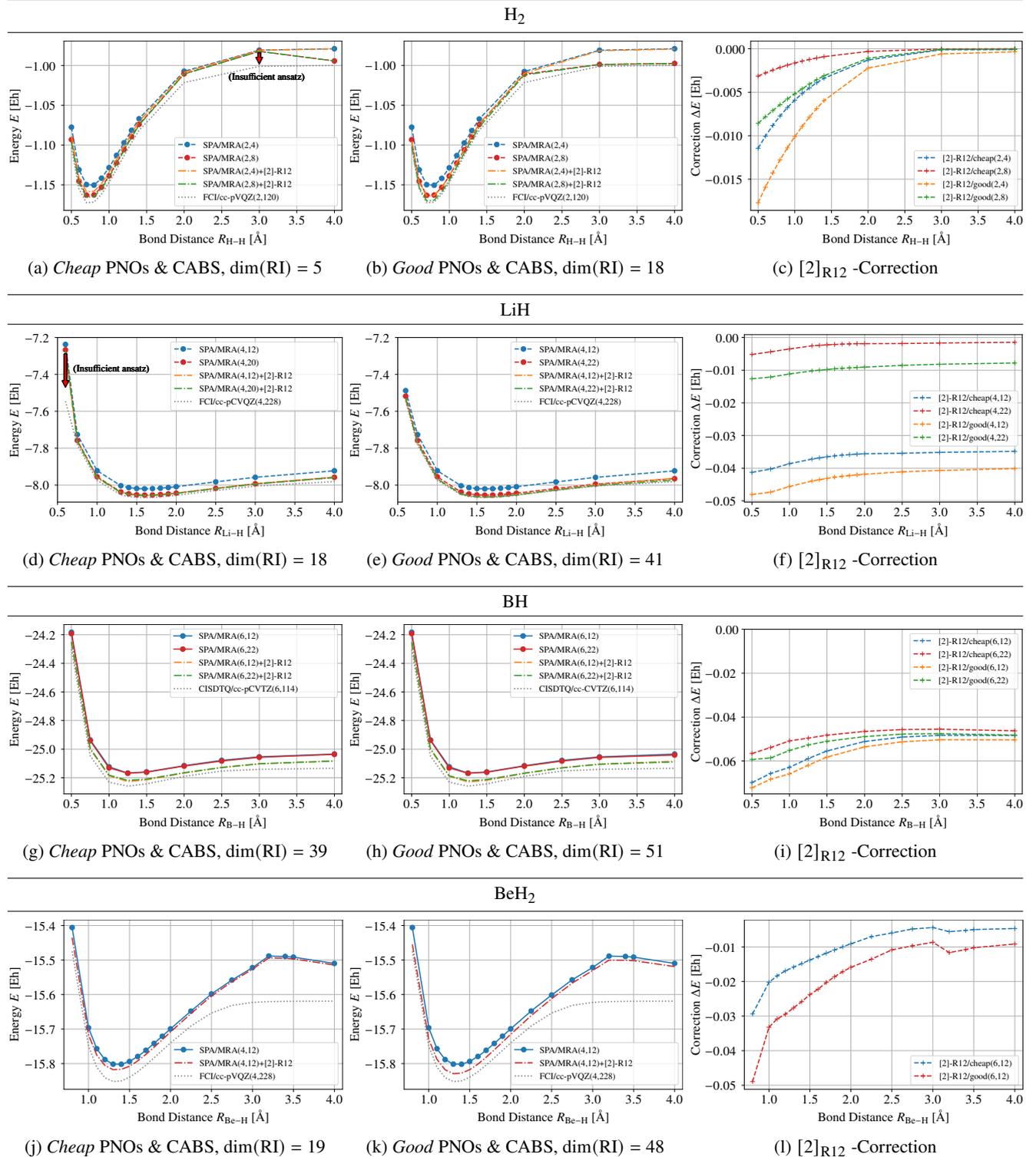

    \centering
    \hhhrule
    {H$_2$}\\
    \hhhrule
    \begin{subfigure}{.33\linewidth}
        \begin{adjustbox}{width=\linewidth}
            \input{figs/h2-pes-cheap.pgf}
        \end{adjustbox}
        \caption{\emph{Cheap} PNOs \& CABS, $\dim(\text{RI})=5$}\label{fig:h21}
    \end{subfigure}
    \hfill
    \begin{subfigure}{.33\linewidth}
        \begin{adjustbox}{width=\linewidth}
            \input{figs/h2-pes-good.pgf}
        \end{adjustbox}
        \caption{\emph{Good} PNOs \& CABS, $\dim(\text{RI})=18$}\label{fig:h22}
    \end{subfigure}
    \hfill
    \begin{subfigure}{.33\linewidth}
        \begin{adjustbox}{width=\linewidth}
            \input{figs/h2-mra-corr.pgf}
        \end{adjustbox}
        \caption{\trot-Correction}\label{fig:h23}
    \end{subfigure}
\hhhrule
    {LiH}\\
    \hhhrule
    \centering
    \begin{subfigure}{.33\linewidth}
        \begin{adjustbox}{width=\linewidth}
            \input{figs/lih-pes-cheap.pgf}
        \end{adjustbox}
        \caption{\emph{Cheap} PNOs \& CABS, $\dim(\text{RI})=18$}\label{fig:lih1}
    \end{subfigure}
    \hfill
    \begin{subfigure}{.33\linewidth}
        \begin{adjustbox}{width=\linewidth}
            \input{figs/lih-pes-good.pgf}
        \end{adjustbox}
        \caption{\emph{Good} PNOs \& CABS, $\dim(\text{RI})=41$}\label{fig:lih2}
    \end{subfigure}
    \hfill
    \begin{subfigure}{.33\linewidth}
        \begin{adjustbox}{width=\linewidth}
            \input{figs/lih-mra-corr.pgf}
        \end{adjustbox}
        \caption{\trot-Correction}\label{fig:lih3}
    \end{subfigure}
\hhhrule
    {BH}\\
    \hhhrule
    \centering
    \begin{subfigure}{.33\linewidth}
        \begin{adjustbox}{width=\linewidth}
            \input{figs/bh-pes-cheap.pgf}
        \end{adjustbox}
        \caption{\emph{Cheap} PNOs \& CABS, $\dim(\text{RI})=39$}\label{fig:bh1}
    \end{subfigure}
    \hfill
    \begin{subfigure}{.33\linewidth}
        \begin{adjustbox}{width=\linewidth}
            \input{figs/bh-pes-good.pgf}
        \end{adjustbox}
        \caption{\emph{Good} PNOs \& CABS, $\dim(\text{RI})=51$}\label{fig:bh2}
    \end{subfigure}
    \hfill
    \begin{subfigure}{.33\linewidth}
        \begin{adjustbox}{width=\linewidth}
            \input{figs/bh-mra-corr.pgf}
        \end{adjustbox}
        \caption{\trot-Correction}\label{fig:bh3}
    \end{subfigure}
\hhhrule
    {BeH$_2$}\\
    \hhhrule
    \centering
    \begin{subfigure}{.33\linewidth}
        \begin{adjustbox}{width=\linewidth}
            \input{figs/beh2-pes-cheap.pgf}
        \end{adjustbox}
        \caption{\emph{Cheap} PNOs \& CABS, $\dim(\text{RI})=19$}\label{fig:beh1}
    \end{subfigure}
    \hfill
    \begin{subfigure}{.33\linewidth}
        \begin{adjustbox}{width=\linewidth}
            \input{figs/beh2-pes-good.pgf}
        \end{adjustbox}
        \caption{\emph{Good} PNOs \& CABS, $\dim(\text{RI})=48$}\label{fig:beh2}
    \end{subfigure}
    \hfill
    \begin{subfigure}{.33\linewidth}
        \begin{adjustbox}{width=\linewidth}
            \input{figs/beh2-mra-corr.pgf}
        \end{adjustbox}
        \caption{\trot-Correction}\label{fig:beh3}
    \end{subfigure}
    \caption{Potential energy surfaces with \trot-correction and PNOs as CABS for H$_2$, LiH, BH, BeH$_2$, following Ref.~\cite{Kottmann2020a}. SPA ansatz with doubles excitations only (SPA-UpCCD~\cite{kottmann2021optimized}). \rv{}{For validation of the SPA ansatz compared to classical techniques such as HF, CC, we refere to Ref.~\cite{kottmann2021optimized}}}.
    \label{fig:pes-mra}
\end{figure*}

\begin{figure*}[t]
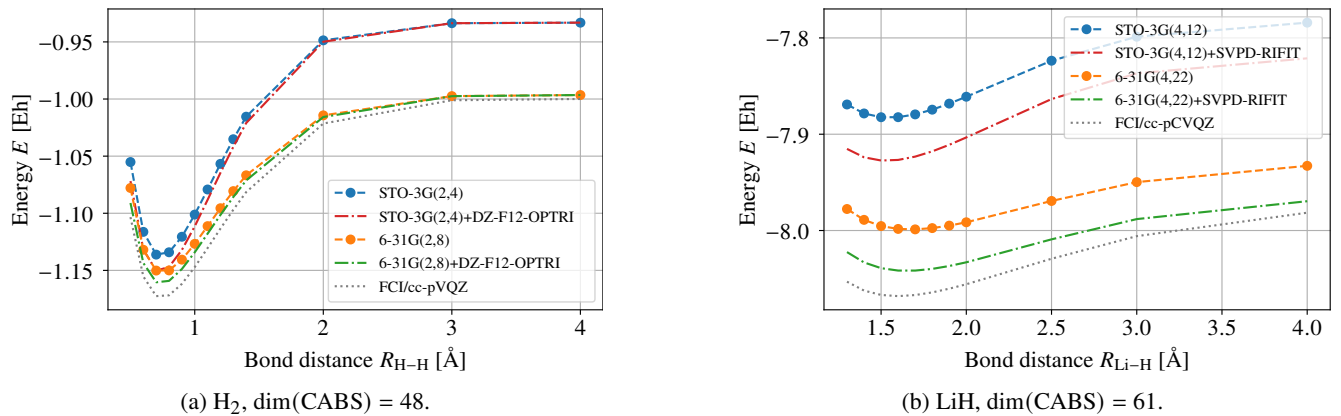

    \centering
    \begin{subfigure}{.45\linewidth}
        \begin{adjustbox}{width=\linewidth}
            \input{figs/h2-enco.pgf}
        \end{adjustbox}
        \caption{{H$_2$}, $\dim(\text{CABS})=48$.}
    \end{subfigure}
    \hfill
    \begin{subfigure}{.45\linewidth}
        \begin{adjustbox}{width=\linewidth}
            \input{figs/lih-enco.pgf}
        \end{adjustbox}
        \caption{{LiH}, $\dim(\text{CABS})=61$.}
    \end{subfigure}
    \caption{Corrected potential energy surfaces using classical FCI and Gaussian basis sets.}
    \label{fig:h2lihgbs}
\end{figure*}

The last set of potential energy surfaces we look at are comprised by the PES for the hydrogen molecule and lithium hydride in Gaussian basis sets in Fig.~\ref{fig:h2lihgbs}. They show potential energy surfaces for H$_2$ and LiH in \bas{STO-3G} and \bas{6-31G} with a specific CABS basis generated by CABS+ \cite{Valeev2004a} to demonstrate that Gaussian basis sets with designated CABS can be used in the same manner.
Here, we do not observe any anomalies with respect to the results before but see only consistent behaviour, e.g. the vanishing correction for large bond distances in the case of H$_2$.


\subsection{NPE + MAX}

\begin{figure*}[t]
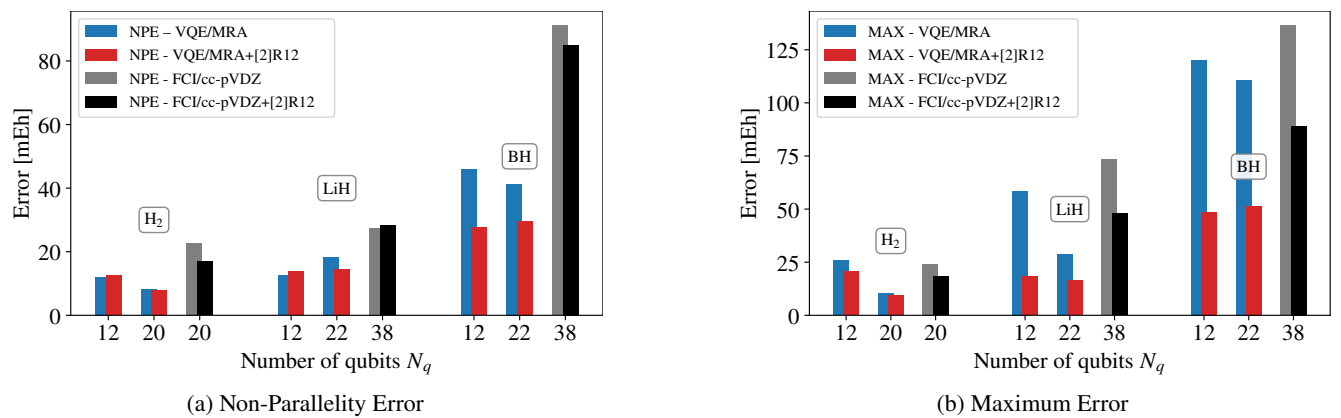

   \centering 
    \begin{subfigure}{.45\linewidth}
        \begin{adjustbox}{width=\linewidth}
            \input{figs/npe_pnos.pgf}
        \end{adjustbox}
        \caption{Non-Parallelity Error }
    \end{subfigure}
    \hfill
    \begin{subfigure}{.45\linewidth}
        \begin{adjustbox}{width=\linewidth}
            \input{figs/max_pnos.pgf}
        \end{adjustbox}
        \caption{Maximum Error }
    \end{subfigure}
    \caption{Performance of MRA-PNOs with and without \trot in comparison to FCI in a DZ-basis. CABS for MRA are generated using the \emph{good} setup ($\dim$(CABS) as in Fig.~\ref{fig:pes-mra}; for the GBS we used cc-pVDZ-F12-OPTRI, amounting to $\dim(\text{CABS})$ of 48 for H$_2$ and 110 for both LiH, BH). }
    \label{fig:errors}
\end{figure*}

As a final result, we consider the non-parallelity error (NPE) and maximum error (MAX) defined over the computed PES for the cases of H$_2$, LiH and BH, following~\cite{Kottmann2020a}.
We compare MRA-PNOs with and without corrections with a corrected DZ-basis (with \bas{cc-pVDZ-F12-OPTRI} as CABS).
We do not use a F12-optimized \bas{cc-pVDZ-F12} for the orbital basis to keep the number of qubits closer to the MRA-PNO computations. 
With $\Delta E(R) = \abs{E(R)-E^{\mathrm{FCI}}_\mathrm{ref}(R)}$, these error metrics are defined as
\begin{align}
    \Delta E_{\mathrm{NPE}} &= \max_{R\in\mathrm{PES}} \Delta E(R)  - \min_{R\in\mathrm{PES}} \Delta E(R)   \\
    \Delta E_{\mathrm{MAX}} &= \max_{R\in\mathrm{PES}} \Delta E(R).
\end{align}
Looking at Fig.~\ref{fig:errors}, we can say that in particular the MAX error is consistently and significantly reduced. Since the explicitly correlated correction is not a constant shift but typically higher for lower bond distances and thus higher electron-electron correlation, the NPE error is not necessarily reduced but also does not increase by much in our computations. This holds in particular for the smaller systems because this effect carries more weight here. Occasional increase in NPE can be traced to higher correction for short bond distances, which holds in particular for smaller systems (compare to discussion of H$_2$ in section~\ref{subsec:res-mol}). This effect is is less severe for larger systems, as witnessed by BH in Fig.~\ref{fig:errors}. For C$_2$, Ref.~\cite{roskop2016application} observed a reduction in NPE as well.

\section{Conclusion and Outlook}
First and foremost, we have proposed a workflow for the combination of the explicitly correlated \trot-correction together with VQE, and provided the software infrastructure to carry out such computations.
Based on a few small test systems, we were able to show a significant increase in accuracy when using this correction. \rv{}{Although \trot is a non-variational perturbation method, within out experiments, we only noticed few and low overshoots beyond FCI as long as the reference method produced a somewhat sensible approximation. For non-sensical or very bad reference inputs, the behaviour of the correction also seemed unpredictable.} In particular in combination with orthogonalized MRA-PNOs~\cite{Kottmann2020,Kottmann2020a} employed in the reference method as well as complementary basis shows considerable promise, and allows to perform the quantum routine at a cost of a minimal basis  yielding accuracies  comparable to large correlation-consistent basis sets currently unfeasible for quantum algorithms. Of course, PNOs generated by a set of Gaussian orbitals can be considered equivalently. To ease the combination of PNOs with a complementary basis composed of Gaussian basis functions, an interface of the MRA part of \textsc{Madness}~\cite{Harrison2016} with e.g. \textsc{MPQC}~\cite{ChongPengCannadaLewisXiaoWangMarjoryClement} can be envisioned.

Additionally, exploiting the PNO-structure for the parametrized quantum circuit turned out to be a powerful tool to reduce parameters within the classical optimization subroutine of VQE. For larger systems such as BeH$_2$ however, this parametrization turned out to loose validity -- ways to overcome this are proposed in Ref.~\cite{Kottmann2020a}. 
Here we used the separable pair ansatz~\cite{kottmann2021optimized} as an affordable way to construct quantum circuits, illustrating also that \trot can not overcome fundamental limitations in the ansatz.

We divided our MRA-PNO computations into  a set of \emph{cheap} and \emph{good} PNOs, based on the quality of the generated PNOs. We note that for small systems, the \emph{cheap} set often does not yield a large enough number of extra PNOs to induce a significant impact, larger systems with more than 1-2 pairs quickly generate a quite rich set of PNOs (see correction for BH in Fig.~\ref{fig:bh3}, where the corrections for both sets of PNOs do not differ much).
Further, known physical behaviour has to be kept in mind and monitored when growing PNOs, such as the pair structure of LiH (off-diagonal PNOs barely contribute) or critical geometries along the potential energy surface that tend to degeneracies (compare to the \emph{cheap} sets of H$_2$ and LiH in Figs.~\ref{fig:h21},\ref{fig:lih1}.). Yet in any case, whenever the reference PNOs are sufficiently good, adding remaining left-over orbitals throughout \trot in the spirit of Ref.~\cite{Takeshita2020} is beneficial.  \rv{}{Beyond that, it would be interesting to consider the constructing geminal-spanning orbitals from Ref.~\cite{pavovsevic2014geminal} instead of MP2-PNOs using MRA.}

While our findings imply a reduction in the number of qubits at the same level of accuracy, due to its perturbative nature we cannot give projections regarding circuit depth, number of entangling gates, or other cost measures. Yet in comparison to the transcorrelated approach~\cite{Motta2020}, the Hamiltonians using a perturbation method remain unchanged and do not exhibit an increased number of terms. On the other hand, the perturbation method does not allow for the same level of flexibility as a regularized Hamiltonian. In the end, the choice between these two approaches depends on a case-by-case basis and might even boil down to a matter of taste.
It remains to investigate which of the methods yield more accurate results. Within a classical benchmark~\cite{Kersten2016}, they performed equally well. 

Additionally it would be interesting to examine the influence of noise on \trot, which has not been considered in this work. To this end, we refer to Ref.~\cite{urbanek2020chemistryqse} who applied the virtual quantum subspace expansion on actual quantum hardware. Beyond that, Refs.~\cite{urbanek2020chemistryqse,booth2012explicitly} proposed a technique to account for measurement noise -- similar approaches can be thought of in this case.

Finally, one might think of combinations of the quantum subspace expansion together with \trot, given that the necessary reduced density matrices are already available. However this does turn out to be not trivial because when applying two different perturbative schemes, one must be careful not to account for the same behaviour twice and ``overcompensate''. One way to approach this might be to follow the idea of a ``CABS~singles'' correction~\cite{Kong2011} to additionally account for one-electron incompleteness.

\section{Acknowledgments}

This work was supported by the U.S. Department of Energy under Award No. DE-SC0019374.

A.A.-G. acknowledges the generous support from Google, Inc.  in the form of a Google Focused Award. A.A.-G. also acknowledges support from the Canada Industrial Research Chairs  Program and the Canada 150 Research Chairs Program. P.S. acknowledges support by a fellowship within the IFI program of the German Academic Exchange Service (DAAD). Simulations were performed with computing resources granted by RWTH Aachen University under project thes0850. This research was enabled in part by support provided by Compute Canada. Computations were performed on the niagara supercomputer at the SciNet HPC Consortium\cite{niagara1, niagara2}. SciNet is funded by: the Canada Foundation for Innovation; the Government of Ontario; Ontario Research Fund - Research Excellence; and the University of Toronto. J.S.K thanks Florian Bischoff for introducing him to explicit correlation and the depths of madness back in the days. We thank the generous support of Anders G. Fr\o{}seth.

\bibliography{main.bib}

\begin{thebibliography}{107}%
\makeatletter
\providecommand \@ifxundefined [1]{%
 \@ifx{#1\undefined}
}%
\providecommand \@ifnum [1]{%
 \ifnum #1\expandafter \@firstoftwo
 \else \expandafter \@secondoftwo
 \fi
}%
\providecommand \@ifx [1]{%
 \ifx #1\expandafter \@firstoftwo
 \else \expandafter \@secondoftwo
 \fi
}%
\providecommand \natexlab [1]{#1}%
\providecommand \enquote  [1]{``#1''}%
\providecommand \bibnamefont  [1]{#1}%
\providecommand \bibfnamefont [1]{#1}%
\providecommand \citenamefont [1]{#1}%
\providecommand \href@noop [0]{\@secondoftwo}%
\providecommand \href [0]{\begingroup \@sanitize@url \@href}%
\providecommand \@href[1]{\@@startlink{#1}\@@href}%
\providecommand \@@href[1]{\endgroup#1\@@endlink}%
\providecommand \@sanitize@url [0]{\catcode `\\12\catcode `\$12\catcode
  `\&12\catcode `\#12\catcode `\^12\catcode `\_12\catcode `\%12\relax}%
\providecommand \@@startlink[1]{}%
\providecommand \@@endlink[0]{}%
\providecommand \url  [0]{\begingroup\@sanitize@url \@url }%
\providecommand \@url [1]{\endgroup\@href {#1}{\urlprefix }}%
\providecommand \urlprefix  [0]{URL }%
\providecommand \Eprint [0]{\href }%
\providecommand \doibase [0]{http://dx.doi.org/}%
\providecommand \selectlanguage [0]{\@gobble}%
\providecommand \bibinfo  [0]{\@secondoftwo}%
\providecommand \bibfield  [0]{\@secondoftwo}%
\providecommand \translation [1]{[#1]}%
\providecommand \BibitemOpen [0]{}%
\providecommand \bibitemStop [0]{}%
\providecommand \bibitemNoStop [0]{.\EOS\space}%
\providecommand \EOS [0]{\spacefactor3000\relax}%
\providecommand \BibitemShut  [1]{\csname bibitem#1\endcsname}%
\let\auto@bib@innerbib\@empty
\bibitem [{\citenamefont {Arute}\ \emph {et~al.}(2019)\citenamefont {Arute},
  \citenamefont {Arya}, \citenamefont {Babbush}, \citenamefont {Bacon},
  \citenamefont {Bardin}, \citenamefont {Barends}, \citenamefont {Biswas},
  \citenamefont {Boixo}, \citenamefont {Brandao}, \citenamefont {Buell},
  \citenamefont {Burkett}, \citenamefont {Chen}, \citenamefont {Chen},
  \citenamefont {Chiaro}, \citenamefont {Collins}, \citenamefont {Courtney},
  \citenamefont {Dunsworth}, \citenamefont {Farhi}, \citenamefont {Foxen},
  \citenamefont {Fowler}, \citenamefont {Gidney}, \citenamefont {Giustina},
  \citenamefont {Graff}, \citenamefont {Guerin}, \citenamefont {Habegger},
  \citenamefont {Harrigan}, \citenamefont {Hartmann}, \citenamefont {Ho},
  \citenamefont {Hoffmann}, \citenamefont {Huang}, \citenamefont {Humble},
  \citenamefont {Isakov}, \citenamefont {Jeffrey}, \citenamefont {Jiang},
  \citenamefont {Kafri}, \citenamefont {Kechedzhi}, \citenamefont {Kelly},
  \citenamefont {Klimov}, \citenamefont {Knysh}, \citenamefont {Korotkov},
  \citenamefont {Kostritsa}, \citenamefont {Landhuis}, \citenamefont
  {Lindmark}, \citenamefont {Lucero}, \citenamefont {Lyakh}, \citenamefont
  {Mandr{\`{a}}}, \citenamefont {McClean}, \citenamefont {McEwen},
  \citenamefont {Megrant}, \citenamefont {Mi}, \citenamefont {Michielsen},
  \citenamefont {Mohseni}, \citenamefont {Mutus}, \citenamefont {Naaman},
  \citenamefont {Neeley}, \citenamefont {Neill}, \citenamefont {Niu},
  \citenamefont {Ostby}, \citenamefont {Petukhov}, \citenamefont {Platt},
  \citenamefont {Quintana}, \citenamefont {Rieffel}, \citenamefont {Roushan},
  \citenamefont {Rubin}, \citenamefont {Sank}, \citenamefont {Satzinger},
  \citenamefont {Smelyanskiy}, \citenamefont {Sung}, \citenamefont
  {Trevithick}, \citenamefont {Vainsencher}, \citenamefont {Villalonga},
  \citenamefont {White}, \citenamefont {Yao}, \citenamefont {Yeh},
  \citenamefont {Zalcman}, \citenamefont {Neven},\ and\ \citenamefont
  {Martinis}}]{Arute2019}%
  \BibitemOpen
  \bibfield  {author} {\bibinfo {author} {\bibfnamefont {F.}~\bibnamefont
  {Arute}}, \bibinfo {author} {\bibfnamefont {K.}~\bibnamefont {Arya}},
  \bibinfo {author} {\bibfnamefont {R.}~\bibnamefont {Babbush}}, \bibinfo
  {author} {\bibfnamefont {D.}~\bibnamefont {Bacon}}, \bibinfo {author}
  {\bibfnamefont {J.~C.}\ \bibnamefont {Bardin}}, \bibinfo {author}
  {\bibfnamefont {R.}~\bibnamefont {Barends}}, \bibinfo {author} {\bibfnamefont
  {R.}~\bibnamefont {Biswas}}, \bibinfo {author} {\bibfnamefont
  {S.}~\bibnamefont {Boixo}}, \bibinfo {author} {\bibfnamefont {F.~G.}\
  \bibnamefont {Brandao}}, \bibinfo {author} {\bibfnamefont {D.~A.}\
  \bibnamefont {Buell}}, \bibinfo {author} {\bibfnamefont {B.}~\bibnamefont
  {Burkett}}, \bibinfo {author} {\bibfnamefont {Y.}~\bibnamefont {Chen}},
  \bibinfo {author} {\bibfnamefont {Z.}~\bibnamefont {Chen}}, \bibinfo {author}
  {\bibfnamefont {B.}~\bibnamefont {Chiaro}}, \bibinfo {author} {\bibfnamefont
  {R.}~\bibnamefont {Collins}}, \bibinfo {author} {\bibfnamefont
  {W.}~\bibnamefont {Courtney}}, \bibinfo {author} {\bibfnamefont
  {A.}~\bibnamefont {Dunsworth}}, \bibinfo {author} {\bibfnamefont
  {E.}~\bibnamefont {Farhi}}, \bibinfo {author} {\bibfnamefont
  {B.}~\bibnamefont {Foxen}}, \bibinfo {author} {\bibfnamefont
  {A.}~\bibnamefont {Fowler}}, \bibinfo {author} {\bibfnamefont
  {C.}~\bibnamefont {Gidney}}, \bibinfo {author} {\bibfnamefont
  {M.}~\bibnamefont {Giustina}}, \bibinfo {author} {\bibfnamefont
  {R.}~\bibnamefont {Graff}}, \bibinfo {author} {\bibfnamefont
  {K.}~\bibnamefont {Guerin}}, \bibinfo {author} {\bibfnamefont
  {S.}~\bibnamefont {Habegger}}, \bibinfo {author} {\bibfnamefont {M.~P.}\
  \bibnamefont {Harrigan}}, \bibinfo {author} {\bibfnamefont {M.~J.}\
  \bibnamefont {Hartmann}}, \bibinfo {author} {\bibfnamefont {A.}~\bibnamefont
  {Ho}}, \bibinfo {author} {\bibfnamefont {M.}~\bibnamefont {Hoffmann}},
  \bibinfo {author} {\bibfnamefont {T.}~\bibnamefont {Huang}}, \bibinfo
  {author} {\bibfnamefont {T.~S.}\ \bibnamefont {Humble}}, \bibinfo {author}
  {\bibfnamefont {S.~V.}\ \bibnamefont {Isakov}}, \bibinfo {author}
  {\bibfnamefont {E.}~\bibnamefont {Jeffrey}}, \bibinfo {author} {\bibfnamefont
  {Z.}~\bibnamefont {Jiang}}, \bibinfo {author} {\bibfnamefont
  {D.}~\bibnamefont {Kafri}}, \bibinfo {author} {\bibfnamefont
  {K.}~\bibnamefont {Kechedzhi}}, \bibinfo {author} {\bibfnamefont
  {J.}~\bibnamefont {Kelly}}, \bibinfo {author} {\bibfnamefont {P.~V.}\
  \bibnamefont {Klimov}}, \bibinfo {author} {\bibfnamefont {S.}~\bibnamefont
  {Knysh}}, \bibinfo {author} {\bibfnamefont {A.}~\bibnamefont {Korotkov}},
  \bibinfo {author} {\bibfnamefont {F.}~\bibnamefont {Kostritsa}}, \bibinfo
  {author} {\bibfnamefont {D.}~\bibnamefont {Landhuis}}, \bibinfo {author}
  {\bibfnamefont {M.}~\bibnamefont {Lindmark}}, \bibinfo {author}
  {\bibfnamefont {E.}~\bibnamefont {Lucero}}, \bibinfo {author} {\bibfnamefont
  {D.}~\bibnamefont {Lyakh}}, \bibinfo {author} {\bibfnamefont
  {S.}~\bibnamefont {Mandr{\`{a}}}}, \bibinfo {author} {\bibfnamefont {J.~R.}\
  \bibnamefont {McClean}}, \bibinfo {author} {\bibfnamefont {M.}~\bibnamefont
  {McEwen}}, \bibinfo {author} {\bibfnamefont {A.}~\bibnamefont {Megrant}},
  \bibinfo {author} {\bibfnamefont {X.}~\bibnamefont {Mi}}, \bibinfo {author}
  {\bibfnamefont {K.}~\bibnamefont {Michielsen}}, \bibinfo {author}
  {\bibfnamefont {M.}~\bibnamefont {Mohseni}}, \bibinfo {author} {\bibfnamefont
  {J.}~\bibnamefont {Mutus}}, \bibinfo {author} {\bibfnamefont
  {O.}~\bibnamefont {Naaman}}, \bibinfo {author} {\bibfnamefont
  {M.}~\bibnamefont {Neeley}}, \bibinfo {author} {\bibfnamefont
  {C.}~\bibnamefont {Neill}}, \bibinfo {author} {\bibfnamefont {M.~Y.}\
  \bibnamefont {Niu}}, \bibinfo {author} {\bibfnamefont {E.}~\bibnamefont
  {Ostby}}, \bibinfo {author} {\bibfnamefont {A.}~\bibnamefont {Petukhov}},
  \bibinfo {author} {\bibfnamefont {J.~C.}\ \bibnamefont {Platt}}, \bibinfo
  {author} {\bibfnamefont {C.}~\bibnamefont {Quintana}}, \bibinfo {author}
  {\bibfnamefont {E.~G.}\ \bibnamefont {Rieffel}}, \bibinfo {author}
  {\bibfnamefont {P.}~\bibnamefont {Roushan}}, \bibinfo {author} {\bibfnamefont
  {N.~C.}\ \bibnamefont {Rubin}}, \bibinfo {author} {\bibfnamefont
  {D.}~\bibnamefont {Sank}}, \bibinfo {author} {\bibfnamefont {K.~J.}\
  \bibnamefont {Satzinger}}, \bibinfo {author} {\bibfnamefont {V.}~\bibnamefont
  {Smelyanskiy}}, \bibinfo {author} {\bibfnamefont {K.~J.}\ \bibnamefont
  {Sung}}, \bibinfo {author} {\bibfnamefont {M.~D.}\ \bibnamefont
  {Trevithick}}, \bibinfo {author} {\bibfnamefont {A.}~\bibnamefont
  {Vainsencher}}, \bibinfo {author} {\bibfnamefont {B.}~\bibnamefont
  {Villalonga}}, \bibinfo {author} {\bibfnamefont {T.}~\bibnamefont {White}},
  \bibinfo {author} {\bibfnamefont {Z.~J.}\ \bibnamefont {Yao}}, \bibinfo
  {author} {\bibfnamefont {P.}~\bibnamefont {Yeh}}, \bibinfo {author}
  {\bibfnamefont {A.}~\bibnamefont {Zalcman}}, \bibinfo {author} {\bibfnamefont
  {H.}~\bibnamefont {Neven}}, \ and\ \bibinfo {author} {\bibfnamefont {J.~M.}\
  \bibnamefont {Martinis}},\ }\href {\doibase 10.1038/s41586-019-1666-5}
  {\bibfield  {journal} {\bibinfo  {journal} {Nature}\ }\textbf {\bibinfo
  {volume} {574}},\ \bibinfo {pages} {505} (\bibinfo {year}
  {2019})}\BibitemShut {NoStop}%
\bibitem [{\citenamefont {{AI Quantum}}(2020)}]{AIQuantum2020}%
  \BibitemOpen
  \bibfield  {author} {\bibinfo {author} {\bibfnamefont {G.}~\bibnamefont {{AI
  Quantum}}},\ }\href {\doibase 10.1126/science.abb9811} {\bibfield  {journal}
  {\bibinfo  {journal} {Science}\ }\textbf {\bibinfo {volume} {369}},\ \bibinfo
  {pages} {1084} (\bibinfo {year} {2020})},\ \Eprint
  {http://arxiv.org/abs/2004.04174} {arXiv:2004.04174} \BibitemShut {NoStop}%
\bibitem [{\citenamefont {Zhong}\ \emph {et~al.}(2021)\citenamefont {Zhong},
  \citenamefont {Wang}, \citenamefont {Deng}, \citenamefont {Chen},
  \citenamefont {Peng}, \citenamefont {Luo}, \citenamefont {Qin}, \citenamefont
  {Wu}, \citenamefont {Ding}, \citenamefont {Hu}, \citenamefont {Hu},
  \citenamefont {Yang}, \citenamefont {Zhang}, \citenamefont {Li},
  \citenamefont {Li}, \citenamefont {Jiang}, \citenamefont {Gan}, \citenamefont
  {Yang}, \citenamefont {You}, \citenamefont {Wang}, \citenamefont {Li},
  \citenamefont {Liu}, \citenamefont {Lu},\ and\ \citenamefont
  {Pan}}]{Zhong2021}%
  \BibitemOpen
  \bibfield  {author} {\bibinfo {author} {\bibfnamefont {H.~S.}\ \bibnamefont
  {Zhong}}, \bibinfo {author} {\bibfnamefont {H.}~\bibnamefont {Wang}},
  \bibinfo {author} {\bibfnamefont {Y.~H.}\ \bibnamefont {Deng}}, \bibinfo
  {author} {\bibfnamefont {M.~C.}\ \bibnamefont {Chen}}, \bibinfo {author}
  {\bibfnamefont {L.~C.}\ \bibnamefont {Peng}}, \bibinfo {author}
  {\bibfnamefont {Y.~H.}\ \bibnamefont {Luo}}, \bibinfo {author} {\bibfnamefont
  {J.}~\bibnamefont {Qin}}, \bibinfo {author} {\bibfnamefont {D.}~\bibnamefont
  {Wu}}, \bibinfo {author} {\bibfnamefont {X.}~\bibnamefont {Ding}}, \bibinfo
  {author} {\bibfnamefont {Y.}~\bibnamefont {Hu}}, \bibinfo {author}
  {\bibfnamefont {P.}~\bibnamefont {Hu}}, \bibinfo {author} {\bibfnamefont
  {X.~Y.}\ \bibnamefont {Yang}}, \bibinfo {author} {\bibfnamefont {W.~J.}\
  \bibnamefont {Zhang}}, \bibinfo {author} {\bibfnamefont {H.}~\bibnamefont
  {Li}}, \bibinfo {author} {\bibfnamefont {Y.}~\bibnamefont {Li}}, \bibinfo
  {author} {\bibfnamefont {X.}~\bibnamefont {Jiang}}, \bibinfo {author}
  {\bibfnamefont {L.}~\bibnamefont {Gan}}, \bibinfo {author} {\bibfnamefont
  {G.}~\bibnamefont {Yang}}, \bibinfo {author} {\bibfnamefont {L.}~\bibnamefont
  {You}}, \bibinfo {author} {\bibfnamefont {Z.}~\bibnamefont {Wang}}, \bibinfo
  {author} {\bibfnamefont {L.}~\bibnamefont {Li}}, \bibinfo {author}
  {\bibfnamefont {N.~L.}\ \bibnamefont {Liu}}, \bibinfo {author} {\bibfnamefont
  {C.~Y.}\ \bibnamefont {Lu}}, \ and\ \bibinfo {author} {\bibfnamefont {J.~W.}\
  \bibnamefont {Pan}},\ }\href {\doibase 10.1126/science.abe8770} {\bibfield
  {journal} {\bibinfo  {journal} {Science}\ }\textbf {\bibinfo {volume}
  {370}},\ \bibinfo {pages} {1460} (\bibinfo {year} {2021})},\ \Eprint
  {http://arxiv.org/abs/2012.01625} {arXiv:2012.01625} \BibitemShut {NoStop}%
\bibitem [{\citenamefont {Preskill}(2018)}]{Preskill2018}%
  \BibitemOpen
  \bibfield  {author} {\bibinfo {author} {\bibfnamefont {J.}~\bibnamefont
  {Preskill}},\ }\href {\doibase 10.22331/q-2018-08-06-79} {\bibfield
  {journal} {\bibinfo  {journal} {Quantum}\ }\textbf {\bibinfo {volume} {2}},\
  \bibinfo {pages} {79} (\bibinfo {year} {2018})},\ \Eprint
  {http://arxiv.org/abs/1801.00862} {arXiv:1801.00862} \BibitemShut {NoStop}%
\bibitem [{\citenamefont {Cerezo}\ \emph {et~al.}(2021)\citenamefont {Cerezo},
  \citenamefont {Arrasmith}, \citenamefont {Babbush}, \citenamefont {Benjamin},
  \citenamefont {Endo}, \citenamefont {Fujii}, \citenamefont {McClean},
  \citenamefont {Mitarai}, \citenamefont {Yuan}, \citenamefont {Cincio} \emph
  {et~al.}}]{cerezo2020variational}%
  \BibitemOpen
  \bibfield  {author} {\bibinfo {author} {\bibfnamefont {M.}~\bibnamefont
  {Cerezo}}, \bibinfo {author} {\bibfnamefont {A.}~\bibnamefont {Arrasmith}},
  \bibinfo {author} {\bibfnamefont {R.}~\bibnamefont {Babbush}}, \bibinfo
  {author} {\bibfnamefont {S.~C.}\ \bibnamefont {Benjamin}}, \bibinfo {author}
  {\bibfnamefont {S.}~\bibnamefont {Endo}}, \bibinfo {author} {\bibfnamefont
  {K.}~\bibnamefont {Fujii}}, \bibinfo {author} {\bibfnamefont {J.~R.}\
  \bibnamefont {McClean}}, \bibinfo {author} {\bibfnamefont {K.}~\bibnamefont
  {Mitarai}}, \bibinfo {author} {\bibfnamefont {X.}~\bibnamefont {Yuan}},
  \bibinfo {author} {\bibfnamefont {L.}~\bibnamefont {Cincio}},  \emph
  {et~al.},\ }\href@noop {} {\bibfield  {journal} {\bibinfo  {journal} {Nature
  Reviews Physics}\ ,\ \bibinfo {pages} {1}} (\bibinfo {year}
  {2021})}\BibitemShut {NoStop}%
\bibitem [{\citenamefont {Bharti}\ \emph {et~al.}(2021)\citenamefont {Bharti},
  \citenamefont {Cervera-Lierta}, \citenamefont {Kyaw}, \citenamefont {Haug},
  \citenamefont {Alperin-Lea}, \citenamefont {Anand}, \citenamefont {Degroote},
  \citenamefont {Heimonen}, \citenamefont {Kottmann}, \citenamefont {Menke}
  \emph {et~al.}}]{bharti2021noisy}%
  \BibitemOpen
  \bibfield  {author} {\bibinfo {author} {\bibfnamefont {K.}~\bibnamefont
  {Bharti}}, \bibinfo {author} {\bibfnamefont {A.}~\bibnamefont
  {Cervera-Lierta}}, \bibinfo {author} {\bibfnamefont {T.~H.}\ \bibnamefont
  {Kyaw}}, \bibinfo {author} {\bibfnamefont {T.}~\bibnamefont {Haug}}, \bibinfo
  {author} {\bibfnamefont {S.}~\bibnamefont {Alperin-Lea}}, \bibinfo {author}
  {\bibfnamefont {A.}~\bibnamefont {Anand}}, \bibinfo {author} {\bibfnamefont
  {M.}~\bibnamefont {Degroote}}, \bibinfo {author} {\bibfnamefont
  {H.}~\bibnamefont {Heimonen}}, \bibinfo {author} {\bibfnamefont {J.~S.}\
  \bibnamefont {Kottmann}}, \bibinfo {author} {\bibfnamefont {T.}~\bibnamefont
  {Menke}},  \emph {et~al.},\ }\href@noop {} {\bibfield  {journal} {\bibinfo
  {journal} {arXiv preprint arXiv:2101.08448}\ } (\bibinfo {year}
  {2021})}\BibitemShut {NoStop}%
\bibitem [{\citenamefont {Endo}\ \emph {et~al.}(2021)\citenamefont {Endo},
  \citenamefont {Cai}, \citenamefont {Benjamin},\ and\ \citenamefont
  {Yuan}}]{endo2021hybrid}%
  \BibitemOpen
  \bibfield  {author} {\bibinfo {author} {\bibfnamefont {S.}~\bibnamefont
  {Endo}}, \bibinfo {author} {\bibfnamefont {Z.}~\bibnamefont {Cai}}, \bibinfo
  {author} {\bibfnamefont {S.~C.}\ \bibnamefont {Benjamin}}, \ and\ \bibinfo
  {author} {\bibfnamefont {X.}~\bibnamefont {Yuan}},\ }\href@noop {} {\bibfield
   {journal} {\bibinfo  {journal} {Journal of the Physical Society of Japan}\
  }\textbf {\bibinfo {volume} {90}},\ \bibinfo {pages} {032001} (\bibinfo
  {year} {2021})}\BibitemShut {NoStop}%
\bibitem [{\citenamefont {McClean}\ \emph {et~al.}(2017)\citenamefont
  {McClean}, \citenamefont {Kimchi-Schwartz}, \citenamefont {Carter},\ and\
  \citenamefont {{De Jong}}}]{McClean2017}%
  \BibitemOpen
  \bibfield  {author} {\bibinfo {author} {\bibfnamefont {J.~R.}\ \bibnamefont
  {McClean}}, \bibinfo {author} {\bibfnamefont {M.~E.}\ \bibnamefont
  {Kimchi-Schwartz}}, \bibinfo {author} {\bibfnamefont {J.}~\bibnamefont
  {Carter}}, \ and\ \bibinfo {author} {\bibfnamefont {W.~A.}\ \bibnamefont {{De
  Jong}}},\ }\href {\doibase 10.1103/PhysRevA.95.042308} {\bibfield  {journal}
  {\bibinfo  {journal} {Physical Review A}\ }\textbf {\bibinfo {volume} {95}},\
  \bibinfo {pages} {1} (\bibinfo {year} {2017})},\ \Eprint
  {http://arxiv.org/abs/1603.05681} {arXiv:1603.05681} \BibitemShut {NoStop}%
\bibitem [{\citenamefont {Takeshita}\ \emph {et~al.}(2020)\citenamefont
  {Takeshita}, \citenamefont {Rubin}, \citenamefont {Jiang}, \citenamefont
  {Lee}, \citenamefont {Babbush},\ and\ \citenamefont
  {McClean}}]{Takeshita2020}%
  \BibitemOpen
  \bibfield  {author} {\bibinfo {author} {\bibfnamefont {T.}~\bibnamefont
  {Takeshita}}, \bibinfo {author} {\bibfnamefont {N.~C.}\ \bibnamefont
  {Rubin}}, \bibinfo {author} {\bibfnamefont {Z.}~\bibnamefont {Jiang}},
  \bibinfo {author} {\bibfnamefont {E.}~\bibnamefont {Lee}}, \bibinfo {author}
  {\bibfnamefont {R.}~\bibnamefont {Babbush}}, \ and\ \bibinfo {author}
  {\bibfnamefont {J.~R.}\ \bibnamefont {McClean}},\ }\href {\doibase
  10.1103/PhysRevX.10.011004} {\bibfield  {journal} {\bibinfo  {journal}
  {Physical Review X}\ }\textbf {\bibinfo {volume} {10}},\ \bibinfo {pages}
  {11004} (\bibinfo {year} {2020})},\ \Eprint {http://arxiv.org/abs/1902.10679}
  {arXiv:1902.10679} \BibitemShut {NoStop}%
\bibitem [{\citenamefont {Motta}\ \emph {et~al.}(2020)\citenamefont {Motta},
  \citenamefont {Gujarati}, \citenamefont {Rice}, \citenamefont {Kumar},
  \citenamefont {Masteran}, \citenamefont {Latone}, \citenamefont {Lee},
  \citenamefont {Valeev},\ and\ \citenamefont {Takeshita}}]{Motta2020}%
  \BibitemOpen
  \bibfield  {author} {\bibinfo {author} {\bibfnamefont {M.}~\bibnamefont
  {Motta}}, \bibinfo {author} {\bibfnamefont {T.~P.}\ \bibnamefont {Gujarati}},
  \bibinfo {author} {\bibfnamefont {J.~E.}\ \bibnamefont {Rice}}, \bibinfo
  {author} {\bibfnamefont {A.}~\bibnamefont {Kumar}}, \bibinfo {author}
  {\bibfnamefont {C.}~\bibnamefont {Masteran}}, \bibinfo {author}
  {\bibfnamefont {J.~A.}\ \bibnamefont {Latone}}, \bibinfo {author}
  {\bibfnamefont {E.}~\bibnamefont {Lee}}, \bibinfo {author} {\bibfnamefont
  {E.~F.}\ \bibnamefont {Valeev}}, \ and\ \bibinfo {author} {\bibfnamefont
  {T.~Y.}\ \bibnamefont {Takeshita}},\ }\href {\doibase
  https://doi.org/10.1039/D0CP04106H} {\bibfield  {journal} {\bibinfo
  {journal} {Physical Chemistry Chemical Physics}\ }\textbf {\bibinfo {volume}
  {22}},\ \bibinfo {pages} {24270} (\bibinfo {year} {2020})}\BibitemShut
  {NoStop}%
\bibitem [{\citenamefont {Kottmann}\ \emph
  {et~al.}(2021{\natexlab{a}})\citenamefont {Kottmann}, \citenamefont
  {Schleich}, \citenamefont {Tamayo-Mendoza},\ and\ \citenamefont
  {Aspuru-Guzik}}]{Kottmann2020a}%
  \BibitemOpen
  \bibfield  {author} {\bibinfo {author} {\bibfnamefont {J.~S.}\ \bibnamefont
  {Kottmann}}, \bibinfo {author} {\bibfnamefont {P.}~\bibnamefont {Schleich}},
  \bibinfo {author} {\bibfnamefont {T.}~\bibnamefont {Tamayo-Mendoza}}, \ and\
  \bibinfo {author} {\bibfnamefont {A.}~\bibnamefont {Aspuru-Guzik}},\ }\href
  {\doibase 10.1021/acs.jpclett.0c03410} {\bibfield  {journal} {\bibinfo
  {journal} {The Journal of Physical Chemistry Letters}\ }\textbf {\bibinfo
  {volume} {12}},\ \bibinfo {pages} {663} (\bibinfo {year}
  {2021}{\natexlab{a}})},\ \Eprint {http://arxiv.org/abs/2008.02819}
  {arXiv:2008.02819} \BibitemShut {NoStop}%
\bibitem [{\citenamefont {Zhang}\ \emph {et~al.}(2021)\citenamefont {Zhang},
  \citenamefont {Kyaw}, \citenamefont {Kottmann}, \citenamefont {Degroote},\
  and\ \citenamefont {Aspuru-Guzik}}]{zhang2021mutual}%
  \BibitemOpen
  \bibfield  {author} {\bibinfo {author} {\bibfnamefont {Z.-J.}\ \bibnamefont
  {Zhang}}, \bibinfo {author} {\bibfnamefont {T.~H.}\ \bibnamefont {Kyaw}},
  \bibinfo {author} {\bibfnamefont {J.}~\bibnamefont {Kottmann}}, \bibinfo
  {author} {\bibfnamefont {M.}~\bibnamefont {Degroote}}, \ and\ \bibinfo
  {author} {\bibfnamefont {A.}~\bibnamefont {Aspuru-Guzik}},\ }\href {\doibase
  10.1088/2058-9565/abdca4} {\bibfield  {journal} {\bibinfo  {journal} {Quantum
  Science and Technology}\ } (\bibinfo {year} {2021}),\
  10.1088/2058-9565/abdca4}\BibitemShut {NoStop}%
\bibitem [{\citenamefont {Boys}\ and\ \citenamefont
  {Handy}(1969{\natexlab{a}})}]{Boys1969a}%
  \BibitemOpen
  \bibfield  {author} {\bibinfo {author} {\bibfnamefont {S.~F.}\ \bibnamefont
  {Boys}}\ and\ \bibinfo {author} {\bibfnamefont {N.~C.}\ \bibnamefont
  {Handy}},\ }\href {\doibase 10.1098/rspa.1969.0061} {\bibfield  {journal}
  {\bibinfo  {journal} {Proceedings of the Royal Society of London. A.
  Mathematical and Physical Sciences}\ }\textbf {\bibinfo {volume} {310}},\
  \bibinfo {pages} {43} (\bibinfo {year} {1969}{\natexlab{a}})}\BibitemShut
  {NoStop}%
\bibitem [{\citenamefont {Boys}\ and\ \citenamefont
  {Handy}(1969{\natexlab{b}})}]{Boys1969b}%
  \BibitemOpen
  \bibfield  {author} {\bibinfo {author} {\bibfnamefont {S.~F.}\ \bibnamefont
  {Boys}}\ and\ \bibinfo {author} {\bibfnamefont {N.~C.}\ \bibnamefont
  {Handy}},\ }in\ \href@noop {} {\emph {\bibinfo {booktitle} {Society}}},\
  Vol.\ \bibinfo {volume} {310}\ (\bibinfo {year} {1969})\ pp.\ \bibinfo
  {pages} {63--78}\BibitemShut {NoStop}%
\bibitem [{\citenamefont {McArdle}\ and\ \citenamefont
  {Tew}(2020)}]{McArdle2020}%
  \BibitemOpen
  \bibfield  {author} {\bibinfo {author} {\bibfnamefont {S.}~\bibnamefont
  {McArdle}}\ and\ \bibinfo {author} {\bibfnamefont {D.~P.}\ \bibnamefont
  {Tew}},\ }\href@noop {} {\bibfield  {journal} {\bibinfo  {journal} {arXiv
  preprint arXiv:2006.11181}\ } (\bibinfo {year} {2020})}\BibitemShut {NoStop}%
\bibitem [{\citenamefont {McArdle}\ \emph {et~al.}(2019)\citenamefont
  {McArdle}, \citenamefont {Jones}, \citenamefont {Endo}, \citenamefont {Li},
  \citenamefont {Benjamin},\ and\ \citenamefont {Yuan}}]{McArdle2019}%
  \BibitemOpen
  \bibfield  {author} {\bibinfo {author} {\bibfnamefont {S.}~\bibnamefont
  {McArdle}}, \bibinfo {author} {\bibfnamefont {T.}~\bibnamefont {Jones}},
  \bibinfo {author} {\bibfnamefont {S.}~\bibnamefont {Endo}}, \bibinfo {author}
  {\bibfnamefont {Y.}~\bibnamefont {Li}}, \bibinfo {author} {\bibfnamefont
  {S.~C.}\ \bibnamefont {Benjamin}}, \ and\ \bibinfo {author} {\bibfnamefont
  {X.}~\bibnamefont {Yuan}},\ }\href {\doibase 10.1038/s41534-019-0187-2}
  {\bibfield  {journal} {\bibinfo  {journal} {npj Quantum Information}\
  }\textbf {\bibinfo {volume} {5}} (\bibinfo {year} {2019}),\
  10.1038/s41534-019-0187-2},\ \Eprint {http://arxiv.org/abs/1804.03023}
  {arXiv:1804.03023} \BibitemShut {NoStop}%
\bibitem [{\citenamefont {Yanai}\ and\ \citenamefont
  {Shiozaki}(2012)}]{Yanai2012}%
  \BibitemOpen
  \bibfield  {author} {\bibinfo {author} {\bibfnamefont {T.}~\bibnamefont
  {Yanai}}\ and\ \bibinfo {author} {\bibfnamefont {T.}~\bibnamefont
  {Shiozaki}},\ }\href {\doibase 10.1063/1.3688225} {\bibfield  {journal}
  {\bibinfo  {journal} {Journal of Chemical Physics}\ }\textbf {\bibinfo
  {volume} {136}} (\bibinfo {year} {2012}),\ 10.1063/1.3688225}\BibitemShut
  {NoStop}%
\bibitem [{\citenamefont {Torheyden}\ and\ \citenamefont
  {Valeev}(2009)}]{Torheyden2009}%
  \BibitemOpen
  \bibfield  {author} {\bibinfo {author} {\bibfnamefont {M.}~\bibnamefont
  {Torheyden}}\ and\ \bibinfo {author} {\bibfnamefont {E.~F.}\ \bibnamefont
  {Valeev}},\ }\href {\doibase 10.1063/1.3254836} {\bibfield  {journal}
  {\bibinfo  {journal} {Journal of Chemical Physics}\ }\textbf {\bibinfo
  {volume} {131}} (\bibinfo {year} {2009}),\ 10.1063/1.3254836}\BibitemShut
  {NoStop}%
\bibitem [{\citenamefont {Kong}\ and\ \citenamefont {Valeev}(2011)}]{Kong2011}%
  \BibitemOpen
  \bibfield  {author} {\bibinfo {author} {\bibfnamefont {L.}~\bibnamefont
  {Kong}}\ and\ \bibinfo {author} {\bibfnamefont {E.~F.}\ \bibnamefont
  {Valeev}},\ }\href {\doibase 10.1063/1.3664729} {\bibfield  {journal}
  {\bibinfo  {journal} {Journal of Chemical Physics}\ }\textbf {\bibinfo
  {volume} {135}} (\bibinfo {year} {2011}),\ 10.1063/1.3664729}\BibitemShut
  {NoStop}%
\bibitem [{\citenamefont {Roskop}\ \emph {et~al.}(2014)\citenamefont {Roskop},
  \citenamefont {Kong}, \citenamefont {Valeev}, \citenamefont {Gordon},\ and\
  \citenamefont {Windus}}]{Roskop2014}%
  \BibitemOpen
  \bibfield  {author} {\bibinfo {author} {\bibfnamefont {L.~B.}\ \bibnamefont
  {Roskop}}, \bibinfo {author} {\bibfnamefont {L.}~\bibnamefont {Kong}},
  \bibinfo {author} {\bibfnamefont {E.~F.}\ \bibnamefont {Valeev}}, \bibinfo
  {author} {\bibfnamefont {M.~S.}\ \bibnamefont {Gordon}}, \ and\ \bibinfo
  {author} {\bibfnamefont {T.~L.}\ \bibnamefont {Windus}},\ }\href {\doibase
  10.1021/ct4006773} {\bibfield  {journal} {\bibinfo  {journal} {Journal of
  Chemical Theory and Computation}\ }\textbf {\bibinfo {volume} {10}},\
  \bibinfo {pages} {90} (\bibinfo {year} {2014})}\BibitemShut {NoStop}%
\bibitem [{\citenamefont {McClean}\ \emph
  {et~al.}(2020{\natexlab{a}})\citenamefont {McClean}, \citenamefont {Jiang},
  \citenamefont {Rubin}, \citenamefont {Babbush},\ and\ \citenamefont
  {Neven}}]{McClean2020}%
  \BibitemOpen
  \bibfield  {author} {\bibinfo {author} {\bibfnamefont {J.~R.}\ \bibnamefont
  {McClean}}, \bibinfo {author} {\bibfnamefont {Z.}~\bibnamefont {Jiang}},
  \bibinfo {author} {\bibfnamefont {N.~C.}\ \bibnamefont {Rubin}}, \bibinfo
  {author} {\bibfnamefont {R.}~\bibnamefont {Babbush}}, \ and\ \bibinfo
  {author} {\bibfnamefont {H.}~\bibnamefont {Neven}},\ }\href {\doibase
  10.1038/s41467-020-14341-w} {\bibfield  {journal} {\bibinfo  {journal}
  {Nature Communications}\ }\textbf {\bibinfo {volume} {11}},\ \bibinfo {pages}
  {1} (\bibinfo {year} {2020}{\natexlab{a}})},\ \Eprint
  {http://arxiv.org/abs/1903.05786} {arXiv:1903.05786} \BibitemShut {NoStop}%
\bibitem [{\citenamefont {Kong}\ and\ \citenamefont
  {Valeev}(2010)}]{Kong2010a}%
  \BibitemOpen
  \bibfield  {author} {\bibinfo {author} {\bibfnamefont {L.}~\bibnamefont
  {Kong}}\ and\ \bibinfo {author} {\bibfnamefont {E.~F.}\ \bibnamefont
  {Valeev}},\ }\href {\doibase 10.1063/1.3499600} {\bibfield  {journal}
  {\bibinfo  {journal} {Journal of Chemical Physics}\ }\textbf {\bibinfo
  {volume} {133}},\ \bibinfo {pages} {174126} (\bibinfo {year}
  {2010})}\BibitemShut {NoStop}%
\bibitem [{\citenamefont {Cao}\ \emph {et~al.}(2019)\citenamefont {Cao},
  \citenamefont {Romero}, \citenamefont {Olson}, \citenamefont {Degroote},
  \citenamefont {Johnson}, \citenamefont {Kieferov{\'{a}}}, \citenamefont
  {Kivlichan}, \citenamefont {Menke}, \citenamefont {Peropadre}, \citenamefont
  {Sawaya}, \citenamefont {Sim}, \citenamefont {Veis},\ and\ \citenamefont
  {Aspuru-Guzik}}]{Cao2018}%
  \BibitemOpen
  \bibfield  {author} {\bibinfo {author} {\bibfnamefont {Y.}~\bibnamefont
  {Cao}}, \bibinfo {author} {\bibfnamefont {J.}~\bibnamefont {Romero}},
  \bibinfo {author} {\bibfnamefont {J.~P.}\ \bibnamefont {Olson}}, \bibinfo
  {author} {\bibfnamefont {M.}~\bibnamefont {Degroote}}, \bibinfo {author}
  {\bibfnamefont {P.~D.}\ \bibnamefont {Johnson}}, \bibinfo {author}
  {\bibfnamefont {M.}~\bibnamefont {Kieferov{\'{a}}}}, \bibinfo {author}
  {\bibfnamefont {I.~D.}\ \bibnamefont {Kivlichan}}, \bibinfo {author}
  {\bibfnamefont {T.}~\bibnamefont {Menke}}, \bibinfo {author} {\bibfnamefont
  {B.}~\bibnamefont {Peropadre}}, \bibinfo {author} {\bibfnamefont {N.~P.}\
  \bibnamefont {Sawaya}}, \bibinfo {author} {\bibfnamefont {S.}~\bibnamefont
  {Sim}}, \bibinfo {author} {\bibfnamefont {L.}~\bibnamefont {Veis}}, \ and\
  \bibinfo {author} {\bibfnamefont {A.}~\bibnamefont {Aspuru-Guzik}},\ }\href
  {\doibase 10.1021/acs.chemrev.8b00803} {\bibfield  {journal} {\bibinfo
  {journal} {Chemical Reviews}\ }\textbf {\bibinfo {volume} {119}},\ \bibinfo
  {pages} {10856} (\bibinfo {year} {2019})},\ \Eprint
  {http://arxiv.org/abs/1812.09976} {arXiv:1812.09976} \BibitemShut {NoStop}%
\bibitem [{\citenamefont {Schleich}(2020)}]{schleich2020regularization}%
  \BibitemOpen
  \bibfield  {author} {\bibinfo {author} {\bibfnamefont {P.}~\bibnamefont
  {Schleich}},\ }\emph {\bibinfo {title} {Regularization of Quantum Chemistry
  on Quantum Computers by means of Explicit Correlation}},\ \href
  {http://www.mathcces.rwth-aachen.de/_media/3teaching/00projects/2020_ma_schleich.pdf}
  {Master's thesis},\ \bibinfo  {school} {RWTH Aachen University,
  \url{http://www.mathcces.rwth-aachen.de/_media/3teaching/00projects/2020_ma_schleich.pdf}}
  (\bibinfo {year} {2020})\BibitemShut {NoStop}%
\bibitem [{\citenamefont {Kutzelnigg}\ and\ \citenamefont
  {Mukherjee}(1997)}]{Kutzelnigg1997}%
  \BibitemOpen
  \bibfield  {author} {\bibinfo {author} {\bibfnamefont {W.}~\bibnamefont
  {Kutzelnigg}}\ and\ \bibinfo {author} {\bibfnamefont {D.}~\bibnamefont
  {Mukherjee}},\ }\href {\doibase 10.1063/1.474405} {\bibfield  {journal}
  {\bibinfo  {journal} {Journal of Chemical Physics}\ }\textbf {\bibinfo
  {volume} {107}},\ \bibinfo {pages} {432} (\bibinfo {year}
  {1997})}\BibitemShut {NoStop}%
\bibitem [{\citenamefont {Kutzelnigg}(2003)}]{Kutzelnigg2003}%
  \BibitemOpen
  \bibfield  {author} {\bibinfo {author} {\bibfnamefont {W.}~\bibnamefont
  {Kutzelnigg}},\ }in\ \href {\doibase 10.1007/978-94-017-0313-0_1} {\emph
  {\bibinfo {booktitle} {Explicitly Correlated Wave Functions in Chemistry and
  Physics}}},\ \bibinfo {editor} {edited by\ \bibinfo {editor} {\bibfnamefont
  {J.}~\bibnamefont {Rychlewski}}}\ (\bibinfo  {publisher} {Springer,
  Dordrecht},\ \bibinfo {year} {2003})\ pp.\ \bibinfo {pages}
  {3--90}\BibitemShut {NoStop}%
\bibitem [{\citenamefont {Kato}(1957)}]{Kato1957}%
  \BibitemOpen
  \bibfield  {author} {\bibinfo {author} {\bibfnamefont {T.}~\bibnamefont
  {Kato}},\ }\href {\doibase 10.1002/cpa.3160100201} {\bibfield  {journal}
  {\bibinfo  {journal} {Communications on Pure and Applied Mathematics}\
  }\textbf {\bibinfo {volume} {10}},\ \bibinfo {pages} {151} (\bibinfo {year}
  {1957})}\BibitemShut {NoStop}%
\bibitem [{\citenamefont {Lakin}(1965)}]{Lakin1965}%
  \BibitemOpen
  \bibfield  {author} {\bibinfo {author} {\bibfnamefont {W.}~\bibnamefont
  {Lakin}},\ }\href {\doibase 10.1063/1.1697255} {\bibfield  {journal}
  {\bibinfo  {journal} {The Journal of Chemical Physics}\ }\textbf {\bibinfo
  {volume} {43}},\ \bibinfo {pages} {2954} (\bibinfo {year}
  {1965})}\BibitemShut {NoStop}%
\bibitem [{\citenamefont {Pack}\ and\ \citenamefont {{Byers
  Brown}}(1966)}]{Pack1966}%
  \BibitemOpen
  \bibfield  {author} {\bibinfo {author} {\bibfnamefont {R.~T.}\ \bibnamefont
  {Pack}}\ and\ \bibinfo {author} {\bibfnamefont {W.}~\bibnamefont {{Byers
  Brown}}},\ }\href {\doibase 10.1063/1.1727605} {\bibfield  {journal}
  {\bibinfo  {journal} {The Journal of Chemical Physics}\ }\textbf {\bibinfo
  {volume} {45}},\ \bibinfo {pages} {625} (\bibinfo {year} {1966})}\BibitemShut
  {NoStop}%
\bibitem [{\citenamefont {Kutzelnigg}\ and\ \citenamefont
  {Morgan}(1992)}]{Kutzelnigg1992}%
  \BibitemOpen
  \bibfield  {author} {\bibinfo {author} {\bibfnamefont {W.}~\bibnamefont
  {Kutzelnigg}}\ and\ \bibinfo {author} {\bibfnamefont {J.~D.}\ \bibnamefont
  {Morgan}},\ }\href {\doibase 10.1063/1.462811} {\bibfield  {journal}
  {\bibinfo  {journal} {The Journal of Chemical Physics}\ }\textbf {\bibinfo
  {volume} {96}},\ \bibinfo {pages} {4484} (\bibinfo {year}
  {1992})}\BibitemShut {NoStop}%
\bibitem [{\citenamefont {Hylleraas}(1929)}]{Hylleraas1929}%
  \BibitemOpen
  \bibfield  {author} {\bibinfo {author} {\bibfnamefont {E.~A.}\ \bibnamefont
  {Hylleraas}},\ }\href {\doibase 10.1007/BF01375457} {\bibfield  {journal}
  {\bibinfo  {journal} {Zeitschrift f{\"{u}}r Physik}\ }\textbf {\bibinfo
  {volume} {54}},\ \bibinfo {pages} {347} (\bibinfo {year} {1929})}\BibitemShut
  {NoStop}%
\bibitem [{\citenamefont {Hirschfelder}(1963)}]{Hirschfelder1963}%
  \BibitemOpen
  \bibfield  {author} {\bibinfo {author} {\bibfnamefont {J.~O.}\ \bibnamefont
  {Hirschfelder}},\ }\href {\doibase 10.1063/1.1734157} {\bibfield  {journal}
  {\bibinfo  {journal} {The Journal of Chemical Physics}\ }\textbf {\bibinfo
  {volume} {39}},\ \bibinfo {pages} {3145} (\bibinfo {year}
  {1963})}\BibitemShut {NoStop}%
\bibitem [{\citenamefont {Schwartz}(1963)}]{Schwartz1963}%
  \BibitemOpen
  \bibfield  {author} {\bibinfo {author} {\bibfnamefont {C.}~\bibnamefont
  {Schwartz}},\ }\href
  {http://scholar.google.com/scholar?hl=en{\&}btnG=Search{\&}q=intitle:Estimating+Convergence+Rates+of+Variational+Calculations{\#}0}
  {\bibfield  {journal} {\bibinfo  {journal} {Methods in Computational
  Physics}\ }\textbf {\bibinfo {volume} {2}},\ \bibinfo {pages} {241} (\bibinfo
  {year} {1963})}\BibitemShut {NoStop}%
\bibitem [{\citenamefont {Hill}(1985)}]{Hill1985}%
  \BibitemOpen
  \bibfield  {author} {\bibinfo {author} {\bibfnamefont {R.~N.}\ \bibnamefont
  {Hill}},\ }\href {\doibase 10.1063/1.449481} {\bibfield  {journal} {\bibinfo
  {journal} {The Journal of Chemical Physics}\ }\textbf {\bibinfo {volume}
  {83}},\ \bibinfo {pages} {1173} (\bibinfo {year} {1985})}\BibitemShut
  {NoStop}%
\bibitem [{\citenamefont {Kutzelnigg}(2008)}]{Kutzelnigg2008}%
  \BibitemOpen
  \bibfield  {author} {\bibinfo {author} {\bibfnamefont {W.}~\bibnamefont
  {Kutzelnigg}},\ }\href {\doibase 10.1039/b805284k} {\bibfield  {journal}
  {\bibinfo  {journal} {Physical Chemistry Chemical Physics}\ }\textbf
  {\bibinfo {volume} {10}},\ \bibinfo {pages} {3460} (\bibinfo {year}
  {2008})}\BibitemShut {NoStop}%
\bibitem [{\citenamefont {Kong}\ \emph {et~al.}(2012)\citenamefont {Kong},
  \citenamefont {Bischoff},\ and\ \citenamefont {Valeev}}]{Kong2012}%
  \BibitemOpen
  \bibfield  {author} {\bibinfo {author} {\bibfnamefont {L.}~\bibnamefont
  {Kong}}, \bibinfo {author} {\bibfnamefont {F.~A.}\ \bibnamefont {Bischoff}},
  \ and\ \bibinfo {author} {\bibfnamefont {E.~F.}\ \bibnamefont {Valeev}},\
  }\href {\doibase 10.1021/cr200204r} {\bibfield  {journal} {\bibinfo
  {journal} {Chemical Reviews}\ }\textbf {\bibinfo {volume} {112}},\ \bibinfo
  {pages} {75} (\bibinfo {year} {2012})}\BibitemShut {NoStop}%
\bibitem [{\citenamefont {Klopper}\ \emph {et~al.}(2006)\citenamefont
  {Klopper}, \citenamefont {Manby}, \citenamefont {Ten-No},\ and\ \citenamefont
  {Valeev}}]{Klopper2006}%
  \BibitemOpen
  \bibfield  {author} {\bibinfo {author} {\bibfnamefont {W.}~\bibnamefont
  {Klopper}}, \bibinfo {author} {\bibfnamefont {F.~R.}\ \bibnamefont {Manby}},
  \bibinfo {author} {\bibfnamefont {S.}~\bibnamefont {Ten-No}}, \ and\ \bibinfo
  {author} {\bibfnamefont {E.~F.}\ \bibnamefont {Valeev}},\ }\href {\doibase
  10.1080/01442350600799921} {\bibfield  {journal} {\bibinfo  {journal}
  {International Reviews in Physical Chemistry}\ }\textbf {\bibinfo {volume}
  {25}},\ \bibinfo {pages} {427} (\bibinfo {year} {2006})}\BibitemShut
  {NoStop}%
\bibitem [{\citenamefont {Ten-no}\ and\ \citenamefont
  {Noga}(2012)}]{Ten-no2012}%
  \BibitemOpen
  \bibfield  {author} {\bibinfo {author} {\bibfnamefont {S.}~\bibnamefont
  {Ten-no}}\ and\ \bibinfo {author} {\bibfnamefont {J.}~\bibnamefont {Noga}},\
  }\href {\doibase 10.1002/wcms.68} {\bibfield  {journal} {\bibinfo  {journal}
  {Wiley Interdisciplinary Reviews: Computational Molecular Science}\ }\textbf
  {\bibinfo {volume} {2}},\ \bibinfo {pages} {114} (\bibinfo {year}
  {2012})}\BibitemShut {NoStop}%
\bibitem [{\citenamefont {Gr{\"{u}}neis}\ \emph {et~al.}(2017)\citenamefont
  {Gr{\"{u}}neis}, \citenamefont {Hirata}, \citenamefont {ya~Ohnishi},\ and\
  \citenamefont {Ten-No}}]{Gruneis2017}%
  \BibitemOpen
  \bibfield  {author} {\bibinfo {author} {\bibfnamefont {A.}~\bibnamefont
  {Gr{\"{u}}neis}}, \bibinfo {author} {\bibfnamefont {S.}~\bibnamefont
  {Hirata}}, \bibinfo {author} {\bibfnamefont {Y.}~\bibnamefont {ya~Ohnishi}},
  \ and\ \bibinfo {author} {\bibfnamefont {S.}~\bibnamefont {Ten-No}},\ }\href
  {\doibase 10.1063/1.4976974} {\enquote {\bibinfo {title} {{Perspective:
  Explicitly correlated electronic structure theory for complex systems}},}\ }
  (\bibinfo {year} {2017})\BibitemShut {NoStop}%
\bibitem [{\citenamefont {Sims}\ and\ \citenamefont
  {Hagstrom}(1971)}]{sims1971hyllci}%
  \BibitemOpen
  \bibfield  {author} {\bibinfo {author} {\bibfnamefont {J.~S.}\ \bibnamefont
  {Sims}}\ and\ \bibinfo {author} {\bibfnamefont {S.}~\bibnamefont
  {Hagstrom}},\ }\href {\doibase 10.1103/PhysRevA.4.908} {\bibfield  {journal}
  {\bibinfo  {journal} {Phys. Rev. A}\ }\textbf {\bibinfo {volume} {4}},\
  \bibinfo {pages} {908} (\bibinfo {year} {1971})}\BibitemShut {NoStop}%
\bibitem [{\citenamefont {Canc{\`{e}}s}\ \emph {et~al.}(2003)\citenamefont
  {Canc{\`{e}}s}, \citenamefont {Defranceschi}, \citenamefont {Kutzelnigg},
  \citenamefont {{Le Bris}},\ and\ \citenamefont {Maday}}]{Cances2003}%
  \BibitemOpen
  \bibfield  {author} {\bibinfo {author} {\bibfnamefont {E.}~\bibnamefont
  {Canc{\`{e}}s}}, \bibinfo {author} {\bibfnamefont {M.}~\bibnamefont
  {Defranceschi}}, \bibinfo {author} {\bibfnamefont {W.}~\bibnamefont
  {Kutzelnigg}}, \bibinfo {author} {\bibfnamefont {C.}~\bibnamefont {{Le
  Bris}}}, \ and\ \bibinfo {author} {\bibfnamefont {Y.}~\bibnamefont {Maday}},\
  }\href {\doibase 10.1016/S1570-8659(03)10003-8} {\bibfield  {journal}
  {\bibinfo  {journal} {Handbook of Numerical Analysis}\ }\textbf {\bibinfo
  {volume} {10}},\ \bibinfo {pages} {3} (\bibinfo {year} {2003})}\BibitemShut
  {NoStop}%
\bibitem [{\citenamefont {Kutzelnigg}(1985)}]{Kutzelnigg1985}%
  \BibitemOpen
  \bibfield  {author} {\bibinfo {author} {\bibfnamefont {W.}~\bibnamefont
  {Kutzelnigg}},\ }\href {\doibase 10.1007/BF00527669} {\bibfield  {journal}
  {\bibinfo  {journal} {Theoretica Chimica Acta}\ }\textbf {\bibinfo {volume}
  {68}},\ \bibinfo {pages} {445} (\bibinfo {year} {1985})}\BibitemShut
  {NoStop}%
\bibitem [{\citenamefont {Klopper}\ and\ \citenamefont
  {Kutzelnigg}(1987)}]{Klopper1987}%
  \BibitemOpen
  \bibfield  {author} {\bibinfo {author} {\bibfnamefont {W.}~\bibnamefont
  {Klopper}}\ and\ \bibinfo {author} {\bibfnamefont {W.}~\bibnamefont
  {Kutzelnigg}},\ }\href {\doibase 10.1016/0009-2614(87)80005-2} {\bibfield
  {journal} {\bibinfo  {journal} {Chemical Physics Letters}\ }\textbf {\bibinfo
  {volume} {134}},\ \bibinfo {pages} {17} (\bibinfo {year} {1987})}\BibitemShut
  {NoStop}%
\bibitem [{\citenamefont {Klopper}\ and\ \citenamefont
  {Kutzelnigg}(1991)}]{Klopper1991}%
  \BibitemOpen
  \bibfield  {author} {\bibinfo {author} {\bibfnamefont {W.}~\bibnamefont
  {Klopper}}\ and\ \bibinfo {author} {\bibfnamefont {W.}~\bibnamefont
  {Kutzelnigg}},\ }\href {\doibase 10.1063/1.459923} {\bibfield  {journal}
  {\bibinfo  {journal} {The Journal of Chemical Physics}\ }\textbf {\bibinfo
  {volume} {94}},\ \bibinfo {pages} {2020} (\bibinfo {year}
  {1991})}\BibitemShut {NoStop}%
\bibitem [{\citenamefont {Kutzelnigg}\ and\ \citenamefont
  {Klopper}(1991)}]{Kutzelnigg1991}%
  \BibitemOpen
  \bibfield  {author} {\bibinfo {author} {\bibfnamefont {W.}~\bibnamefont
  {Kutzelnigg}}\ and\ \bibinfo {author} {\bibfnamefont {W.}~\bibnamefont
  {Klopper}},\ }\href {\doibase 10.1063/1.459921} {\bibfield  {journal}
  {\bibinfo  {journal} {The Journal of Chemical Physics}\ }\textbf {\bibinfo
  {volume} {94}},\ \bibinfo {pages} {1985} (\bibinfo {year}
  {1991})}\BibitemShut {NoStop}%
\bibitem [{\citenamefont {Ten-No}(2007)}]{Ten-no2007correction}%
  \BibitemOpen
  \bibfield  {author} {\bibinfo {author} {\bibfnamefont {S.}~\bibnamefont
  {Ten-No}},\ }\href {\doibase 10.1016/j.cplett.2007.09.006} {\bibfield
  {journal} {\bibinfo  {journal} {Chemical Physics Letters}\ }\textbf {\bibinfo
  {volume} {447}},\ \bibinfo {pages} {175} (\bibinfo {year}
  {2007})}\BibitemShut {NoStop}%
\bibitem [{\citenamefont {Boys}\ and\ \citenamefont
  {Handy}(1969{\natexlab{c}})}]{Boys1969}%
  \BibitemOpen
  \bibfield  {author} {\bibinfo {author} {\bibfnamefont {S.~F.}\ \bibnamefont
  {Boys}}\ and\ \bibinfo {author} {\bibfnamefont {N.~C.}\ \bibnamefont
  {Handy}},\ }\href {\doibase 10.1098/rspa.1969.0038} {\bibfield  {journal}
  {\bibinfo  {journal} {Proceedings of the Royal Society of London. A.
  Mathematical and Physical Sciences}\ }\textbf {\bibinfo {volume} {309}},\
  \bibinfo {pages} {209} (\bibinfo {year} {1969}{\natexlab{c}})}\BibitemShut
  {NoStop}%
\bibitem [{\citenamefont {Luo}(2010)}]{Luo2010}%
  \BibitemOpen
  \bibfield  {author} {\bibinfo {author} {\bibfnamefont {H.}~\bibnamefont
  {Luo}},\ }\href {\doibase 10.1063/1.3505037} {\bibfield  {journal} {\bibinfo
  {journal} {Journal of Chemical Physics}\ }\textbf {\bibinfo {volume} {133}}
  (\bibinfo {year} {2010}),\ 10.1063/1.3505037}\BibitemShut {NoStop}%
\bibitem [{\citenamefont {Kersten}\ \emph {et~al.}(2016)\citenamefont
  {Kersten}, \citenamefont {Booth},\ and\ \citenamefont {Alavi}}]{Kersten2016}%
  \BibitemOpen
  \bibfield  {author} {\bibinfo {author} {\bibfnamefont {J.~A.}\ \bibnamefont
  {Kersten}}, \bibinfo {author} {\bibfnamefont {G.~H.}\ \bibnamefont {Booth}},
  \ and\ \bibinfo {author} {\bibfnamefont {A.}~\bibnamefont {Alavi}},\ }\href
  {\doibase 10.1063/1.4959245} {\bibfield  {journal} {\bibinfo  {journal}
  {Journal of Chemical Physics}\ }\textbf {\bibinfo {volume} {145}},\ \bibinfo
  {pages} {054117} (\bibinfo {year} {2016})},\ \Eprint
  {http://arxiv.org/abs/1605.07065} {arXiv:1605.07065} \BibitemShut {NoStop}%
\bibitem [{\citenamefont {Boys}\ and\ \citenamefont
  {Wilkes}(1960)}]{boys1960integral}%
  \BibitemOpen
  \bibfield  {author} {\bibinfo {author} {\bibfnamefont {S.~F.}\ \bibnamefont
  {Boys}}\ and\ \bibinfo {author} {\bibfnamefont {M.~V.}\ \bibnamefont
  {Wilkes}},\ }\href {\doibase 10.1098/rspa.1960.0195} {\bibfield  {journal}
  {\bibinfo  {journal} {Proceedings of the Royal Society of London. Series A.
  Mathematical and Physical Sciences}\ }\textbf {\bibinfo {volume} {258}},\
  \bibinfo {pages} {402} (\bibinfo {year} {1960})}\BibitemShut {NoStop}%
\bibitem [{\citenamefont {Singer}\ and\ \citenamefont
  {Wilkes}(1960)}]{singer1960gaussian}%
  \BibitemOpen
  \bibfield  {author} {\bibinfo {author} {\bibfnamefont {K.}~\bibnamefont
  {Singer}}\ and\ \bibinfo {author} {\bibfnamefont {M.~V.}\ \bibnamefont
  {Wilkes}},\ }\href {\doibase 10.1098/rspa.1960.0196} {\bibfield  {journal}
  {\bibinfo  {journal} {Proceedings of the Royal Society of London. Series A.
  Mathematical and Physical Sciences}\ }\textbf {\bibinfo {volume} {258}},\
  \bibinfo {pages} {412} (\bibinfo {year} {1960})}\BibitemShut {NoStop}%
\bibitem [{\citenamefont {Ten-No}(2004{\natexlab{a}})}]{Ten-no2004slater}%
  \BibitemOpen
  \bibfield  {author} {\bibinfo {author} {\bibfnamefont {S.}~\bibnamefont
  {Ten-No}},\ }\href {\doibase 10.1016/j.cplett.2004.09.041} {\bibfield
  {journal} {\bibinfo  {journal} {Chemical Physics Letters}\ }\textbf {\bibinfo
  {volume} {398}},\ \bibinfo {pages} {56} (\bibinfo {year}
  {2004}{\natexlab{a}})}\BibitemShut {NoStop}%
\bibitem [{\citenamefont {Johnson}\ \emph {et~al.}(2017)\citenamefont
  {Johnson}, \citenamefont {Hirata},\ and\ \citenamefont
  {Ten-no}}]{Johnson2017}%
  \BibitemOpen
  \bibfield  {author} {\bibinfo {author} {\bibfnamefont {C.~M.}\ \bibnamefont
  {Johnson}}, \bibinfo {author} {\bibfnamefont {S.}~\bibnamefont {Hirata}}, \
  and\ \bibinfo {author} {\bibfnamefont {S.}~\bibnamefont {Ten-no}},\ }\href
  {\doibase 10.1016/j.cplett.2017.02.072} {\bibfield  {journal} {\bibinfo
  {journal} {Chemical Physics Letters}\ }\textbf {\bibinfo {volume} {683}},\
  \bibinfo {pages} {247} (\bibinfo {year} {2017})}\BibitemShut {NoStop}%
\bibitem [{\citenamefont {Smith}\ \emph {et~al.}(2020)\citenamefont {Smith},
  \citenamefont {Burns}, \citenamefont {Simmonett}, \citenamefont {Parrish},
  \citenamefont {Schieber}, \citenamefont {Galvelis}, \citenamefont {Kraus},
  \citenamefont {Kruse}, \citenamefont {{Di Remigio}}, \citenamefont
  {Alenaizan}, \citenamefont {James}, \citenamefont {Lehtola}, \citenamefont
  {Misiewicz}, \citenamefont {Scheurer}, \citenamefont {Shaw}, \citenamefont
  {Schriber}, \citenamefont {Xie}, \citenamefont {Glick}, \citenamefont
  {Sirianni}, \citenamefont {O'Brien}, \citenamefont {Waldrop}, \citenamefont
  {Kumar}, \citenamefont {Hohenstein}, \citenamefont {Pritchard}, \citenamefont
  {Brooks}, \citenamefont {Schaefer}, \citenamefont {Sokolov}, \citenamefont
  {Patkowski}, \citenamefont {Deprince}, \citenamefont {Bozkaya}, \citenamefont
  {King}, \citenamefont {Evangelista}, \citenamefont {Turney}, \citenamefont
  {Crawford},\ and\ \citenamefont {Sherrill}}]{Smith2020}%
  \BibitemOpen
  \bibfield  {author} {\bibinfo {author} {\bibfnamefont {D.~G.}\ \bibnamefont
  {Smith}}, \bibinfo {author} {\bibfnamefont {L.~A.}\ \bibnamefont {Burns}},
  \bibinfo {author} {\bibfnamefont {A.~C.}\ \bibnamefont {Simmonett}}, \bibinfo
  {author} {\bibfnamefont {R.~M.}\ \bibnamefont {Parrish}}, \bibinfo {author}
  {\bibfnamefont {M.~C.}\ \bibnamefont {Schieber}}, \bibinfo {author}
  {\bibfnamefont {R.}~\bibnamefont {Galvelis}}, \bibinfo {author}
  {\bibfnamefont {P.}~\bibnamefont {Kraus}}, \bibinfo {author} {\bibfnamefont
  {H.}~\bibnamefont {Kruse}}, \bibinfo {author} {\bibfnamefont
  {R.}~\bibnamefont {{Di Remigio}}}, \bibinfo {author} {\bibfnamefont
  {A.}~\bibnamefont {Alenaizan}}, \bibinfo {author} {\bibfnamefont {A.~M.}\
  \bibnamefont {James}}, \bibinfo {author} {\bibfnamefont {S.}~\bibnamefont
  {Lehtola}}, \bibinfo {author} {\bibfnamefont {J.~P.}\ \bibnamefont
  {Misiewicz}}, \bibinfo {author} {\bibfnamefont {M.}~\bibnamefont {Scheurer}},
  \bibinfo {author} {\bibfnamefont {R.~A.}\ \bibnamefont {Shaw}}, \bibinfo
  {author} {\bibfnamefont {J.~B.}\ \bibnamefont {Schriber}}, \bibinfo {author}
  {\bibfnamefont {Y.}~\bibnamefont {Xie}}, \bibinfo {author} {\bibfnamefont
  {Z.~L.}\ \bibnamefont {Glick}}, \bibinfo {author} {\bibfnamefont {D.~A.}\
  \bibnamefont {Sirianni}}, \bibinfo {author} {\bibfnamefont {J.~S.}\
  \bibnamefont {O'Brien}}, \bibinfo {author} {\bibfnamefont {J.~M.}\
  \bibnamefont {Waldrop}}, \bibinfo {author} {\bibfnamefont {A.}~\bibnamefont
  {Kumar}}, \bibinfo {author} {\bibfnamefont {E.~G.}\ \bibnamefont
  {Hohenstein}}, \bibinfo {author} {\bibfnamefont {B.~P.}\ \bibnamefont
  {Pritchard}}, \bibinfo {author} {\bibfnamefont {B.~R.}\ \bibnamefont
  {Brooks}}, \bibinfo {author} {\bibfnamefont {H.~F.}\ \bibnamefont
  {Schaefer}}, \bibinfo {author} {\bibfnamefont {A.~Y.}\ \bibnamefont
  {Sokolov}}, \bibinfo {author} {\bibfnamefont {K.}~\bibnamefont {Patkowski}},
  \bibinfo {author} {\bibfnamefont {A.~E.}\ \bibnamefont {Deprince}}, \bibinfo
  {author} {\bibfnamefont {U.}~\bibnamefont {Bozkaya}}, \bibinfo {author}
  {\bibfnamefont {R.~A.}\ \bibnamefont {King}}, \bibinfo {author}
  {\bibfnamefont {F.~A.}\ \bibnamefont {Evangelista}}, \bibinfo {author}
  {\bibfnamefont {J.~M.}\ \bibnamefont {Turney}}, \bibinfo {author}
  {\bibfnamefont {T.~D.}\ \bibnamefont {Crawford}}, \ and\ \bibinfo {author}
  {\bibfnamefont {C.~D.}\ \bibnamefont {Sherrill}},\ }\href {\doibase
  10.1063/5.0006002} {\bibfield  {journal} {\bibinfo  {journal} {Journal of
  Chemical Physics}\ }\textbf {\bibinfo {volume} {152}},\ \bibinfo {pages}
  {184108} (\bibinfo {year} {2020})}\BibitemShut {NoStop}%
\bibitem [{\citenamefont {Kutzelnigg}\ and\ \citenamefont
  {Mukherjee}(1999)}]{Kutzelnigg1999}%
  \BibitemOpen
  \bibfield  {author} {\bibinfo {author} {\bibfnamefont {W.}~\bibnamefont
  {Kutzelnigg}}\ and\ \bibinfo {author} {\bibfnamefont {D.}~\bibnamefont
  {Mukherjee}},\ }\href {\doibase 10.1063/1.478189} {\bibfield  {journal}
  {\bibinfo  {journal} {Journal of Chemical Physics}\ }\textbf {\bibinfo
  {volume} {110}},\ \bibinfo {pages} {2800} (\bibinfo {year}
  {1999})}\BibitemShut {NoStop}%
\bibitem [{\citenamefont {Kutzelnigg}\ \emph {et~al.}(2010)\citenamefont
  {Kutzelnigg}, \citenamefont {Shamasundar},\ and\ \citenamefont
  {Mukherjee}}]{Kutzelnigg2010}%
  \BibitemOpen
  \bibfield  {author} {\bibinfo {author} {\bibfnamefont {W.}~\bibnamefont
  {Kutzelnigg}}, \bibinfo {author} {\bibfnamefont {K.~R.}\ \bibnamefont
  {Shamasundar}}, \ and\ \bibinfo {author} {\bibfnamefont {D.}~\bibnamefont
  {Mukherjee}},\ }\href {\doibase 10.1080/00268970903547926} {\bibfield
  {journal} {\bibinfo  {journal} {Molecular Physics}\ }\textbf {\bibinfo
  {volume} {108}},\ \bibinfo {pages} {433} (\bibinfo {year}
  {2010})}\BibitemShut {NoStop}%
\bibitem [{\citenamefont {Ten-No}(2004{\natexlab{b}})}]{Ten-No2004}%
  \BibitemOpen
  \bibfield  {author} {\bibinfo {author} {\bibfnamefont {S.}~\bibnamefont
  {Ten-No}},\ }\href {\doibase 10.1063/1.1757439} {\bibfield  {journal}
  {\bibinfo  {journal} {Journal of Chemical Physics}\ }\textbf {\bibinfo
  {volume} {121}},\ \bibinfo {pages} {117} (\bibinfo {year}
  {2004}{\natexlab{b}})}\BibitemShut {NoStop}%
\bibitem [{\citenamefont {Peruzzo}\ \emph {et~al.}(2014)\citenamefont
  {Peruzzo}, \citenamefont {McClean}, \citenamefont {Shadbolt}, \citenamefont
  {Yung}, \citenamefont {Zhou}, \citenamefont {Love}, \citenamefont
  {Aspuru-Guzik},\ and\ \citenamefont {O'Brien}}]{Peruzzo2014}%
  \BibitemOpen
  \bibfield  {author} {\bibinfo {author} {\bibfnamefont {A.}~\bibnamefont
  {Peruzzo}}, \bibinfo {author} {\bibfnamefont {J.}~\bibnamefont {McClean}},
  \bibinfo {author} {\bibfnamefont {P.}~\bibnamefont {Shadbolt}}, \bibinfo
  {author} {\bibfnamefont {M.~H.}\ \bibnamefont {Yung}}, \bibinfo {author}
  {\bibfnamefont {X.~Q.}\ \bibnamefont {Zhou}}, \bibinfo {author}
  {\bibfnamefont {P.~J.}\ \bibnamefont {Love}}, \bibinfo {author}
  {\bibfnamefont {A.}~\bibnamefont {Aspuru-Guzik}}, \ and\ \bibinfo {author}
  {\bibfnamefont {J.~L.}\ \bibnamefont {O'Brien}},\ }\href {\doibase
  10.1038/ncomms5213} {\bibfield  {journal} {\bibinfo  {journal} {Nature
  Communications}\ }\textbf {\bibinfo {volume} {5}} (\bibinfo {year} {2014}),\
  10.1038/ncomms5213}\BibitemShut {NoStop}%
\bibitem [{\citenamefont {Kottmann}\ \emph {et~al.}(2020)\citenamefont
  {Kottmann}, \citenamefont {Bischoff},\ and\ \citenamefont
  {Valeev}}]{Kottmann2020}%
  \BibitemOpen
  \bibfield  {author} {\bibinfo {author} {\bibfnamefont {J.~S.}\ \bibnamefont
  {Kottmann}}, \bibinfo {author} {\bibfnamefont {F.~A.}\ \bibnamefont
  {Bischoff}}, \ and\ \bibinfo {author} {\bibfnamefont {E.~F.}\ \bibnamefont
  {Valeev}},\ }\href {\doibase 10.1063/1.5141880} {\bibfield  {journal}
  {\bibinfo  {journal} {Journal of Chemical Physics}\ }\textbf {\bibinfo
  {volume} {152}},\ \bibinfo {pages} {074105} (\bibinfo {year}
  {2020})}\BibitemShut {NoStop}%
\bibitem [{\citenamefont {Harrison}\ \emph {et~al.}(2016)\citenamefont
  {Harrison}, \citenamefont {Beylkin}, \citenamefont {Bischoff}, \citenamefont
  {Calvin}, \citenamefont {Fann}, \citenamefont {Fosso-Tande}, \citenamefont
  {Galindo}, \citenamefont {Hammond}, \citenamefont {Hartman-Baker},
  \citenamefont {Hill}, \citenamefont {Jia}, \citenamefont {Kottmann},
  \citenamefont {Ou}, \citenamefont {Pei}, \citenamefont {Ratcliff},
  \citenamefont {Reuter}, \citenamefont {Richie-Halford}, \citenamefont
  {Romero}, \citenamefont {Sekino}, \citenamefont {Shelton}, \citenamefont
  {Sundahl}, \citenamefont {Thornton}, \citenamefont {Valeev}, \citenamefont
  {V{\'{a}}zquez-Mayagoitia}, \citenamefont {Vence}, \citenamefont {Yanai},\
  and\ \citenamefont {Yokoi}}]{Harrison2016}%
  \BibitemOpen
  \bibfield  {author} {\bibinfo {author} {\bibfnamefont {R.~J.}\ \bibnamefont
  {Harrison}}, \bibinfo {author} {\bibfnamefont {G.}~\bibnamefont {Beylkin}},
  \bibinfo {author} {\bibfnamefont {F.~A.}\ \bibnamefont {Bischoff}}, \bibinfo
  {author} {\bibfnamefont {J.~A.}\ \bibnamefont {Calvin}}, \bibinfo {author}
  {\bibfnamefont {G.~I.}\ \bibnamefont {Fann}}, \bibinfo {author}
  {\bibfnamefont {J.}~\bibnamefont {Fosso-Tande}}, \bibinfo {author}
  {\bibfnamefont {D.}~\bibnamefont {Galindo}}, \bibinfo {author} {\bibfnamefont
  {J.~R.}\ \bibnamefont {Hammond}}, \bibinfo {author} {\bibfnamefont
  {R.}~\bibnamefont {Hartman-Baker}}, \bibinfo {author} {\bibfnamefont {J.~C.}\
  \bibnamefont {Hill}}, \bibinfo {author} {\bibfnamefont {J.}~\bibnamefont
  {Jia}}, \bibinfo {author} {\bibfnamefont {J.~S.}\ \bibnamefont {Kottmann}},
  \bibinfo {author} {\bibfnamefont {M.~J.}\ \bibnamefont {Ou}}, \bibinfo
  {author} {\bibfnamefont {J.}~\bibnamefont {Pei}}, \bibinfo {author}
  {\bibfnamefont {L.~E.}\ \bibnamefont {Ratcliff}}, \bibinfo {author}
  {\bibfnamefont {M.~G.}\ \bibnamefont {Reuter}}, \bibinfo {author}
  {\bibfnamefont {A.~C.}\ \bibnamefont {Richie-Halford}}, \bibinfo {author}
  {\bibfnamefont {N.~A.}\ \bibnamefont {Romero}}, \bibinfo {author}
  {\bibfnamefont {H.}~\bibnamefont {Sekino}}, \bibinfo {author} {\bibfnamefont
  {W.~A.}\ \bibnamefont {Shelton}}, \bibinfo {author} {\bibfnamefont {B.~E.}\
  \bibnamefont {Sundahl}}, \bibinfo {author} {\bibfnamefont {W.~S.}\
  \bibnamefont {Thornton}}, \bibinfo {author} {\bibfnamefont {E.~F.}\
  \bibnamefont {Valeev}}, \bibinfo {author} {\bibfnamefont
  {{\'{A}}.}~\bibnamefont {V{\'{a}}zquez-Mayagoitia}}, \bibinfo {author}
  {\bibfnamefont {N.}~\bibnamefont {Vence}}, \bibinfo {author} {\bibfnamefont
  {T.}~\bibnamefont {Yanai}}, \ and\ \bibinfo {author} {\bibfnamefont
  {Y.}~\bibnamefont {Yokoi}},\ }in\ \href {\doibase 10.1137/15M1026171} {\emph
  {\bibinfo {booktitle} {{Madness: A multiresolution, adaptive numerical
  environment for scientific simulation}}}},\ Vol.~\bibinfo {volume} {38}\
  (\bibinfo  {publisher} {Society for Industrial and Applied Mathematics
  Publications},\ \bibinfo {year} {2016})\ pp.\ \bibinfo {pages}
  {S123--S142}\BibitemShut {NoStop}%
\bibitem [{\citenamefont {Valeev}(2004)}]{Valeev2004a}%
  \BibitemOpen
  \bibfield  {author} {\bibinfo {author} {\bibfnamefont {E.~F.}\ \bibnamefont
  {Valeev}},\ }\href {\doibase 10.1016/j.cplett.2004.07.061} {\bibfield
  {journal} {\bibinfo  {journal} {Chemical Physics Letters}\ }\textbf {\bibinfo
  {volume} {395}},\ \bibinfo {pages} {190} (\bibinfo {year}
  {2004})}\BibitemShut {NoStop}%
\bibitem [{Note1()}]{Note1}%
  \BibitemOpen
  \bibinfo {note} {This is one way to derive it. One could also blindly start
  with some perturbation and the second-order Hylleraas functional,
  interpreting the RI-approximation in the usual sense to break down
  higher-order integrals. But since this approach well showcases where the
  correction for \protect \emph {basis-set incompleteness} comes from, we
  choose to sketch it here. A more thorough outline can be found in Ref.~\cite
  {schleich2020regularization}.}\BibitemShut {Stop}%
\bibitem [{\citenamefont {L{\"{o}}wdin}(1962)}]{Lowdin1962}%
  \BibitemOpen
  \bibfield  {author} {\bibinfo {author} {\bibfnamefont {P.~O.}\ \bibnamefont
  {L{\"{o}}wdin}},\ }\href {\doibase 10.1063/1.1724312} {\bibfield  {journal}
  {\bibinfo  {journal} {Journal of Mathematical Physics}\ }\textbf {\bibinfo
  {volume} {3}},\ \bibinfo {pages} {969} (\bibinfo {year} {1962})}\BibitemShut
  {NoStop}%
\bibitem [{\citenamefont {Valeev}(2008)}]{Valeev2008}%
  \BibitemOpen
  \bibfield  {author} {\bibinfo {author} {\bibfnamefont {E.~F.}\ \bibnamefont
  {Valeev}},\ }\href {\doibase 10.1039/b713938a} {\bibfield  {journal}
  {\bibinfo  {journal} {Physical Chemistry Chemical Physics}\ }\textbf
  {\bibinfo {volume} {10}},\ \bibinfo {pages} {106} (\bibinfo {year}
  {2008})}\BibitemShut {NoStop}%
\bibitem [{\citenamefont {Torheyden}\ and\ \citenamefont
  {Valeev}(2008)}]{Torheyden2008}%
  \BibitemOpen
  \bibfield  {author} {\bibinfo {author} {\bibfnamefont {M.}~\bibnamefont
  {Torheyden}}\ and\ \bibinfo {author} {\bibfnamefont {E.~F.}\ \bibnamefont
  {Valeev}},\ }\href {\doibase 10.1039/b803620a} {\bibfield  {journal}
  {\bibinfo  {journal} {Physical Chemistry Chemical Physics}\ }\textbf
  {\bibinfo {volume} {10}},\ \bibinfo {pages} {3410} (\bibinfo {year}
  {2008})}\BibitemShut {NoStop}%
\bibitem [{Note2()}]{Note2}%
  \BibitemOpen
  \bibinfo {note} {Cumulant approximations~\cite {Kutzelnigg1999} as well as
  the standard, extended and generalized Brillouin conditions~\cite
  {Kutzelnigg1991} and a set of so called screening approximations~\cite
  {Valeev2008,Torheyden2009,Kong2011}.}\BibitemShut {Stop}%
\bibitem [{\citenamefont {Ked{\v{z}}uch}\ \emph {et~al.}(2005)\citenamefont
  {Ked{\v{z}}uch}, \citenamefont {Milko},\ and\ \citenamefont
  {Noga}}]{Kedzuch2005}%
  \BibitemOpen
  \bibfield  {author} {\bibinfo {author} {\bibfnamefont {S.}~\bibnamefont
  {Ked{\v{z}}uch}}, \bibinfo {author} {\bibfnamefont {M.}~\bibnamefont
  {Milko}}, \ and\ \bibinfo {author} {\bibfnamefont {J.}~\bibnamefont {Noga}},\
  }\href {\doibase 10.1002/qua.20744} {\bibfield  {journal} {\bibinfo
  {journal} {International Journal of Quantum Chemistry}\ }\textbf {\bibinfo
  {volume} {105}},\ \bibinfo {pages} {929} (\bibinfo {year}
  {2005})}\BibitemShut {NoStop}%
\bibitem [{\citenamefont {Pavo{\v{s}}evi{\'c}}\ \emph
  {et~al.}(2016)\citenamefont {Pavo{\v{s}}evi{\'c}}, \citenamefont {Pinski},
  \citenamefont {Riplinger}, \citenamefont {Neese},\ and\ \citenamefont
  {Valeev}}]{pavovsevic2016sparsemaps}%
  \BibitemOpen
  \bibfield  {author} {\bibinfo {author} {\bibfnamefont {F.}~\bibnamefont
  {Pavo{\v{s}}evi{\'c}}}, \bibinfo {author} {\bibfnamefont {P.}~\bibnamefont
  {Pinski}}, \bibinfo {author} {\bibfnamefont {C.}~\bibnamefont {Riplinger}},
  \bibinfo {author} {\bibfnamefont {F.}~\bibnamefont {Neese}}, \ and\ \bibinfo
  {author} {\bibfnamefont {E.~F.}\ \bibnamefont {Valeev}},\ }\href@noop {}
  {\bibfield  {journal} {\bibinfo  {journal} {The Journal of Chemical Physics}\
  }\textbf {\bibinfo {volume} {144}},\ \bibinfo {pages} {144109} (\bibinfo
  {year} {2016})}\BibitemShut {NoStop}%
\bibitem [{\citenamefont {McClean}\ \emph {et~al.}(2016)\citenamefont
  {McClean}, \citenamefont {Romero}, \citenamefont {Babbush},\ and\
  \citenamefont {Aspuru-Guzik}}]{McClean2016}%
  \BibitemOpen
  \bibfield  {author} {\bibinfo {author} {\bibfnamefont {J.~R.}\ \bibnamefont
  {McClean}}, \bibinfo {author} {\bibfnamefont {J.}~\bibnamefont {Romero}},
  \bibinfo {author} {\bibfnamefont {R.}~\bibnamefont {Babbush}}, \ and\
  \bibinfo {author} {\bibfnamefont {A.}~\bibnamefont {Aspuru-Guzik}},\ }\href
  {\doibase 10.1088/1367-2630/18/2/023023} {\bibfield  {journal} {\bibinfo
  {journal} {New Journal of Physics}\ }\textbf {\bibinfo {volume} {18}}
  (\bibinfo {year} {2016}),\ 10.1088/1367-2630/18/2/023023},\ \Eprint
  {http://arxiv.org/abs/1509.04279} {arXiv:1509.04279} \BibitemShut {NoStop}%
\bibitem [{\citenamefont {McArdle}\ \emph {et~al.}(2020)\citenamefont
  {McArdle}, \citenamefont {Endo}, \citenamefont {Aspuru-Guzik}, \citenamefont
  {Benjamin},\ and\ \citenamefont {Yuan}}]{McArdle2018}%
  \BibitemOpen
  \bibfield  {author} {\bibinfo {author} {\bibfnamefont {S.}~\bibnamefont
  {McArdle}}, \bibinfo {author} {\bibfnamefont {S.}~\bibnamefont {Endo}},
  \bibinfo {author} {\bibfnamefont {A.}~\bibnamefont {Aspuru-Guzik}}, \bibinfo
  {author} {\bibfnamefont {S.~C.}\ \bibnamefont {Benjamin}}, \ and\ \bibinfo
  {author} {\bibfnamefont {X.}~\bibnamefont {Yuan}},\ }\href {\doibase
  10.1103/RevModPhys.92.015003} {\bibfield  {journal} {\bibinfo  {journal}
  {Reviews of Modern Physics}\ }\textbf {\bibinfo {volume} {92}} (\bibinfo
  {year} {2020}),\ 10.1103/RevModPhys.92.015003},\ \Eprint
  {http://arxiv.org/abs/1808.10402} {arXiv:1808.10402} \BibitemShut {NoStop}%
\bibitem [{\citenamefont {Romero}\ \emph {et~al.}(2019)\citenamefont {Romero},
  \citenamefont {Babbush}, \citenamefont {McClean}, \citenamefont {Hempel},
  \citenamefont {Love},\ and\ \citenamefont {Aspuru-Guzik}}]{Romero2019}%
  \BibitemOpen
  \bibfield  {author} {\bibinfo {author} {\bibfnamefont {J.}~\bibnamefont
  {Romero}}, \bibinfo {author} {\bibfnamefont {R.}~\bibnamefont {Babbush}},
  \bibinfo {author} {\bibfnamefont {J.~R.}\ \bibnamefont {McClean}}, \bibinfo
  {author} {\bibfnamefont {C.}~\bibnamefont {Hempel}}, \bibinfo {author}
  {\bibfnamefont {P.~J.}\ \bibnamefont {Love}}, \ and\ \bibinfo {author}
  {\bibfnamefont {A.}~\bibnamefont {Aspuru-Guzik}},\ }\href {\doibase
  10.1088/2058-9565/aad3e4} {\bibfield  {journal} {\bibinfo  {journal} {Quantum
  Science and Technology}\ }\textbf {\bibinfo {volume} {4}} (\bibinfo {year}
  {2019}),\ 10.1088/2058-9565/aad3e4},\ \Eprint
  {http://arxiv.org/abs/1701.02691} {arXiv:1701.02691} \BibitemShut {NoStop}%
\bibitem [{\citenamefont {Anand}\ \emph {et~al.}(2022)\citenamefont {Anand},
  \citenamefont {Schleich}, \citenamefont {Alperin-Lea}, \citenamefont
  {Jensen}, \citenamefont {Sim}, \citenamefont {D{\'\i}az-Tinoco},
  \citenamefont {Kottmann}, \citenamefont {Degroote}, \citenamefont
  {Izmaylov},\ and\ \citenamefont {Aspuru-Guzik}}]{anand2021quantum}%
  \BibitemOpen
  \bibfield  {author} {\bibinfo {author} {\bibfnamefont {A.}~\bibnamefont
  {Anand}}, \bibinfo {author} {\bibfnamefont {P.}~\bibnamefont {Schleich}},
  \bibinfo {author} {\bibfnamefont {S.}~\bibnamefont {Alperin-Lea}}, \bibinfo
  {author} {\bibfnamefont {P.~W.}\ \bibnamefont {Jensen}}, \bibinfo {author}
  {\bibfnamefont {S.}~\bibnamefont {Sim}}, \bibinfo {author} {\bibfnamefont
  {M.}~\bibnamefont {D{\'\i}az-Tinoco}}, \bibinfo {author} {\bibfnamefont
  {J.~S.}\ \bibnamefont {Kottmann}}, \bibinfo {author} {\bibfnamefont
  {M.}~\bibnamefont {Degroote}}, \bibinfo {author} {\bibfnamefont {A.~F.}\
  \bibnamefont {Izmaylov}}, \ and\ \bibinfo {author} {\bibfnamefont
  {A.}~\bibnamefont {Aspuru-Guzik}},\ }\href@noop {} {\bibfield  {journal}
  {\bibinfo  {journal} {Chemical Society Reviews}\ } (\bibinfo {year}
  {2022})}\BibitemShut {NoStop}%
\bibitem [{\citenamefont {Lee}\ \emph {et~al.}(2019)\citenamefont {Lee},
  \citenamefont {Huggins}, \citenamefont {Head-Gordon},\ and\ \citenamefont
  {Whaley}}]{Lee2019}%
  \BibitemOpen
  \bibfield  {author} {\bibinfo {author} {\bibfnamefont {J.}~\bibnamefont
  {Lee}}, \bibinfo {author} {\bibfnamefont {W.~J.}\ \bibnamefont {Huggins}},
  \bibinfo {author} {\bibfnamefont {M.}~\bibnamefont {Head-Gordon}}, \ and\
  \bibinfo {author} {\bibfnamefont {K.~B.}\ \bibnamefont {Whaley}},\ }\href
  {\doibase 10.1021/acs.jctc.8b01004} {\bibfield  {journal} {\bibinfo
  {journal} {Journal of Chemical Theory and Computation}\ }\textbf {\bibinfo
  {volume} {15}},\ \bibinfo {pages} {311} (\bibinfo {year} {2019})},\ \Eprint
  {http://arxiv.org/abs/1810.02327} {arXiv:1810.02327} \BibitemShut {NoStop}%
\bibitem [{\citenamefont {Grimsley}\ \emph {et~al.}(2019)\citenamefont
  {Grimsley}, \citenamefont {Economou}, \citenamefont {Barnes},\ and\
  \citenamefont {Mayhall}}]{Grimsley2019}%
  \BibitemOpen
  \bibfield  {author} {\bibinfo {author} {\bibfnamefont {H.~R.}\ \bibnamefont
  {Grimsley}}, \bibinfo {author} {\bibfnamefont {S.~E.}\ \bibnamefont
  {Economou}}, \bibinfo {author} {\bibfnamefont {E.}~\bibnamefont {Barnes}}, \
  and\ \bibinfo {author} {\bibfnamefont {N.~J.}\ \bibnamefont {Mayhall}},\
  }\href {\doibase 10.1038/s41467-019-10988-2} {\bibfield  {journal} {\bibinfo
  {journal} {Nature Communications}\ }\textbf {\bibinfo {volume} {10}}
  (\bibinfo {year} {2019}),\ 10.1038/s41467-019-10988-2},\ \Eprint
  {http://arxiv.org/abs/1812.11173} {arXiv:1812.11173} \BibitemShut {NoStop}%
\bibitem [{\citenamefont {Grimsley}\ \emph {et~al.}(2020)\citenamefont
  {Grimsley}, \citenamefont {Claudino}, \citenamefont {Economou}, \citenamefont
  {Barnes},\ and\ \citenamefont {Mayhall}}]{Grimsley2020}%
  \BibitemOpen
  \bibfield  {author} {\bibinfo {author} {\bibfnamefont {H.~R.}\ \bibnamefont
  {Grimsley}}, \bibinfo {author} {\bibfnamefont {D.}~\bibnamefont {Claudino}},
  \bibinfo {author} {\bibfnamefont {S.~E.}\ \bibnamefont {Economou}}, \bibinfo
  {author} {\bibfnamefont {E.}~\bibnamefont {Barnes}}, \ and\ \bibinfo {author}
  {\bibfnamefont {N.~J.}\ \bibnamefont {Mayhall}},\ }\href {\doibase
  10.1021/acs.jctc.9b01083} {\bibfield  {journal} {\bibinfo  {journal} {Journal
  of Chemical Theory and Computation}\ }\textbf {\bibinfo {volume} {16}},\
  \bibinfo {pages} {1} (\bibinfo {year} {2020})},\ \Eprint
  {http://arxiv.org/abs/1910.10329} {arXiv:1910.10329} \BibitemShut {NoStop}%
\bibitem [{\citenamefont {Izmaylov}\ \emph {et~al.}(2020)\citenamefont
  {Izmaylov}, \citenamefont {D{\'{i}}az-Tinoco},\ and\ \citenamefont
  {Lang}}]{Izmaylov2020}%
  \BibitemOpen
  \bibfield  {author} {\bibinfo {author} {\bibfnamefont {A.~F.}\ \bibnamefont
  {Izmaylov}}, \bibinfo {author} {\bibfnamefont {M.}~\bibnamefont
  {D{\'{i}}az-Tinoco}}, \ and\ \bibinfo {author} {\bibfnamefont {R.~A.}\
  \bibnamefont {Lang}},\ }\href {\doibase 10.1039/d0cp01707h} {\bibfield
  {journal} {\bibinfo  {journal} {Physical Chemistry Chemical Physics}\
  }\textbf {\bibinfo {volume} {22}},\ \bibinfo {pages} {12980} (\bibinfo {year}
  {2020})},\ \Eprint {http://arxiv.org/abs/2003.07351} {arXiv:2003.07351}
  \BibitemShut {NoStop}%
\bibitem [{\citenamefont {Ryabinkin}\ \emph {et~al.}(2018)\citenamefont
  {Ryabinkin}, \citenamefont {Yen}, \citenamefont {Genin},\ and\ \citenamefont
  {Izmaylov}}]{ryabinkin2018qubit}%
  \BibitemOpen
  \bibfield  {author} {\bibinfo {author} {\bibfnamefont {I.~G.}\ \bibnamefont
  {Ryabinkin}}, \bibinfo {author} {\bibfnamefont {T.-C.}\ \bibnamefont {Yen}},
  \bibinfo {author} {\bibfnamefont {S.~N.}\ \bibnamefont {Genin}}, \ and\
  \bibinfo {author} {\bibfnamefont {A.~F.}\ \bibnamefont {Izmaylov}},\ }\href
  {\doibase 10.1021/acs.jctc.8b00932} {\bibfield  {journal} {\bibinfo
  {journal} {Journal of Chemical Theory and Computation}\ }\textbf {\bibinfo
  {volume} {14}},\ \bibinfo {pages} {6317} (\bibinfo {year}
  {2018})}\BibitemShut {NoStop}%
\bibitem [{\citenamefont {Kottmann}\ \emph
  {et~al.}(2021{\natexlab{b}})\citenamefont {Kottmann}, \citenamefont
  {Alperin-Lea}, \citenamefont {Tamayo-Mendoza}, \citenamefont
  {Cervera-Lierta}, \citenamefont {Lavigne}, \citenamefont {Yen}, \citenamefont
  {Verteletskyi}, \citenamefont {Schleich}, \citenamefont {Anand},
  \citenamefont {Degroote},\ and\ \citenamefont {et~al.}}]{Kottmann2020b}%
  \BibitemOpen
  \bibfield  {author} {\bibinfo {author} {\bibfnamefont {J.}~\bibnamefont
  {Kottmann}}, \bibinfo {author} {\bibfnamefont {S.}~\bibnamefont
  {Alperin-Lea}}, \bibinfo {author} {\bibfnamefont {T.}~\bibnamefont
  {Tamayo-Mendoza}}, \bibinfo {author} {\bibfnamefont {A.}~\bibnamefont
  {Cervera-Lierta}}, \bibinfo {author} {\bibfnamefont {C.}~\bibnamefont
  {Lavigne}}, \bibinfo {author} {\bibfnamefont {T.-C.}\ \bibnamefont {Yen}},
  \bibinfo {author} {\bibfnamefont {V.}~\bibnamefont {Verteletskyi}}, \bibinfo
  {author} {\bibfnamefont {P.}~\bibnamefont {Schleich}}, \bibinfo {author}
  {\bibfnamefont {A.}~\bibnamefont {Anand}}, \bibinfo {author} {\bibfnamefont
  {M.}~\bibnamefont {Degroote}}, \ and\ \bibinfo {author} {\bibnamefont
  {et~al.}},\ }\href {\doibase 10.1088/2058-9565/abe567} {\bibfield  {journal}
  {\bibinfo  {journal} {Quantum Science and Technology}\ } (\bibinfo {year}
  {2021}{\natexlab{b}}),\ 10.1088/2058-9565/abe567}\BibitemShut {NoStop}%
\bibitem [{\citenamefont {Kottmann}\ \emph
  {et~al.}(2021{\natexlab{c}})\citenamefont {Kottmann}, \citenamefont {Anand},\
  and\ \citenamefont {Aspuru-Guzik}}]{kottmann2021feasible}%
  \BibitemOpen
  \bibfield  {author} {\bibinfo {author} {\bibfnamefont {J.~S.}\ \bibnamefont
  {Kottmann}}, \bibinfo {author} {\bibfnamefont {A.}~\bibnamefont {Anand}}, \
  and\ \bibinfo {author} {\bibfnamefont {A.}~\bibnamefont {Aspuru-Guzik}},\
  }\href {\doibase 10.1039/D0SC06627C} {\bibfield  {journal} {\bibinfo
  {journal} {Chem. Sci.}\ }\textbf {\bibinfo {volume} {12}},\ \bibinfo {pages}
  {3497} (\bibinfo {year} {2021}{\natexlab{c}})}\BibitemShut {NoStop}%
\bibitem [{\citenamefont {Izmaylov}\ \emph {et~al.}(2019)\citenamefont
  {Izmaylov}, \citenamefont {Yen}, \citenamefont {Lang},\ and\ \citenamefont
  {Verteletskyi}}]{izmaylov2019unitary}%
  \BibitemOpen
  \bibfield  {author} {\bibinfo {author} {\bibfnamefont {A.~F.}\ \bibnamefont
  {Izmaylov}}, \bibinfo {author} {\bibfnamefont {T.-C.}\ \bibnamefont {Yen}},
  \bibinfo {author} {\bibfnamefont {R.~A.}\ \bibnamefont {Lang}}, \ and\
  \bibinfo {author} {\bibfnamefont {V.}~\bibnamefont {Verteletskyi}},\
  }\href@noop {} {\bibfield  {journal} {\bibinfo  {journal} {Journal of
  chemical theory and computation}\ }\textbf {\bibinfo {volume} {16}},\
  \bibinfo {pages} {190} (\bibinfo {year} {2019})}\BibitemShut {NoStop}%
\bibitem [{\citenamefont {Verteletskyi}\ \emph {et~al.}(2020)\citenamefont
  {Verteletskyi}, \citenamefont {Yen},\ and\ \citenamefont
  {Izmaylov}}]{Verteletskyi2020}%
  \BibitemOpen
  \bibfield  {author} {\bibinfo {author} {\bibfnamefont {V.}~\bibnamefont
  {Verteletskyi}}, \bibinfo {author} {\bibfnamefont {T.~C.}\ \bibnamefont
  {Yen}}, \ and\ \bibinfo {author} {\bibfnamefont {A.~F.}\ \bibnamefont
  {Izmaylov}},\ }\href {\doibase 10.1063/1.5141458} {\bibfield  {journal}
  {\bibinfo  {journal} {Journal of Chemical Physics}\ }\textbf {\bibinfo
  {volume} {152}} (\bibinfo {year} {2020}),\ 10.1063/1.5141458},\ \Eprint
  {http://arxiv.org/abs/1907.03358} {arXiv:1907.03358} \BibitemShut {NoStop}%
\bibitem [{\citenamefont {Crawford}\ \emph {et~al.}(2021)\citenamefont
  {Crawford}, \citenamefont {Straaten}, \citenamefont {Wang}, \citenamefont
  {Parks}, \citenamefont {Campbell},\ and\ \citenamefont
  {Brierley}}]{crawford2021measurement}%
  \BibitemOpen
  \bibfield  {author} {\bibinfo {author} {\bibfnamefont {O.}~\bibnamefont
  {Crawford}}, \bibinfo {author} {\bibfnamefont {B.~v.}\ \bibnamefont
  {Straaten}}, \bibinfo {author} {\bibfnamefont {D.}~\bibnamefont {Wang}},
  \bibinfo {author} {\bibfnamefont {T.}~\bibnamefont {Parks}}, \bibinfo
  {author} {\bibfnamefont {E.}~\bibnamefont {Campbell}}, \ and\ \bibinfo
  {author} {\bibfnamefont {S.}~\bibnamefont {Brierley}},\ }\href {\doibase
  10.22331/q-2021-01-20-385} {\bibfield  {journal} {\bibinfo  {journal}
  {Quantum}\ }\textbf {\bibinfo {volume} {5}},\ \bibinfo {pages} {385}
  (\bibinfo {year} {2021})}\BibitemShut {NoStop}%
\bibitem [{\citenamefont {Huggins}\ \emph {et~al.}(2021)\citenamefont
  {Huggins}, \citenamefont {McClean}, \citenamefont {Rubin}, \citenamefont
  {Jiang}, \citenamefont {Wiebe}, \citenamefont {Whaley},\ and\ \citenamefont
  {Babbush}}]{Huggins2019a}%
  \BibitemOpen
  \bibfield  {author} {\bibinfo {author} {\bibfnamefont {W.~J.}\ \bibnamefont
  {Huggins}}, \bibinfo {author} {\bibfnamefont {J.~R.}\ \bibnamefont
  {McClean}}, \bibinfo {author} {\bibfnamefont {N.~C.}\ \bibnamefont {Rubin}},
  \bibinfo {author} {\bibfnamefont {Z.}~\bibnamefont {Jiang}}, \bibinfo
  {author} {\bibfnamefont {N.}~\bibnamefont {Wiebe}}, \bibinfo {author}
  {\bibfnamefont {K.~B.}\ \bibnamefont {Whaley}}, \ and\ \bibinfo {author}
  {\bibfnamefont {R.}~\bibnamefont {Babbush}},\ }\href@noop {} {\bibfield
  {journal} {\bibinfo  {journal} {npj Quantum Information}\ }\textbf {\bibinfo
  {volume} {7}},\ \bibinfo {pages} {1} (\bibinfo {year} {2021})}\BibitemShut
  {NoStop}%
\bibitem [{\citenamefont {Yen}\ and\ \citenamefont {Izmaylov}(2020)}]{Yen2020}%
  \BibitemOpen
  \bibfield  {author} {\bibinfo {author} {\bibfnamefont {T.-C.}\ \bibnamefont
  {Yen}}\ and\ \bibinfo {author} {\bibfnamefont {A.~F.}\ \bibnamefont
  {Izmaylov}},\ }\href@noop {} {\bibfield  {journal} {\bibinfo  {journal}
  {arXiv preprint arXiv:2007.01234}\ } (\bibinfo {year} {2020})}\BibitemShut
  {NoStop}%
\bibitem [{\citenamefont {Bonet-Monroig}\ \emph {et~al.}(2020)\citenamefont
  {Bonet-Monroig}, \citenamefont {Babbush},\ and\ \citenamefont
  {O'Brien}}]{bonet2020measurement}%
  \BibitemOpen
  \bibfield  {author} {\bibinfo {author} {\bibfnamefont {X.}~\bibnamefont
  {Bonet-Monroig}}, \bibinfo {author} {\bibfnamefont {R.}~\bibnamefont
  {Babbush}}, \ and\ \bibinfo {author} {\bibfnamefont {T.~E.}\ \bibnamefont
  {O'Brien}},\ }\href {\doibase 10.1103/PhysRevX.10.031064} {\bibfield
  {journal} {\bibinfo  {journal} {Phys. Rev. X}\ }\textbf {\bibinfo {volume}
  {10}},\ \bibinfo {pages} {031064} (\bibinfo {year} {2020})}\BibitemShut
  {NoStop}%
\bibitem [{\citenamefont {Rubin}\ \emph {et~al.}(2018)\citenamefont {Rubin},
  \citenamefont {Babbush},\ and\ \citenamefont {McClean}}]{Rubin2018}%
  \BibitemOpen
  \bibfield  {author} {\bibinfo {author} {\bibfnamefont {N.~C.}\ \bibnamefont
  {Rubin}}, \bibinfo {author} {\bibfnamefont {R.}~\bibnamefont {Babbush}}, \
  and\ \bibinfo {author} {\bibfnamefont {J.}~\bibnamefont {McClean}},\ }\href
  {\doibase 10.1088/1367-2630/aab919} {\bibfield  {journal} {\bibinfo
  {journal} {New Journal of Physics}\ }\textbf {\bibinfo {volume} {20}},\
  \bibinfo {pages} {053020} (\bibinfo {year} {2018})},\ \Eprint
  {http://arxiv.org/abs/1801.03524} {arXiv:1801.03524} \BibitemShut {NoStop}%
\bibitem [{\citenamefont {Gonthier}\ \emph {et~al.}(2020)\citenamefont
  {Gonthier}, \citenamefont {Radin}, \citenamefont {Buda}, \citenamefont
  {Doskocil}, \citenamefont {Abuan},\ and\ \citenamefont
  {Romero}}]{gonthier2020identifying}%
  \BibitemOpen
  \bibfield  {author} {\bibinfo {author} {\bibfnamefont {J.~F.}\ \bibnamefont
  {Gonthier}}, \bibinfo {author} {\bibfnamefont {M.~D.}\ \bibnamefont {Radin}},
  \bibinfo {author} {\bibfnamefont {C.}~\bibnamefont {Buda}}, \bibinfo {author}
  {\bibfnamefont {E.~J.}\ \bibnamefont {Doskocil}}, \bibinfo {author}
  {\bibfnamefont {C.~M.}\ \bibnamefont {Abuan}}, \ and\ \bibinfo {author}
  {\bibfnamefont {J.}~\bibnamefont {Romero}},\ }\href@noop {} {\bibfield
  {journal} {\bibinfo  {journal} {arXiv preprint arXiv:2012.04001}\ } (\bibinfo
  {year} {2020})}\BibitemShut {NoStop}%
\bibitem [{\citenamefont {Booth}\ \emph {et~al.}(2012)\citenamefont {Booth},
  \citenamefont {Cleland}, \citenamefont {Alavi},\ and\ \citenamefont
  {Tew}}]{booth2012explicitly}%
  \BibitemOpen
  \bibfield  {author} {\bibinfo {author} {\bibfnamefont {G.~H.}\ \bibnamefont
  {Booth}}, \bibinfo {author} {\bibfnamefont {D.}~\bibnamefont {Cleland}},
  \bibinfo {author} {\bibfnamefont {A.}~\bibnamefont {Alavi}}, \ and\ \bibinfo
  {author} {\bibfnamefont {D.~P.}\ \bibnamefont {Tew}},\ }\href@noop {}
  {\bibfield  {journal} {\bibinfo  {journal} {The Journal of chemical physics}\
  }\textbf {\bibinfo {volume} {137}},\ \bibinfo {pages} {164112} (\bibinfo
  {year} {2012})}\BibitemShut {NoStop}%
\bibitem [{\citenamefont {Urbanek}\ \emph {et~al.}(2020)\citenamefont
  {Urbanek}, \citenamefont {Camps}, \citenamefont {Van~Beeumen},\ and\
  \citenamefont {de~Jong}}]{urbanek2020chemistryqse}%
  \BibitemOpen
  \bibfield  {author} {\bibinfo {author} {\bibfnamefont {M.}~\bibnamefont
  {Urbanek}}, \bibinfo {author} {\bibfnamefont {D.}~\bibnamefont {Camps}},
  \bibinfo {author} {\bibfnamefont {R.}~\bibnamefont {Van~Beeumen}}, \ and\
  \bibinfo {author} {\bibfnamefont {W.~A.}\ \bibnamefont {de~Jong}},\ }\href
  {\doibase 10.1021/acs.jctc.0c00447} {\bibfield  {journal} {\bibinfo
  {journal} {Journal of Chemical Theory and Computation}\ }\textbf {\bibinfo
  {volume} {16}},\ \bibinfo {pages} {5425} (\bibinfo {year}
  {2020})}\BibitemShut {NoStop}%
\bibitem [{\citenamefont {Tew}\ \emph {et~al.}(2011)\citenamefont {Tew},
  \citenamefont {Helmich},\ and\ \citenamefont {H{\"a}ttig}}]{tew2011local}%
  \BibitemOpen
  \bibfield  {author} {\bibinfo {author} {\bibfnamefont {D.~P.}\ \bibnamefont
  {Tew}}, \bibinfo {author} {\bibfnamefont {B.}~\bibnamefont {Helmich}}, \ and\
  \bibinfo {author} {\bibfnamefont {C.}~\bibnamefont {H{\"a}ttig}},\
  }\href@noop {} {\bibfield  {journal} {\bibinfo  {journal} {The Journal of
  chemical physics}\ }\textbf {\bibinfo {volume} {135}},\ \bibinfo {pages}
  {074107} (\bibinfo {year} {2011})}\BibitemShut {NoStop}%
\bibitem [{\citenamefont {Schmitz}\ \emph {et~al.}(2014)\citenamefont
  {Schmitz}, \citenamefont {H{\"a}ttig},\ and\ \citenamefont
  {Tew}}]{schmitz2014explicitly}%
  \BibitemOpen
  \bibfield  {author} {\bibinfo {author} {\bibfnamefont {G.}~\bibnamefont
  {Schmitz}}, \bibinfo {author} {\bibfnamefont {C.}~\bibnamefont {H{\"a}ttig}},
  \ and\ \bibinfo {author} {\bibfnamefont {D.~P.}\ \bibnamefont {Tew}},\
  }\href@noop {} {\bibfield  {journal} {\bibinfo  {journal} {Physical Chemistry
  Chemical Physics}\ }\textbf {\bibinfo {volume} {16}},\ \bibinfo {pages}
  {22167} (\bibinfo {year} {2014})}\BibitemShut {NoStop}%
\bibitem [{\citenamefont {L{\"{o}}wdin}(1955)}]{Lowdin1955}%
  \BibitemOpen
  \bibfield  {author} {\bibinfo {author} {\bibfnamefont {P.~O.}\ \bibnamefont
  {L{\"{o}}wdin}},\ }\href {\doibase 10.1103/PhysRev.97.1474} {\bibfield
  {journal} {\bibinfo  {journal} {Physical Review}\ }\textbf {\bibinfo {volume}
  {97}},\ \bibinfo {pages} {1474} (\bibinfo {year} {1955})}\BibitemShut
  {NoStop}%
\bibitem [{\citenamefont {Kottmann}\ and\ \citenamefont
  {Aspuru-Guzik}(2022)}]{kottmann2021optimized}%
  \BibitemOpen
  \bibfield  {author} {\bibinfo {author} {\bibfnamefont {J.~S.}\ \bibnamefont
  {Kottmann}}\ and\ \bibinfo {author} {\bibfnamefont {A.}~\bibnamefont
  {Aspuru-Guzik}},\ }\href@noop {} {\bibfield  {journal} {\bibinfo  {journal}
  {Physical Review A}\ }\textbf {\bibinfo {volume} {105}},\ \bibinfo {pages}
  {032449} (\bibinfo {year} {2022})}\BibitemShut {NoStop}%
\bibitem [{Note3()}]{Note3}%
  \BibitemOpen
  \bibinfo {note} {\protect \textsc {Psi4}: Currently, CABS-functionality is
  not available in the main repository; a hacky but ready-to-use implementation
  is accessible at \protect \url
  {https://github.com/philipp-q/psi4/tree/ri_space}. For \protect \textsc
  {Madness}, we use the forked branch \protect \url
  {https://github.com/kottmanj/madness/tree/pno_integrals_cabs}, which is
  described in more detail in Ref.~\cite {Kottmann2020a}. Installation
  instructions for \protect \textsc {Madness} together with \protect \textsc
  {Tequila} can be found at \protect \url
  {https://github.com/kottmanj/madness/tree/tequila}. Integration into the main
  repository is planned here.}\BibitemShut {Stop}%
\bibitem [{\citenamefont {McClean}\ \emph
  {et~al.}(2020{\natexlab{b}})\citenamefont {McClean}, \citenamefont {Rubin},
  \citenamefont {Sung}, \citenamefont {Kivlichan}, \citenamefont
  {Bonet-Monroig}, \citenamefont {Cao}, \citenamefont {Dai}, \citenamefont
  {Fried}, \citenamefont {Gidney}, \citenamefont {Gimby}, \citenamefont
  {Gokhale}, \citenamefont {Haner}, \citenamefont {Hardikar}, \citenamefont
  {Havl{\'{i}}{\v{c}}ek}, \citenamefont {Higgott}, \citenamefont {Huang},
  \citenamefont {Izaac}, \citenamefont {Jiang}, \citenamefont {Liu},
  \citenamefont {Mcardle}, \citenamefont {Neeley}, \citenamefont {O'Brien},
  \citenamefont {O'Gorman}, \citenamefont {Ozfidan}, \citenamefont {Radin},
  \citenamefont {Romero}, \citenamefont {Sawaya}, \citenamefont {Senjean},
  \citenamefont {Setia}, \citenamefont {Sim}, \citenamefont {Steiger},
  \citenamefont {Steudtner}, \citenamefont {Sun}, \citenamefont {Sun},
  \citenamefont {Wang}, \citenamefont {Zhang},\ and\ \citenamefont
  {Babbush}}]{McClean2017a}%
  \BibitemOpen
  \bibfield  {author} {\bibinfo {author} {\bibfnamefont {J.~R.}\ \bibnamefont
  {McClean}}, \bibinfo {author} {\bibfnamefont {N.~C.}\ \bibnamefont {Rubin}},
  \bibinfo {author} {\bibfnamefont {K.~J.}\ \bibnamefont {Sung}}, \bibinfo
  {author} {\bibfnamefont {I.~D.}\ \bibnamefont {Kivlichan}}, \bibinfo {author}
  {\bibfnamefont {X.}~\bibnamefont {Bonet-Monroig}}, \bibinfo {author}
  {\bibfnamefont {Y.}~\bibnamefont {Cao}}, \bibinfo {author} {\bibfnamefont
  {C.}~\bibnamefont {Dai}}, \bibinfo {author} {\bibfnamefont {E.~S.}\
  \bibnamefont {Fried}}, \bibinfo {author} {\bibfnamefont {C.}~\bibnamefont
  {Gidney}}, \bibinfo {author} {\bibfnamefont {B.}~\bibnamefont {Gimby}},
  \bibinfo {author} {\bibfnamefont {P.}~\bibnamefont {Gokhale}}, \bibinfo
  {author} {\bibfnamefont {T.}~\bibnamefont {Haner}}, \bibinfo {author}
  {\bibfnamefont {T.}~\bibnamefont {Hardikar}}, \bibinfo {author}
  {\bibfnamefont {V.}~\bibnamefont {Havl{\'{i}}{\v{c}}ek}}, \bibinfo {author}
  {\bibfnamefont {O.}~\bibnamefont {Higgott}}, \bibinfo {author} {\bibfnamefont
  {C.}~\bibnamefont {Huang}}, \bibinfo {author} {\bibfnamefont
  {J.}~\bibnamefont {Izaac}}, \bibinfo {author} {\bibfnamefont
  {Z.}~\bibnamefont {Jiang}}, \bibinfo {author} {\bibfnamefont
  {X.}~\bibnamefont {Liu}}, \bibinfo {author} {\bibfnamefont {S.}~\bibnamefont
  {Mcardle}}, \bibinfo {author} {\bibfnamefont {M.}~\bibnamefont {Neeley}},
  \bibinfo {author} {\bibfnamefont {T.}~\bibnamefont {O'Brien}}, \bibinfo
  {author} {\bibfnamefont {B.}~\bibnamefont {O'Gorman}}, \bibinfo {author}
  {\bibfnamefont {I.}~\bibnamefont {Ozfidan}}, \bibinfo {author} {\bibfnamefont
  {M.~D.}\ \bibnamefont {Radin}}, \bibinfo {author} {\bibfnamefont
  {J.}~\bibnamefont {Romero}}, \bibinfo {author} {\bibfnamefont {N.~P.}\
  \bibnamefont {Sawaya}}, \bibinfo {author} {\bibfnamefont {B.}~\bibnamefont
  {Senjean}}, \bibinfo {author} {\bibfnamefont {K.}~\bibnamefont {Setia}},
  \bibinfo {author} {\bibfnamefont {S.}~\bibnamefont {Sim}}, \bibinfo {author}
  {\bibfnamefont {D.~S.}\ \bibnamefont {Steiger}}, \bibinfo {author}
  {\bibfnamefont {M.}~\bibnamefont {Steudtner}}, \bibinfo {author}
  {\bibfnamefont {Q.}~\bibnamefont {Sun}}, \bibinfo {author} {\bibfnamefont
  {W.}~\bibnamefont {Sun}}, \bibinfo {author} {\bibfnamefont {D.}~\bibnamefont
  {Wang}}, \bibinfo {author} {\bibfnamefont {F.}~\bibnamefont {Zhang}}, \ and\
  \bibinfo {author} {\bibfnamefont {R.}~\bibnamefont {Babbush}},\ }\href
  {\doibase 10.1088/2058-9565/ab8ebc} {\bibfield  {journal} {\bibinfo
  {journal} {Quantum Science and Technology}\ }\textbf {\bibinfo {volume} {5}}
  (\bibinfo {year} {2020}{\natexlab{b}}),\ 10.1088/2058-9565/ab8ebc},\ \Eprint
  {http://arxiv.org/abs/1710.07629} {arXiv:1710.07629} \BibitemShut {NoStop}%
\bibitem [{\citenamefont {Suzuki}\ \emph {et~al.}(2021)\citenamefont {Suzuki},
  \citenamefont {Kawase}, \citenamefont {Masumura}, \citenamefont {Hiraga},
  \citenamefont {Nakadai}, \citenamefont {Chen}, \citenamefont {Nakanishi},
  \citenamefont {Mitarai}, \citenamefont {Imai}, \citenamefont {Tamiya} \emph
  {et~al.}}]{suzuki2020qulacs}%
  \BibitemOpen
  \bibfield  {author} {\bibinfo {author} {\bibfnamefont {Y.}~\bibnamefont
  {Suzuki}}, \bibinfo {author} {\bibfnamefont {Y.}~\bibnamefont {Kawase}},
  \bibinfo {author} {\bibfnamefont {Y.}~\bibnamefont {Masumura}}, \bibinfo
  {author} {\bibfnamefont {Y.}~\bibnamefont {Hiraga}}, \bibinfo {author}
  {\bibfnamefont {M.}~\bibnamefont {Nakadai}}, \bibinfo {author} {\bibfnamefont
  {J.}~\bibnamefont {Chen}}, \bibinfo {author} {\bibfnamefont {K.~M.}\
  \bibnamefont {Nakanishi}}, \bibinfo {author} {\bibfnamefont {K.}~\bibnamefont
  {Mitarai}}, \bibinfo {author} {\bibfnamefont {R.}~\bibnamefont {Imai}},
  \bibinfo {author} {\bibfnamefont {S.}~\bibnamefont {Tamiya}},  \emph
  {et~al.},\ }\href@noop {} {\bibfield  {journal} {\bibinfo  {journal}
  {Quantum}\ }\textbf {\bibinfo {volume} {5}},\ \bibinfo {pages} {559}
  (\bibinfo {year} {2021})}\BibitemShut {NoStop}%
\bibitem [{\citenamefont {Bradbury}\ \emph {et~al.}(2020)\citenamefont
  {Bradbury}, \citenamefont {Frostig}, \citenamefont {Hawkins}, \citenamefont
  {Johnson}, \citenamefont {Leary}, \citenamefont {Maclaurin},\ and\
  \citenamefont {Wanderman-Milne}}]{jax2018github}%
  \BibitemOpen
  \bibfield  {author} {\bibinfo {author} {\bibfnamefont {J.}~\bibnamefont
  {Bradbury}}, \bibinfo {author} {\bibfnamefont {R.}~\bibnamefont {Frostig}},
  \bibinfo {author} {\bibfnamefont {P.}~\bibnamefont {Hawkins}}, \bibinfo
  {author} {\bibfnamefont {M.~J.}\ \bibnamefont {Johnson}}, \bibinfo {author}
  {\bibfnamefont {C.}~\bibnamefont {Leary}}, \bibinfo {author} {\bibfnamefont
  {D.}~\bibnamefont {Maclaurin}}, \ and\ \bibinfo {author} {\bibfnamefont
  {S.}~\bibnamefont {Wanderman-Milne}},\ }\href@noop {} {\bibfield  {journal}
  {\bibinfo  {journal} {URL http://github. com/google/jax}\ }\textbf {\bibinfo
  {volume} {4}},\ \bibinfo {pages} {16} (\bibinfo {year} {2020})}\BibitemShut
  {NoStop}%
\bibitem [{\citenamefont {Virtanen}\ \emph {et~al.}(2020)\citenamefont
  {Virtanen}, \citenamefont {Gommers}, \citenamefont {Oliphant}, \citenamefont
  {Haberland}, \citenamefont {Reddy}, \citenamefont {Cournapeau}, \citenamefont
  {Burovski}, \citenamefont {Peterson}, \citenamefont {Weckesser},
  \citenamefont {Bright}, \citenamefont {van~der Walt}, \citenamefont {Brett},
  \citenamefont {Wilson}, \citenamefont {Millman}, \citenamefont {Mayorov},
  \citenamefont {Nelson}, \citenamefont {Jones}, \citenamefont {Kern},
  \citenamefont {Larson}, \citenamefont {Carey}, \citenamefont {Polat},
  \citenamefont {Feng}, \citenamefont {Moore}, \citenamefont {VanderPlas},
  \citenamefont {Laxalde}, \citenamefont {Perktold}, \citenamefont {Cimrman},
  \citenamefont {Henriksen}, \citenamefont {Quintero}, \citenamefont {Harris},
  \citenamefont {Archibald}, \citenamefont {Ribeiro}, \citenamefont
  {Pedregosa}, \citenamefont {van Mulbregt}, \citenamefont {Vijaykumar},
  \citenamefont {Bardelli}, \citenamefont {Rothberg}, \citenamefont {Hilboll},
  \citenamefont {Kloeckner}, \citenamefont {Scopatz}, \citenamefont {Lee},
  \citenamefont {Rokem}, \citenamefont {Woods}, \citenamefont {Fulton},
  \citenamefont {Masson}, \citenamefont {H{\"a}ggstr{\"o}m}, \citenamefont
  {Fitzgerald}, \citenamefont {Nicholson}, \citenamefont {Hagen}, \citenamefont
  {Pasechnik}, \citenamefont {Olivetti}, \citenamefont {Martin}, \citenamefont
  {Wieser}, \citenamefont {Silva}, \citenamefont {Lenders}, \citenamefont
  {Wilhelm}, \citenamefont {Young}, \citenamefont {Price}, \citenamefont
  {Ingold}, \citenamefont {Allen}, \citenamefont {Lee}, \citenamefont {Audren},
  \citenamefont {Probst}, \citenamefont {Dietrich}, \citenamefont {Silterra},
  \citenamefont {Webber}, \citenamefont {Slavi{\v{c}}}, \citenamefont
  {Nothman}, \citenamefont {Buchner}, \citenamefont {Kulick}, \citenamefont
  {Sch{\"o}nberger}, \citenamefont {de~Miranda~Cardoso}, \citenamefont
  {Reimer}, \citenamefont {Harrington}, \citenamefont {Rodr{\'i}guez},
  \citenamefont {Nunez-Iglesias}, \citenamefont {Kuczynski}, \citenamefont
  {Tritz}, \citenamefont {Thoma}, \citenamefont {Newville}, \citenamefont
  {K{\"u}mmerer}, \citenamefont {Bolingbroke}, \citenamefont {Tartre},
  \citenamefont {Pak}, \citenamefont {Smith}, \citenamefont {Nowaczyk},
  \citenamefont {Shebanov}, \citenamefont {Pavlyk}, \citenamefont {Brodtkorb},
  \citenamefont {Lee}, \citenamefont {McGibbon}, \citenamefont {Feldbauer},
  \citenamefont {Lewis}, \citenamefont {Tygier}, \citenamefont {Sievert},
  \citenamefont {Vigna}, \citenamefont {Peterson}, \citenamefont {More},
  \citenamefont {Pudlik}, \citenamefont {Oshima}, \citenamefont {Pingel},
  \citenamefont {Robitaille}, \citenamefont {Spura}, \citenamefont {Jones},
  \citenamefont {Cera}, \citenamefont {Leslie}, \citenamefont {Zito},
  \citenamefont {Krauss}, \citenamefont {Upadhyay}, \citenamefont {Halchenko},
  \citenamefont {V{\'a}zquez-Baeza},\ and\ \citenamefont
  {1.0~Contributors}}]{virtanen2020scipy}%
  \BibitemOpen
  \bibfield  {author} {\bibinfo {author} {\bibfnamefont {P.}~\bibnamefont
  {Virtanen}}, \bibinfo {author} {\bibfnamefont {R.}~\bibnamefont {Gommers}},
  \bibinfo {author} {\bibfnamefont {T.~E.}\ \bibnamefont {Oliphant}}, \bibinfo
  {author} {\bibfnamefont {M.}~\bibnamefont {Haberland}}, \bibinfo {author}
  {\bibfnamefont {T.}~\bibnamefont {Reddy}}, \bibinfo {author} {\bibfnamefont
  {D.}~\bibnamefont {Cournapeau}}, \bibinfo {author} {\bibfnamefont
  {E.}~\bibnamefont {Burovski}}, \bibinfo {author} {\bibfnamefont
  {P.}~\bibnamefont {Peterson}}, \bibinfo {author} {\bibfnamefont
  {W.}~\bibnamefont {Weckesser}}, \bibinfo {author} {\bibfnamefont
  {J.}~\bibnamefont {Bright}}, \bibinfo {author} {\bibfnamefont {S.~J.}\
  \bibnamefont {van~der Walt}}, \bibinfo {author} {\bibfnamefont
  {M.}~\bibnamefont {Brett}}, \bibinfo {author} {\bibfnamefont
  {J.}~\bibnamefont {Wilson}}, \bibinfo {author} {\bibfnamefont {K.~J.}\
  \bibnamefont {Millman}}, \bibinfo {author} {\bibfnamefont {N.}~\bibnamefont
  {Mayorov}}, \bibinfo {author} {\bibfnamefont {A.~R.~J.}\ \bibnamefont
  {Nelson}}, \bibinfo {author} {\bibfnamefont {E.}~\bibnamefont {Jones}},
  \bibinfo {author} {\bibfnamefont {R.}~\bibnamefont {Kern}}, \bibinfo {author}
  {\bibfnamefont {E.}~\bibnamefont {Larson}}, \bibinfo {author} {\bibfnamefont
  {C.~J.}\ \bibnamefont {Carey}}, \bibinfo {author} {\bibfnamefont
  {{\.{I}}.}~\bibnamefont {Polat}}, \bibinfo {author} {\bibfnamefont
  {Y.}~\bibnamefont {Feng}}, \bibinfo {author} {\bibfnamefont {E.~W.}\
  \bibnamefont {Moore}}, \bibinfo {author} {\bibfnamefont {J.}~\bibnamefont
  {VanderPlas}}, \bibinfo {author} {\bibfnamefont {D.}~\bibnamefont {Laxalde}},
  \bibinfo {author} {\bibfnamefont {J.}~\bibnamefont {Perktold}}, \bibinfo
  {author} {\bibfnamefont {R.}~\bibnamefont {Cimrman}}, \bibinfo {author}
  {\bibfnamefont {I.}~\bibnamefont {Henriksen}}, \bibinfo {author}
  {\bibfnamefont {E.~A.}\ \bibnamefont {Quintero}}, \bibinfo {author}
  {\bibfnamefont {C.~R.}\ \bibnamefont {Harris}}, \bibinfo {author}
  {\bibfnamefont {A.~M.}\ \bibnamefont {Archibald}}, \bibinfo {author}
  {\bibfnamefont {A.~H.}\ \bibnamefont {Ribeiro}}, \bibinfo {author}
  {\bibfnamefont {F.}~\bibnamefont {Pedregosa}}, \bibinfo {author}
  {\bibfnamefont {P.}~\bibnamefont {van Mulbregt}}, \bibinfo {author}
  {\bibfnamefont {A.}~\bibnamefont {Vijaykumar}}, \bibinfo {author}
  {\bibfnamefont {A.~P.}\ \bibnamefont {Bardelli}}, \bibinfo {author}
  {\bibfnamefont {A.}~\bibnamefont {Rothberg}}, \bibinfo {author}
  {\bibfnamefont {A.}~\bibnamefont {Hilboll}}, \bibinfo {author} {\bibfnamefont
  {A.}~\bibnamefont {Kloeckner}}, \bibinfo {author} {\bibfnamefont
  {A.}~\bibnamefont {Scopatz}}, \bibinfo {author} {\bibfnamefont
  {A.}~\bibnamefont {Lee}}, \bibinfo {author} {\bibfnamefont {A.}~\bibnamefont
  {Rokem}}, \bibinfo {author} {\bibfnamefont {C.~N.}\ \bibnamefont {Woods}},
  \bibinfo {author} {\bibfnamefont {C.}~\bibnamefont {Fulton}}, \bibinfo
  {author} {\bibfnamefont {C.}~\bibnamefont {Masson}}, \bibinfo {author}
  {\bibfnamefont {C.}~\bibnamefont {H{\"a}ggstr{\"o}m}}, \bibinfo {author}
  {\bibfnamefont {C.}~\bibnamefont {Fitzgerald}}, \bibinfo {author}
  {\bibfnamefont {D.~A.}\ \bibnamefont {Nicholson}}, \bibinfo {author}
  {\bibfnamefont {D.~R.}\ \bibnamefont {Hagen}}, \bibinfo {author}
  {\bibfnamefont {D.~V.}\ \bibnamefont {Pasechnik}}, \bibinfo {author}
  {\bibfnamefont {E.}~\bibnamefont {Olivetti}}, \bibinfo {author}
  {\bibfnamefont {E.}~\bibnamefont {Martin}}, \bibinfo {author} {\bibfnamefont
  {E.}~\bibnamefont {Wieser}}, \bibinfo {author} {\bibfnamefont
  {F.}~\bibnamefont {Silva}}, \bibinfo {author} {\bibfnamefont
  {F.}~\bibnamefont {Lenders}}, \bibinfo {author} {\bibfnamefont
  {F.}~\bibnamefont {Wilhelm}}, \bibinfo {author} {\bibfnamefont
  {G.}~\bibnamefont {Young}}, \bibinfo {author} {\bibfnamefont {G.~A.}\
  \bibnamefont {Price}}, \bibinfo {author} {\bibfnamefont {G.-L.}\ \bibnamefont
  {Ingold}}, \bibinfo {author} {\bibfnamefont {G.~E.}\ \bibnamefont {Allen}},
  \bibinfo {author} {\bibfnamefont {G.~R.}\ \bibnamefont {Lee}}, \bibinfo
  {author} {\bibfnamefont {H.}~\bibnamefont {Audren}}, \bibinfo {author}
  {\bibfnamefont {I.}~\bibnamefont {Probst}}, \bibinfo {author} {\bibfnamefont
  {J.~P.}\ \bibnamefont {Dietrich}}, \bibinfo {author} {\bibfnamefont
  {J.}~\bibnamefont {Silterra}}, \bibinfo {author} {\bibfnamefont {J.~T.}\
  \bibnamefont {Webber}}, \bibinfo {author} {\bibfnamefont {J.}~\bibnamefont
  {Slavi{\v{c}}}}, \bibinfo {author} {\bibfnamefont {J.}~\bibnamefont
  {Nothman}}, \bibinfo {author} {\bibfnamefont {J.}~\bibnamefont {Buchner}},
  \bibinfo {author} {\bibfnamefont {J.}~\bibnamefont {Kulick}}, \bibinfo
  {author} {\bibfnamefont {J.~L.}\ \bibnamefont {Sch{\"o}nberger}}, \bibinfo
  {author} {\bibfnamefont {J.~V.}\ \bibnamefont {de~Miranda~Cardoso}}, \bibinfo
  {author} {\bibfnamefont {J.}~\bibnamefont {Reimer}}, \bibinfo {author}
  {\bibfnamefont {J.}~\bibnamefont {Harrington}}, \bibinfo {author}
  {\bibfnamefont {J.~L.~C.}\ \bibnamefont {Rodr{\'i}guez}}, \bibinfo {author}
  {\bibfnamefont {J.}~\bibnamefont {Nunez-Iglesias}}, \bibinfo {author}
  {\bibfnamefont {J.}~\bibnamefont {Kuczynski}}, \bibinfo {author}
  {\bibfnamefont {K.}~\bibnamefont {Tritz}}, \bibinfo {author} {\bibfnamefont
  {M.}~\bibnamefont {Thoma}}, \bibinfo {author} {\bibfnamefont
  {M.}~\bibnamefont {Newville}}, \bibinfo {author} {\bibfnamefont
  {M.}~\bibnamefont {K{\"u}mmerer}}, \bibinfo {author} {\bibfnamefont
  {M.}~\bibnamefont {Bolingbroke}}, \bibinfo {author} {\bibfnamefont
  {M.}~\bibnamefont {Tartre}}, \bibinfo {author} {\bibfnamefont
  {M.}~\bibnamefont {Pak}}, \bibinfo {author} {\bibfnamefont {N.~J.}\
  \bibnamefont {Smith}}, \bibinfo {author} {\bibfnamefont {N.}~\bibnamefont
  {Nowaczyk}}, \bibinfo {author} {\bibfnamefont {N.}~\bibnamefont {Shebanov}},
  \bibinfo {author} {\bibfnamefont {O.}~\bibnamefont {Pavlyk}}, \bibinfo
  {author} {\bibfnamefont {P.~A.}\ \bibnamefont {Brodtkorb}}, \bibinfo {author}
  {\bibfnamefont {P.}~\bibnamefont {Lee}}, \bibinfo {author} {\bibfnamefont
  {R.~T.}\ \bibnamefont {McGibbon}}, \bibinfo {author} {\bibfnamefont
  {R.}~\bibnamefont {Feldbauer}}, \bibinfo {author} {\bibfnamefont
  {S.}~\bibnamefont {Lewis}}, \bibinfo {author} {\bibfnamefont
  {S.}~\bibnamefont {Tygier}}, \bibinfo {author} {\bibfnamefont
  {S.}~\bibnamefont {Sievert}}, \bibinfo {author} {\bibfnamefont
  {S.}~\bibnamefont {Vigna}}, \bibinfo {author} {\bibfnamefont
  {S.}~\bibnamefont {Peterson}}, \bibinfo {author} {\bibfnamefont
  {S.}~\bibnamefont {More}}, \bibinfo {author} {\bibfnamefont {T.}~\bibnamefont
  {Pudlik}}, \bibinfo {author} {\bibfnamefont {T.}~\bibnamefont {Oshima}},
  \bibinfo {author} {\bibfnamefont {T.~J.}\ \bibnamefont {Pingel}}, \bibinfo
  {author} {\bibfnamefont {T.~P.}\ \bibnamefont {Robitaille}}, \bibinfo
  {author} {\bibfnamefont {T.}~\bibnamefont {Spura}}, \bibinfo {author}
  {\bibfnamefont {T.~R.}\ \bibnamefont {Jones}}, \bibinfo {author}
  {\bibfnamefont {T.}~\bibnamefont {Cera}}, \bibinfo {author} {\bibfnamefont
  {T.}~\bibnamefont {Leslie}}, \bibinfo {author} {\bibfnamefont
  {T.}~\bibnamefont {Zito}}, \bibinfo {author} {\bibfnamefont {T.}~\bibnamefont
  {Krauss}}, \bibinfo {author} {\bibfnamefont {U.}~\bibnamefont {Upadhyay}},
  \bibinfo {author} {\bibfnamefont {Y.~O.}\ \bibnamefont {Halchenko}}, \bibinfo
  {author} {\bibfnamefont {Y.}~\bibnamefont {V{\'a}zquez-Baeza}}, \ and\
  \bibinfo {author} {\bibfnamefont {S.}~\bibnamefont {1.0~Contributors}},\
  }\href {\doibase 10.1038/s41592-019-0686-2} {\bibfield  {journal} {\bibinfo
  {journal} {Nature Methods}\ }\textbf {\bibinfo {volume} {17}},\ \bibinfo
  {pages} {261} (\bibinfo {year} {2020})}\BibitemShut {NoStop}%
\bibitem [{\citenamefont {Dehesa}\ \emph {et~al.}(1992)\citenamefont {Dehesa},
  \citenamefont {Angulo}, \citenamefont {Koga},\ and\ \citenamefont
  {Matsui}}]{dehesa1992study}%
  \BibitemOpen
  \bibfield  {author} {\bibinfo {author} {\bibfnamefont {J.~S.}\ \bibnamefont
  {Dehesa}}, \bibinfo {author} {\bibfnamefont {J.~C.}\ \bibnamefont {Angulo}},
  \bibinfo {author} {\bibfnamefont {T.}~\bibnamefont {Koga}}, \ and\ \bibinfo
  {author} {\bibfnamefont {K.}~\bibnamefont {Matsui}},\ }\href {\doibase
  10.1007/BF01437514} {\bibfield  {journal} {\bibinfo  {journal} {Zeitschrift
  f{\"u}r Physik D Atoms, Molecules and Clusters}\ }\textbf {\bibinfo {volume}
  {25}},\ \bibinfo {pages} {9} (\bibinfo {year} {1992})}\BibitemShut {NoStop}%
\bibitem [{\citenamefont {Nakatsuji}(2012)}]{nakatsuji2012discovery}%
  \BibitemOpen
  \bibfield  {author} {\bibinfo {author} {\bibfnamefont {H.}~\bibnamefont
  {Nakatsuji}},\ }\href {\doibase 10.1021/ar200340j} {\bibfield  {journal}
  {\bibinfo  {journal} {Accounts of Chemical Research}\ }\textbf {\bibinfo
  {volume} {45}},\ \bibinfo {pages} {1480} (\bibinfo {year}
  {2012})}\BibitemShut {NoStop}%
\bibitem [{\citenamefont {Bischoff}(2014)}]{Bischoff2014a}%
  \BibitemOpen
  \bibfield  {author} {\bibinfo {author} {\bibfnamefont {F.~A.}\ \bibnamefont
  {Bischoff}},\ }\href {\doibase 10.1063/1.4901021} {\bibfield  {journal}
  {\bibinfo  {journal} {Journal of Chemical Physics}\ }\textbf {\bibinfo
  {volume} {141}},\ \bibinfo {pages} {184105} (\bibinfo {year}
  {2014})}\BibitemShut {NoStop}%
\bibitem [{\citenamefont {Roskop}\ \emph {et~al.}(2016)\citenamefont {Roskop},
  \citenamefont {Valeev}, \citenamefont {Carter}, \citenamefont {Gordon},\ and\
  \citenamefont {Windus}}]{roskop2016application}%
  \BibitemOpen
  \bibfield  {author} {\bibinfo {author} {\bibfnamefont {L.~B.}\ \bibnamefont
  {Roskop}}, \bibinfo {author} {\bibfnamefont {E.~F.}\ \bibnamefont {Valeev}},
  \bibinfo {author} {\bibfnamefont {E.~A.}\ \bibnamefont {Carter}}, \bibinfo
  {author} {\bibfnamefont {M.~S.}\ \bibnamefont {Gordon}}, \ and\ \bibinfo
  {author} {\bibfnamefont {T.~L.}\ \bibnamefont {Windus}},\ }\href {\doibase
  10.1021/acs.jctc.6b00315} {\bibfield  {journal} {\bibinfo  {journal} {Journal
  of Chemical Theory and Computation}\ }\textbf {\bibinfo {volume} {12}},\
  \bibinfo {pages} {3176} (\bibinfo {year} {2016})},\ \bibinfo {note} {pMID:
  27281508},\ \Eprint
  {http://arxiv.org/abs/https://doi.org/10.1021/acs.jctc.6b00315}
  {https://doi.org/10.1021/acs.jctc.6b00315} \BibitemShut {NoStop}%
\bibitem [{\citenamefont {Peng}\ \emph {et~al.}()\citenamefont {Peng},
  \citenamefont {Lewis}, \citenamefont {Wang}, \citenamefont {Clement},
  \citenamefont {Pavosevic}, \citenamefont {Zhang}, \citenamefont {Rishi},
  \citenamefont {Teke}, \citenamefont {Pierce}, \citenamefont {Calvin},
  \citenamefont {Kenny}, \citenamefont {Seidl}, \citenamefont {Janssen},\ and\
  \citenamefont {Valeev}}]{ChongPengCannadaLewisXiaoWangMarjoryClement}%
  \BibitemOpen
  \bibfield  {author} {\bibinfo {author} {\bibfnamefont {C.}~\bibnamefont
  {Peng}}, \bibinfo {author} {\bibfnamefont {C.}~\bibnamefont {Lewis}},
  \bibinfo {author} {\bibfnamefont {X.}~\bibnamefont {Wang}}, \bibinfo {author}
  {\bibfnamefont {M.}~\bibnamefont {Clement}}, \bibinfo {author} {\bibfnamefont
  {F.}~\bibnamefont {Pavosevic}}, \bibinfo {author} {\bibfnamefont
  {J.}~\bibnamefont {Zhang}}, \bibinfo {author} {\bibfnamefont
  {V.}~\bibnamefont {Rishi}}, \bibinfo {author} {\bibfnamefont
  {N.}~\bibnamefont {Teke}}, \bibinfo {author} {\bibfnamefont {K.}~\bibnamefont
  {Pierce}}, \bibinfo {author} {\bibfnamefont {J.}~\bibnamefont {Calvin}},
  \bibinfo {author} {\bibfnamefont {J.}~\bibnamefont {Kenny}}, \bibinfo
  {author} {\bibfnamefont {E.}~\bibnamefont {Seidl}}, \bibinfo {author}
  {\bibfnamefont {C.}~\bibnamefont {Janssen}}, \ and\ \bibinfo {author}
  {\bibfnamefont {E.}~\bibnamefont {Valeev}},\ }\href
  {http://github.com/ValeevGroup/mpqc} {\enquote {\bibinfo {title} {{The
  Massively Parallel Quantum Chemistry Program (MPQC)}},}\ }\BibitemShut
  {NoStop}%
\bibitem [{\citenamefont {Pavo{\v{s}}evi{\'c}}\ \emph
  {et~al.}(2014)\citenamefont {Pavo{\v{s}}evi{\'c}}, \citenamefont {Neese},\
  and\ \citenamefont {Valeev}}]{pavovsevic2014geminal}%
  \BibitemOpen
  \bibfield  {author} {\bibinfo {author} {\bibfnamefont {F.}~\bibnamefont
  {Pavo{\v{s}}evi{\'c}}}, \bibinfo {author} {\bibfnamefont {F.}~\bibnamefont
  {Neese}}, \ and\ \bibinfo {author} {\bibfnamefont {E.~F.}\ \bibnamefont
  {Valeev}},\ }\href@noop {} {\bibfield  {journal} {\bibinfo  {journal} {The
  Journal of chemical physics}\ }\textbf {\bibinfo {volume} {141}},\ \bibinfo
  {pages} {054106} (\bibinfo {year} {2014})}\BibitemShut {NoStop}%
\bibitem [{\citenamefont {Ponce}\ \emph {et~al.}(2019)\citenamefont {Ponce},
  \citenamefont {van Zon}, \citenamefont {Northrup}, \citenamefont {Gruner},
  \citenamefont {Chen}, \citenamefont {Ertinaz}, \citenamefont {Fedoseev},
  \citenamefont {Groer}, \citenamefont {Mao}, \citenamefont {Mundim} \emph
  {et~al.}}]{niagara1}%
  \BibitemOpen
  \bibfield  {author} {\bibinfo {author} {\bibfnamefont {M.}~\bibnamefont
  {Ponce}}, \bibinfo {author} {\bibfnamefont {R.}~\bibnamefont {van Zon}},
  \bibinfo {author} {\bibfnamefont {S.}~\bibnamefont {Northrup}}, \bibinfo
  {author} {\bibfnamefont {D.}~\bibnamefont {Gruner}}, \bibinfo {author}
  {\bibfnamefont {J.}~\bibnamefont {Chen}}, \bibinfo {author} {\bibfnamefont
  {F.}~\bibnamefont {Ertinaz}}, \bibinfo {author} {\bibfnamefont
  {A.}~\bibnamefont {Fedoseev}}, \bibinfo {author} {\bibfnamefont
  {L.}~\bibnamefont {Groer}}, \bibinfo {author} {\bibfnamefont
  {F.}~\bibnamefont {Mao}}, \bibinfo {author} {\bibfnamefont {B.~C.}\
  \bibnamefont {Mundim}},  \emph {et~al.},\ }in\ \href@noop {} {\emph {\bibinfo
  {booktitle} {Proceedings of the Practice and Experience in Advanced Research
  Computing on Rise of the Machines (learning)}}}\ (\bibinfo {year} {2019})\
  pp.\ \bibinfo {pages} {1--8}\BibitemShut {NoStop}%
\bibitem [{\citenamefont {Loken}\ \emph {et~al.}(2010)\citenamefont {Loken},
  \citenamefont {Gruner}, \citenamefont {Groer}, \citenamefont {Peltier},
  \citenamefont {Bunn}, \citenamefont {Craig}, \citenamefont {Henriques},
  \citenamefont {Dempsey}, \citenamefont {Yu}, \citenamefont {Chen} \emph
  {et~al.}}]{niagara2}%
  \BibitemOpen
  \bibfield  {author} {\bibinfo {author} {\bibfnamefont {C.}~\bibnamefont
  {Loken}}, \bibinfo {author} {\bibfnamefont {D.}~\bibnamefont {Gruner}},
  \bibinfo {author} {\bibfnamefont {L.}~\bibnamefont {Groer}}, \bibinfo
  {author} {\bibfnamefont {R.}~\bibnamefont {Peltier}}, \bibinfo {author}
  {\bibfnamefont {N.}~\bibnamefont {Bunn}}, \bibinfo {author} {\bibfnamefont
  {M.}~\bibnamefont {Craig}}, \bibinfo {author} {\bibfnamefont
  {T.}~\bibnamefont {Henriques}}, \bibinfo {author} {\bibfnamefont
  {J.}~\bibnamefont {Dempsey}}, \bibinfo {author} {\bibfnamefont {C.-H.}\
  \bibnamefont {Yu}}, \bibinfo {author} {\bibfnamefont {J.}~\bibnamefont
  {Chen}},  \emph {et~al.},\ }in\ \href@noop {} {\emph {\bibinfo {booktitle}
  {Journal of Physics-Conference Series}}},\ Vol.\ \bibinfo {volume} {256}\
  (\bibinfo {year} {2010})\ p.\ \bibinfo {pages} {012026}\BibitemShut {NoStop}%
\bibitem [{\citenamefont {May}\ \emph {et~al.}(2005)\citenamefont {May},
  \citenamefont {Valeev}, \citenamefont {Polly},\ and\ \citenamefont
  {Manby}}]{May2005}%
  \BibitemOpen
  \bibfield  {author} {\bibinfo {author} {\bibfnamefont {A.~J.}\ \bibnamefont
  {May}}, \bibinfo {author} {\bibfnamefont {E.}~\bibnamefont {Valeev}},
  \bibinfo {author} {\bibfnamefont {R.}~\bibnamefont {Polly}}, \ and\ \bibinfo
  {author} {\bibfnamefont {F.~R.}\ \bibnamefont {Manby}},\ }\href {\doibase
  10.1039/b507781h} {\bibfield  {journal} {\bibinfo  {journal} {Physical
  Chemistry Chemical Physics}\ }\textbf {\bibinfo {volume} {7}},\ \bibinfo
  {pages} {2710} (\bibinfo {year} {2005})}\BibitemShut {NoStop}%
\end{thebibliography}%


%

\appendix

\section{Code-Sample to calculate \trot with \textsc{Tequila}}\label{sec:codesample}
\begin{lstlisting}[language=python]
import tequila as tq

def compute_with_mra_pnos(act, ri, geometry):
    madroot = '''link to madness executable'''
    madnessinput = {"pnoint": {'cabs_option': 'pno', 'n_pno': act, 'gamma': 1.4, 'orthog': 'cholesky'}}
    # We use "act" orbitals for the reference and an RI space of "ri" orbitals
    # The n_pno in mol sets the overall number of PNOs to be generated, while the n_pno in the madnessinput sets the size of the reference space in the CABS+ procedure
    mol = tq.Molecule(name='molecule', geometry=geometry, basis_set='madness', n_pno=ri, active_orbitals=[i for i in range(act)], executable=executable, **madnessinput)
    H = mol.make_hamiltonian().simplify()
    U = mol.make_upccgsd_ansatz(name='SPA-UCCD')
    E = tq.ExpectationValue(H=H, U=U)
    # HF as initial values
    initial_values = {k: 0.0 for k in E.extract_variables()}
    result = tq.minimize(objective=E, method="bfgs", initial_values=initial_values)
    energy = result.energy
    angles = result.angles
    # Prepare information to build rdms -- this can also be handed over explicitly 
    rdminfo = {"U": U, "variables": angles}
    # Compute f12 correction, use full CABS
    dE = mol.perturbative_f12_correction(**rdminfo)

    return energy, dE


# Quick test with GBS and MRA-PNOs
res_pno = compute_with_mra_pnos(2, 5, 'H 0.0 0.0 0.0\nH 0.0 0.0 1.0')
print(res_pno) # (-1.12866, -0.00638), dim(RI)=6
\end{lstlisting}
\begin{lstlisting}[language=python]
import tequila as tq
import psi4

def compute_with_gbs(obs, cabs, geometry):
    # Currently: Need C1 symmetry for [2]R12 
    mol = tq.Molecule(geometry=geometry, basis_set = obs, point_group='c1')
    # Let's use psi4's "detci" to determine the RDMs -- generally, only CI-methods work right now
    psi4_method = 'detci'
    # Set the CABS-basis set for the correction
    cabs_opt= {'cabs_name': cabs}
    # Here, let's compute the RDMs explicity (but one can do the same as above and hand over this information via keyword arguments)
    mol.compute_rdms(psi4_method=psi4_method)
    energy = mol.logs[psi4_method].variables["CI TOTAL ENERGY"]
    dE = mol.perturbative_f12_correction(rdm1=mol.rdm1, rdm2=mol.rdm2, cabs_type='cabs+', gamma=1.4, cabs_options=cabs_opt)

    return energy, dE


# Quick test
res_gbs = compute_with_gbs('sto-3g', '6-31g', 'H 0.0 0.0 0.0\nH 0.0 0.0 1.0')
print(res_gbs) # (-1.10115, -0.00174), dim(RI)=6
\end{lstlisting}

\begin{figure*}[ht]
    \centering
    \begin{subfigure}{.38\linewidth}
        \begin{adjustbox}{width=\linewidth}
            \input{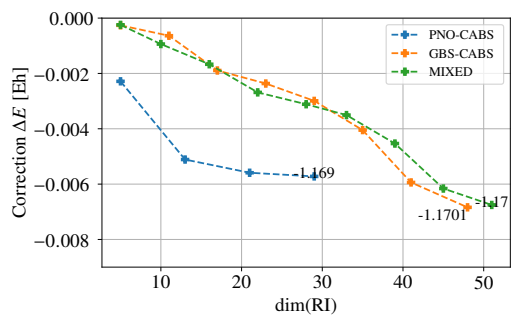}
        \end{adjustbox}
        \caption{H$_2$ with MP2 as surrogate model}
    \end{subfigure}
    \hfill
    \begin{subfigure}{.38\linewidth}
        \begin{adjustbox}{width=\linewidth}
            \input{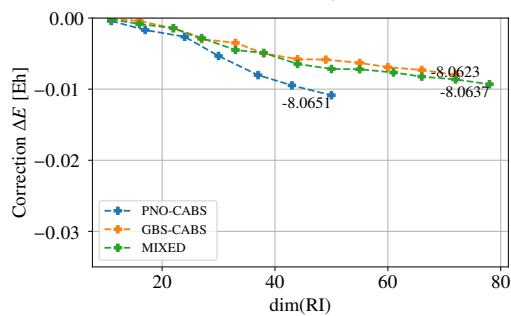}
        \end{adjustbox}
        \caption{H$_2$ with MP2-F12 as surrogate model}
    \end{subfigure}
    \begin{subfigure}{.38\linewidth}
        \begin{adjustbox}{width=\linewidth}
            \input{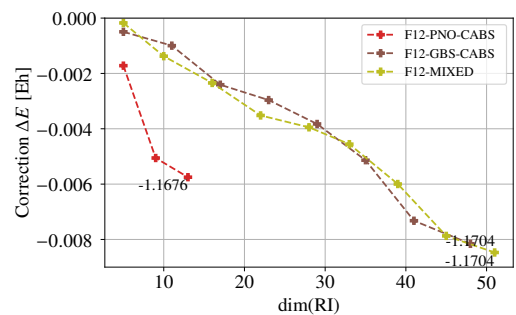}
        \end{adjustbox}
        \caption{LiH with MP2 as surrogate model}
    \end{subfigure}
    \hfill
    \begin{subfigure}{.38\linewidth}
        \begin{adjustbox}{width=\linewidth}
            \input{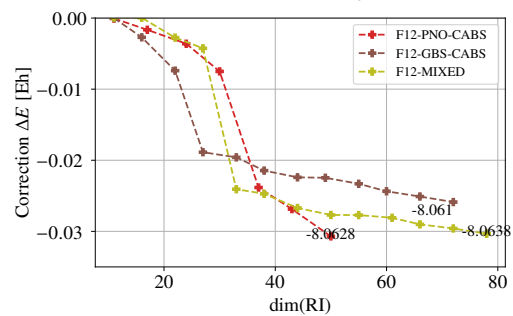}
        \end{adjustbox}
        \caption{LiH with MP2-R12 as surrogate model}
    \end{subfigure}
    \caption{Explicitly correlated \trot-correction over the dimensionality of RI basis for H$_2$ and LiH given a fixed OBS of MRA-PNOs, determined according to the \emph{good} procedure.  The final point is annotated with the corrected energy. H$_2$: \bas{cc-pVDZ-F12-OPTRI} for GBS-CABS and 3~PNOs added to the CABS in the mixed case. LiH:  \bas{def2-SVP-RIFIT} as GBS-CABS, the mixed procedure adds another 6~PNOs to the CABS.} 
        \label{fig:cabsconv}
\end{figure*}

\section{PNOs as CABS}\label{sec:pnocabs}
The following data comes from \cite{schleich2020regularization}. 

First, we take a look at the ``convergence'' of the \trot-correction with respect to the dimensionality of the RI space to allow a judgement whether additional F12-regularization of the MRA-PNOs is beneficial, and how MRA-PNOs perform as a CABS in comparison to designated Gaussian basis sets. Note that we do not consider the RI-error here as this is rather a proof-of-principle for the convergence behaviour of MRA-PNOs vs. GBS for CABS than actual production calculations. Results therefor are depicted in Fig.~\ref{fig:cabsconv}.

In the case of the hydrogen molecule, the best energies are achieved by the GBS-CABS, while the mixed-approach does not improve the result further.
This means that the GBS-CABS, \bas{cc-pVDZ-F12-OPTRI}, is likely to be very well suited as RI in this case, which is expected because it is optimized for this use.
The regularized equivalent yields a barely improved result.
Looking at the PNO-CABS, we see that the unregularized surrogate performs almost as well as the much larger GBS-CABS, while the PNO-CABS grown by MP2-R12 give a worse result despite a correction of almost the same magnitude using even less orbitals as CABS.
The reason thereof is \rv{-- and this is a crucial point --}{} that the reference is different.
In the case of MP2 as surrogate, both the PNOs for OBS and CABS are unregularized, and accordingly, for the MP2-F12 CABS, we use OBS that have been generated within the same process coming from MP2-F12.
We chose this approach because this way, the PNOs are generated within one consistent procedure and then easily combined by CABS+ \cite{Valeev2004a}. Different combinations, mixing MP2 and MP2-F12 for OBS and CABS, can be thought of as well but lead to a less efficient procedure that creates a fair amount of unnecessary orbitals that are to be projected out. 

Taking a look at the converse result for lithium hydride, we see that here, the MP2-F12 options do considerably worse than before. The higher magnitude of the correction is again related to the worse reference energy using PNOs built by MP2-F12. \rv{}{Building a higher-quality CABS in this case might account for that but comes at additional cost.} Further the choice of GBS-CABS (\bas{def2-SVP-RIFIT}) leads to generally low-performing GBS-CABS in this case, which is much less extensive than the we chose for the hydrogen molecule. The noticable improvement by the mixed approach in the case of MP2-F12 supports this. Yet still, the same basis set for LiH would amount to 110 basis functions, which is more than three times than for the PNO-CABS in this case.

To formulate an overall recommendation, we come back to the \trot-correction, which in the framework of \rv{}{Ref.~}\cite{Roskop2014} scales \rv{cubically}{cubically within the framework of approximation C~\cite{Kedzuch2005} or quadratically within approximation D~\cite{May2005,pavovsevic2016sparsemaps}} in the RI-dimensionality. \rv{}{Ref.~\cite{pavovsevic2016sparsemaps} proposes a method using PNOs to reduce the cost of MP2-F12 to quasi-linear dependence in the RI-basis except for some inexpensive quadratic terms; a similar approach could be employed to reduce this cost for \trot, too.} Given the seemingly almost equivalent performance of unregularized PNOs for OBS and CABS and the fact, that there are significantly less CABS functions in this case, this approach seems most desirable within this context. 



\end{document}